\documentclass[aps,prd,reprint,preprintnumbers,superscriptaddress,amsmath,amssymb,nofootinbib]{revtex4-2}
\pdfoutput=1
\usepackage{enumitem}
\usepackage{graphicx}
\usepackage{multirow}
\usepackage{color}
\usepackage{array}
\usepackage{hyperref}
\hypersetup{colorlinks=true,linkcolor=blue,citecolor=blue,urlcolor=blue}
\def\gev{\,\text{Ge\hspace{-0.1em}V}}
\def\mev{\,\text{Me\hspace{-0.1em}V}}
\def\fm{\,\text{fm}}

\newcommand{\avg}[1]{\left< #1 \right>}

\newcommand{\abs}[1]{\left| #1 \right|}
\def\w{\omega}
\def\qsqmax{q^2_\text{max}}
\def\qsqmin{q^2_\text{min}}
\newcommand{\matrixel}[3]{\left< #1 \vphantom{#2#3} \right| #2 \left| #3 \vphantom{#1#2} \right>} 
\newcommand{\mc}[1]{\ensuremath{\mathcal{ #1 }}}

\newcommand{\eqnref}[1]{(\ref{#1})}
\newcommand{\BstoKlnu}{B_s\to K\ell\nu}

\newcommand{\Fmpi}{268}

\begin{document}

\preprint{CERN-TH-2023-046~~~FERMILAB-PUB-23-115-V~~~P3H-23-017~~~SI-HEP-2023-06}

\title{Exclusive semileptonic \texorpdfstring{$\BstoKlnu$ }{Bs->K l nu}
 decays on the lattice}

\author{J.M.~Flynn}
\email{j.m.flynn@soton.ac.uk}
\affiliation{Physics and Astronomy, University of Southampton, Southampton SO17 1BJ, UK}
\affiliation{STAG Research Centre, University of Southampton, Southampton SO17 1BJ, UK}
\author{R.C.~Hill}
\email{ryan.hill@ed.ac.uk}
\affiliation{School of Physics and Astronomy, University of Edinburgh, Edinburgh EH9 3FD, United Kingdom}
\author{A.~J\"uttner}
\email{andreas.juttner@cern.ch}
\affiliation{Physics and Astronomy, University of Southampton, Southampton SO17 1BJ, UK}
\affiliation{Theoretical Physics Department, CERN, Geneva, Switzerland}
\affiliation{STAG Research Centre, University of Southampton, Southampton SO17 1BJ, UK}
\author{A.~Soni}
\email{adlersoni@gmail.com}
\affiliation{Physics Department, Brookhaven National Laboratory,
  Upton, NY 11973, USA}
\author{J.T.~Tsang}\thanks{Corresponding author}
\email{j.t.tsang@cern.ch}
\affiliation{Theoretical Physics Department, CERN, Geneva, Switzerland}
\author{O.~Witzel}
\email{oliver.witzel@uni-siegen.de}
\affiliation{Center for Particle Physics Siegen, Theoretische Physik 1, Naturwissenschaftlich-Technische Fakult\"at, Universit\"at Siegen, 57068 Siegen, Germany}

\collaboration{RBC/UKQCD}
\date{\today}

\begin{abstract}
Semileptonic $\BstoKlnu$ decays provide an alternative $b$-decay channel to
determine the CKM matrix element $|V_{ub}|$, and to obtain a $R$-ratio to
investigate lepton-flavor-universality violations. Results for the CKM matrix
element may also shed light on the discrepancies seen between analyses of
inclusive or exclusive decays. We calculate the decay form factors using lattice
QCD with domain-wall light quarks and a relativistic $b$-quark. We analyze data
at three lattice spacings with unitary pion masses down to $\Fmpi\mev$. Our
numerical results are interpolated/extrapolated to physical quark masses and to
the continuum to obtain the vector and scalar form factors $f_+(q^2)$ and
$f_0(q^2)$ with full error budgets at $q^2$ values spanning the range accessible
in our simulations.  We provide a possible explanation of tensions found between
results for the form factor from different lattice collaborations.  Model- and
truncation-independent $z$-parameterization fits following a recently proposed
Bayesian-inference approach extend our results to the entire allowed kinematic
range. Our results can be combined with experimental measurements of $B_s\to
D_s$ and $B_s\to K$ semileptonic decays to determine $|V_{ub}|=3.8(6)\times
10^{-3}$. The error is currently dominated by experiment. We compute differential branching fractions and two types of $R$
ratios, the one commonly used as well as a variant better suited to test
lepton-flavor universality.
\end{abstract}

\maketitle

\section{Introduction}

High-precision tests of the Standard Model (SM) from flavor physics are an
important complement to direct searches at colliders for new physics. The
absence of tree-level flavor-changing neutral currents in the SM provides one
window where new physics effects could be seen, though it is important to test
high-precision calculations of both tree- and loop-level SM processes against
experiment.

New physics is expected to occur at higher energy scales and seeing its effects
is more likely if the decaying particle can release large amounts of
energy. Decays of mesons containing a heavy $b$-quark provide many opportunities
because the $b$-quark lives long enough for experimental investigation but also
delivers more than $4\gev$ of energy. The large $b$-quark mass allows a plethora
of decay channels and correspondingly many tests of the SM. Tantalizing
deviations between SM predictions and experimental measurements have been
reported~\cite{LHCb:2013ghj, LHCb:2014vgu, BaBar:2012obs, BaBar:2013mob,
  Belle:2015qfa, LHCb:2015gmp}.  While the latest experimental LHCb
results~\cite{LHCb:2022zom} confirm the universality of lepton families under
the weak interaction, the high-energy reach of flavor physics continues to make
it a primary probe for the search of beyond-SM physics.

Here we focus on $B_s$-meson decay arising at tree level in the SM with a kaon
in the final state. Lattice results over the full $q^2$ range for $\BstoKlnu$
form factors have been obtained by HPQCD~\cite{Bouchard:2014ypa,
  Monahan:2018lzv}, RBC-UKQCD~\cite{Flynn:2015mha}, and
Fermilab-MILC~\cite{Bazavov:2019aom}, as well as at a single $q^2$ value by the
Alpha collaboration~\cite{Bahr:2016ayy}. Recent lattice results for $B_s\to
K\ell\nu$ semileptonic decays are summarized in Refs.~\cite{Witzel:2020msp,
  Lytle:2020tbe} and~\cite{FlavourLatticeAveragingGroupFLAG:2021npn}.

$\BstoKlnu$ decays have been observed by LHCb~\cite{Aaij:2020nvo} and
can be used to obtain the magnitudes of the CKM matrix elements $|V_{ub}|$.
Using these alternative $b$-decay channels may shed light on the use of
different kinematical parameterizations for the exclusive decay form factors and
on tensions within and between extractions of $|V_{ub}|$ from inclusive or
exclusive decays.

The remainder of this paper is structured as follows. In
Section~\ref{sec:latticecalc} we describe the form factors we will compute and
provide the details of our lattice computation. In Section~\ref{sec:analysis} we
describe the statistical analysis of the lattice data. In
Section~\ref{sec:syserrors} we determine estimates for all sources of
uncertainties and thereby assemble a complete error
budget. Section~\ref{sec:pheno} discusses the extrapolation of the obtained
form factors over the full kinematical range and provides a wealth of quantities
of phenomenological interest before we conclude in Section~\ref{sec:conclusion}.

\section{Lattice calculation}
\label{sec:latticecalc}

\subsection{Form factors}
We work in the $B_s$-meson rest frame for our calculations. For $\bar B^0_s\to
K^+\ell^-\bar\nu_\ell$, the differential decay rate in this frame is given by
\begin{align}
  \label{eq:B_semileptonic_rate}
  \frac{d\Gamma(B_s{\to} K\ell\nu)}{dq^2} =&
  \eta_\text{EW} \frac{G_F^2 |V_{ub}|^2}{24\pi^3} \,
  \frac{(q^2{-}m_\ell^2)^2 |{\bf  p}_K|}{(q^2)^2}\notag\\
    &\times\bigg[ \Big(1{+}\frac{m_\ell^2}{2q^2}\Big)
    |{\bf p}_K|^{2}|f_+(q^2)|^2\nonumber\\ 
    &+    \frac{3m_\ell^2}{8q^2}
      \frac{(M_{B_s}^2{-}M_K^2)^2}{M_{B_s}^2}|f_0(q^2)|^2
    \bigg]\,.
\end{align}
The $4$-momenta of the $B_s$ and the final-state kaon are denoted by $p_{B_s}$
and $p_K$ respectively. $M_{B_s}$ and $M_K$ denote the corresponding meson
masses, $E_K$ is the $K$-meson energy, $|{\bf p}_K| = (E_K^2-M_K^2)^{1/2}$, and
$q = p_{B_s}-p_K$ is the momentum transfer between the $B_s$ and $K$
mesons. $\eta_\text{EW}$ is an electroweak correction factor\footnote{We follow
  Ref.~\cite{Na:2015kha} and take $\eta_\text{EW}=1.011(5)$ by combining the
  factor computed by Sirlin~\cite{Sirlin:1981ie} with an estimate of final-state
  electromagnetic corrections using the ratio of signal yields from charged and
  neutral decay channels.}. The form factors $f_+$ and $f_0$ arise in the
decomposition of the QCD matrix element
\begin{multline}
  \langle K(p_K) |\mathcal{V}^\mu | B_s(p_{B_s}) \rangle=\\
  f_+(q^2)\bigg(p_K^\mu+p_{B_s}^\mu -q^\mu\frac{M_{B_s}^2-M_K^2}{q^2}\bigg)\\
    + f_0(q^2)\frac{M_{B_s}^2-M^2_K}{q^2}q^\mu.
\end{multline}
In the SM, $\mathcal V^\mu = \bar u \gamma^\mu b$ is the continuum charged
current operator. For lattice simulations it is convenient to use the
alternative parameterization
\begin{equation}
  \langle K |\mathcal{V}^\mu | B_s\rangle  = \sqrt{2M_{B_s}}\,
   \big[v^\mu f_\parallel(E_K)+p^\mu_\perp f_\perp(E_K)\big]\,,
\end{equation}
and the relations
\begin{align}
  \label{eq:f0fromfparfperp}
f_0(q^2) &= \frac{\sqrt{2M_{B_s}}}{M^2_{B_s}{-}M^2_K}\big[ (M_{B_s} {-}
  E_K)f_\parallel(E_K) \nonumber\\&\hspace{12ex}+ (E_K^2 {-} M^2_K)f_\perp(E_K)\big],\\
\label{eq:f+fromfparfperp}
f_+(q^2) &= \frac{1}{\sqrt{2M_{B_s}}}\left[f_\parallel(E_K) + (M_{B_s} -
  E_K)f_\perp(E_K)\right]\,,
\end{align}
where $v$ is the $B_s$ meson $4$-velocity, ${(p_{B_s})}^\mu_\perp\equiv p_K^\mu
- (p_K\cdot v)v^\mu$ and, in the $B_s$ meson rest frame,
\begin{equation}
  f_\parallel(E_K) =\frac{\langle K | \mathcal{V}^0 |
B_s\rangle}{\sqrt{2M_{B_s}}},\quad 
  f_\perp(E_K) = \frac1{p_K^i}\,\frac{\langle K | \mathcal{V}^i |
    B_s\rangle}{\sqrt{2M_{B_s}}}.
  \label{Eq.fpar_fperp}
\end{equation}
We note that no Einstein summation convention is applied in the second
equation; it holds component-by-component for the non-zero components of
${p}_K^i$.

\subsection{Action and parameters}
\begin{table*}
  \caption{RBC/UKQCD coarse (C), medium (M) and fine (F) gauge field
    ensembles~\cite{Allton:2008pn, Aoki:2010dy, Blum:2014tka, Boyle:2017jwu}
    used in this calculation, with $2+1$-flavor domain-wall fermions and
    Iwasaki-gauge action.  The domain-wall height for light and strange quarks
    is $M_5=1.8$.  The ensembles are generated using the Shamir domain-wall
    kernel~\cite{Shamir:1993zy, Furman:1994ky}. $am_l$ denotes the light
    sea-quark mass and $am_s^\text{sea}$ the strange sea-quark mass. The lattice
    spacing and physical strange quark were obtained in the combined analysis of
    Refs.~\cite{Blum:2014tka, Boyle:2017jwu, Boyle:2018knm}. The valence
    strange-quark masses used in our simulations on the C(M) ensembles are
    $am_s^\text{sim} = 0.03224\,(0.025)$, while on F1S we used $am_s^\text{sim}
    = am_s^\text{sea}$.}
  \label{tab:ensembles}
  \[
  \begin{array}{cccccccccccc}
  \hline\hline
  & L/a & T/a & L_s &  a^{-1}\!/\!\gev & am_l & am_s^\text{sea} & am_s^\text{phys}
  & M_\pi/\!\mev & \text{\# cfgs} & \text{\# sources}\\\hline
\text{C1}&24&64&16 &1.7848(50) & 0.005 & 0.040 & 0.03224(18) & 340 &1636 & 1\\
\text{C2}&24&64&16 &1.7848(50) & 0.010 & 0.040 & 0.03224(18) & 434 &1419 & 1\\[1.2ex]
\text{M1}&32&64&16 &2.3833(86) & 0.004 & 0.030 & 0.02477(18) & 301 &628  & 2\\
\text{M2}&32&64&16 &2.3833(86) & 0.006 & 0.030 & 0.02477(18) & 363 &889  & 2\\
\text{M3}&32&64&16 &2.3833(86) & 0.008 & 0.030 & 0.02477(18) & 411 &544  & 2\\[1.2ex]
\text{F1S}&48&96&12 &2.785(11) & 0.002144 & 0.02144 & 0.02167(20) & \Fmpi &98 & 24\\
\hline\hline
  \end{array}
  \]
\end{table*}

Our calculations are based on a subset of RBC-UKQCD's $2+1$ flavor domain-wall
fermion and Iwasaki gauge-field ensembles~\cite{Allton:2008pn, Aoki:2010dy,
  Blum:2014tka, Boyle:2017jwu} which we summarize in
Table~\ref{tab:ensembles}. Our dataset includes six ensembles at three different
lattice spacings, with pion masses down to $\Fmpi\mev$. Light and strange quarks
are simulated using domain-wall fermions~\cite{Kaplan:1992bt, Shamir:1993zy,
  Furman:1994ky, Blum:1996jf, Blum:1997mz, Brower:2012vk}.  The light (up and
down) sea-quark masses are degenerate and correspond to pion masses in the range
$\Fmpi\mev \le M_\pi \le 434\mev$. The strange sea-quark mass is within $20$ to
$25\%$ of its physical value on the~C and~M ensembles, and is tuned to a
deviation of $1\%$ on the F1S ensemble.

Bottom quarks are simulated using the relativistic heavy quark (RHQ)
action~\cite{Christ:2006us, Lin:2006ur}, a variant of the Fermilab
action~\cite{ElKhadra:1996mp} with three non-perturbatively tuned
parameters~\cite{Aoki:2012xaa}. The RHQ tuning determines the bare-quark mass,
$m_0a$, clover coefficient, $c_P$, and anisotropy parameter, $\zeta$, in the RHQ
action using as inputs the experimentally measured $B_s$-meson mass and
hyperfine splitting, together with the constraint that the lattice rest mass
(measured from the exponential decay of meson correlation functions) equals the
kinetic mass (measured from the meson dispersion relation). The outputs of the
RHQ tuning for the ensembles used here are given in
Table~\ref{tab:RHQparameter}, with details of the procedure in
appendix~\ref{appx:rhqtuning}.

Light and strange quarks are generated using point sources, whereas $b$ quarks
are generated using a Gaussian smeared source~\cite{Alford:1995dm,
  Lichtl:2006dt} in order to reduce excited-state contamination. We use the
smearing parameters that were determined in Ref.~\cite{Aoki:2012xaa} and list
them in Tab.~\ref{tab:TuneInputs} in appendix~\ref{appx:rhqtuning}.

Our data is generated using the lattice QCD software packages
\texttt{Qlua}~\cite{Pochinsky:2008zz}
and
\texttt{Chroma}~\cite{Edwards:2004sx}.
Further
details of the set-up can be found in Refs.~\cite{Christ:2014uea, Flynn:2015mha,
  Flynn:2015xna, Boyle:2017jwu, Boyle:2018knm}.

\subsection{Operator renormalization and improvement}
\label{sec:renormimprove}

Matrix elements from our simulations are matched to continuum ones using the
relation
\begin{equation}
  \langle K | \mathcal{V}_\mu | B_s \rangle = Z^{bl}_{V_{\mu}}\langle K |
  V_\mu | B_s\rangle\,,
  \label{eq:rendef}
\end{equation}
where $l$ stands for the light quark and we denote the continuum and lattice
currents by $\mathcal{V}_\mu$ and $V_\mu$, respectively. The renormalization
factor
\begin{equation}
  Z^{bl}_{V_\mu} = \rho^{bl}_{V_\mu}\sqrt{Z^{ll}_VZ^{bb}_V}\,,
  \label{eq:renorm}
\end{equation}
is obtained following Refs.~\cite{Hashimoto:1999yp, ElKhadra:2001rv}. The
flavor-conserving renormalization factors $Z^{ll}_V$ and $Z^{bb}_V$ account for
most of the operator renormalization, while $\rho^{bl}_{V_\mu}$ is a residual
correction, expected to be close to unity because most of the radiative
corrections, including tadpoles, cancel~\cite{Harada:2001fi}. We obtain
$Z^{bb}_V$ non-perturbatively from the matrix element of the $b\to b$ vector
current between two $B_s$ mesons:
\begin{equation}
  Z_V^{bb}\langle B_s|V_0|B_s \rangle = 2M_{B_s}.
  \label{eq:renbb}
\end{equation}
Details of the calculation are given in appendix~A
of Ref.~\cite{Christ:2014uea}. The light-light renormalization factor was
calculated from $Z_A^{ll}$~\cite{Aoki:2010dy} using the relation
$Z_V^{ll} =Z_A^{ll}$ to $O(am_{\text{res}})$ for domain-wall fermions.

Following our previous work~\cite{Flynn:2015mha}, we improve the heavy-light
vector current to $O(\alpha_s a)$ using the following operators
\begin{align}
 V_0^{\rm imp}(x) &= V_0^{0}(x) + c_t^{3} V_0^{3}(x) + c_t^{4}
 V_0^{4}(x), \label{eq:V0_imp} \\
 V_i^{\rm imp}(x) &= V_i^{0}(x) \label{eq:Vi_imp}\\
 &\phantom{={}}+ c_s^{1} V_i^{1}(x) + c_s^{2} V_i^{2}(x) 
 + c_s^{3} V_i^{3}(x) + c_s^{4} V_i^{4}(x) \,.\notag 
\end{align}
where $c_t^n$ and $c_s^n$ are coefficients and
\begin{align}
 V_\mu^{0}(x) &= \bar\psi(x)\gamma^\mu Q(x), \label{eq_V0}\\
 V_\mu^{1}(x) &= \bar\psi(x) 2\overrightarrow{D}_\mu Q(x), \\
 V_\mu^{2}(x) &= \bar\psi(x) 2\overleftarrow{D}_\mu Q(x), \\
 V_\mu^{3}(x) &= \bar\psi(x) 2\gamma_\mu \gamma_i \overrightarrow{D}_i Q(x), \\
 V_\mu^{4}(x) &= \bar\psi(x) 2\gamma_\mu \gamma_i \overleftarrow{D}_i Q(x),
    \label{eq_Vi}
\end{align}
with $\psi$ a light quark and $Q$ an RHQ quark. The covariant derivatives are given
by
\begin{align}
\overrightarrow{D}_\mu Q(x) &=  \frac12\big(U_\mu(x)Q(x+\hat\mu)
   - U_\mu^\dagger(x-\hat\mu) Q(x-\hat{\mu})\big), \\
   \bar\psi(x)\overleftarrow{D}_\mu &=
   \frac12\big(\bar\psi(x+\hat\mu)U_\mu^\dagger(x)
    - \bar\psi(x-\hat\mu)U_\mu(x-\hat\mu)\big).
\end{align}
Both the residual renormalization factor $\rho^{bl}_V$ and the coefficients
$c_t^n$ and $c_s^n$ were computed at one-loop order~\cite{CLehnerPT} in
mean-field improved lattice perturbation theory~\cite{Lepage:1992xa}, with the
results given in Table~\ref{tab:coeff} in Appendix~\ref{appx:PT-coeffts}.

\subsection{Setup of the calculation}
\begin{figure}
  \centering
  \includegraphics[width=0.7\columnwidth]{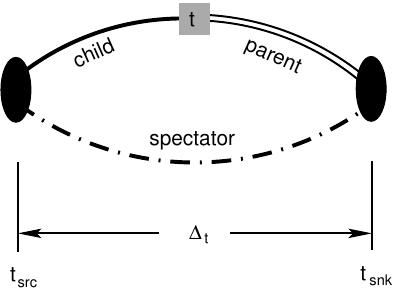}
  
  \caption{Sketch of the construction of the three-point function in the
    ``forward direction''.  The strange spectator quark (dash-dotted line)
    originates from time slice $t_\mathrm{src}$ and propagates forward to time
    slice $t_\text{snk}$ where we create a point sink and turn it into a
    sequential source for the parent $b$ quark (double line) propagating
    backwards.  This sequential propagator is contracted with the child light
    quark (solid line) also originating from $t_\mathrm{src}$. The contraction
    of child and parent quark (gray box) is calculated inserting the operators
    for the vector current and varied over the time slices $t_\mathrm{src}$ to
    $t_\text{snk}$.}
  \label{Fig.sketch3pt}
\end{figure}

In Fig.~\ref{Fig.sketch3pt} we sketch our calculation of three-point correlation
functions using sequential propagators separating initial and final state by
$\Delta_t$~\cite{Gottlieb:1983rh}.  The initial $B_s$ meson is located at
$t_\text{snk}=t_\mathrm{src}+\Delta_t$, whereas the final state kaon is at
$t_\mathrm{src}$. We proceed by letting the strange spectator quark propagate
from $t=t_\mathrm{src}$ to $t=t_\text{snk}$ where we give it a sink which is
turned into the sequential source for the parent $b$ quark. The $b$ quark is
contracted with a child light quark also starting at $t=t_\mathrm{src}$ over all time
slices in the range $t_\mathrm{src}\le t \le t_\text{snk}$.  The contraction is
calculated inserting the operators for the vector current defined in
Eqs.~(\ref{eq_V0})-(\ref{eq_Vi}). In addition to the shown setup in the
``forward'' direction, we effectively double statistics by calculating the
``backward'' direction, \emph{i.e.}~we use a second sink location at
$t'_\text{snk}=t_\mathrm{src}-\Delta_t$. For each configuration, we first
average forward and backward three-point correlators before proceeding with the
analysis. This step is similar to ``folding'' two-point correlators about the
central time slice in order to take advantage of the symmetry of forward- and
backward-propagating states.

We use one time source at $t_\mathrm{src}/a=0$ for the coarse C1 and C2
ensembles, two time sources at $t_\mathrm{src}/a=0,\,32$ for the medium M1, M2,
and M3 ensembles, and 24 time sources separated by four time slices on the F1S
ensemble.  The separation $\Delta_t$ on ensembles C1, C2, M1, M2 and M3 is the
same as in our previous work~\cite{Flynn:2015mha} where we carefully studied
several source-sink separations to cleanly identify the ground-state signal. On
the F1S ensemble we generated data for multiple source-sink separations
($\Delta_t/a = 30, 32, 34$) to study the effects of excited states. We found
that the signal saturates for $\Delta_t/a = 32$ with the centre of the signal
region being ground-state dominated and compatible with the $\Delta_t/a = 34$
dataset. We therefore chose $\Delta_t/a = 32$ for our final analysis. We keep
$\Delta_t$ nearly constant in physical units we choose $\Delta_t/a = 20$, 26,
and 32 for the three lattice spacings corresponding to the C, M, and F
ensembles, respectively. In order to further decorrelate measurements, we
perform a random 4-vector shift on the gauge field prior to placing any
source. This 4-vector shift is equivalent to randomly choosing the first source
position on each configuration but simplifies the bookkeeping.

The initial $B_s$ meson is kept at rest and momentum is inserted in the final
state through the current operator. The simulations use strange spectator quarks
with a mass close to the physical value (see Tab.~\ref{tab:ensembles}) and we
tune the RHQ parameters such that the parent quark corresponds to a physical $b$
quark. The child light quark is unitary and has the same mass as the light sea
quarks.

\section{Analysis}\label{sec:analysis}

The analysis to extract form factor results is implemented ensemble-by-ensemble
as a simultaneous correlated frequentist fit over two-point and three-point
correlation functions to obtain masses and the lattice form factors $f_\|$ and
$f_\perp$ for all simulated momentum transfers, \emph{i.e.}~one single fit per
ensemble.  These form factors are then renormalized and converted to $f_+$ and
$f_0$ using Eqs.~\eqref{eq:f0fromfparfperp} and \eqref{eq:f+fromfparfperp}. They
are subsequently interpolated/extrapolated to physical quark masses and to the
continuum limit in a single step.  We use bootstrap
resampling~\cite{Efron:1979bxm} with 1000 samples.

\subsection{Two-point correlation function fits}
Here we describe individual correlated fits to the $\pi$, $K$ and $B_s$
two-point data. The results for the pion mass determined here will enter
subsequent analyses.  For $K$ and $B_s$ this section only serves to determine
optimal fit-ranges, which we use in the next section for the combined fit with
ratios of three-point and two-point correlators.

The functional form of the two-point correlation functions $C_2$ is given by
\begin{align}
  C^P_2(t, T, {\bf p}_P) &=
  \sum_{{\bf x}} e^{i {\bf p}_P \cdot {\bf x}}
  \langle \mathcal{O}_\mathrm{snk}({\bf x}, t)
  \mathcal{O}^\dagger_\mathrm{src}({\bf 0}, 0) \rangle \notag\\
  &= \sum_{n=0}^N \langle 0 | \mathcal{O}_\mathrm{snk} | X^{(n)} \rangle
  \langle X^{(n)} | \mathcal{O}^\dagger_\mathrm{src} | 0 \rangle \notag \\
  &\quad\times \left( \frac{e^{-E_{P,n} t} + e^{-E_{P,n} (T - t)}}{2E_{P,n}} \right),
\label{eq:meson_fit}
\end{align}
where the interpolating operators $\mc{O}_{\mathrm{src}}$,
$\mc{O}_{\mathrm{snk}}$ are given by $\bar{l}\gamma^5l$, $\bar{s}\gamma^5l$ and
$\bar{b}\gamma^5s$, and are chosen to induce states with the quantum numbers of
the $\pi$, $K$ and $B_s$, respectively.

In a first step, we extract the energies $E_{P,n}$ and amplitudes
$\matrixel{X^{(n)}}{\mathcal{O}}{0}$ for $n=0,1$ separately for the $\pi$, $K$
and $B_s$ meson by fitting the correlation functions to the functional form
given in Eq.~\eqref{eq:meson_fit} with $N=1$.  For the $B_s$ mesons we
simultaneously fit the smeared-sink and point-sink correlation functions under
the constraint that they both describe the same meson energy.  The fit ranges
are determined in such a way that the inclusion of the ground and excited state
visibly describes the data well, while also providing an acceptable
$p$-value. Furthermore, we check that the results are stable under variations of
the fit range.

The energies of the final-state kaon can be related to its rest mass via the 
continuum ($E_K^2 = M_K^2 + {\bf p}_K^2$) or the lattice dispersion relation,
\begin{multline}
  \sinh^2\Big(\frac{aE_K(aM_K,a{\bf p}_K)}2\Big) =\\
    \sinh^2\Big(\frac{aM_K}2\Big) +
    \sum_{i=1}^3 \sin^2\Big(\frac{ap_{K,i}}2\Big)\,.
    \label{eq:disprel}
\end{multline}
We have tested that the data is described by the dispersion relation using
lattice momenta ${\bf p}_K=2\pi {\bf n}/L$ with ${\bf n}^2 = 0,1,2,3,4$, where
equivalent three-momenta are averaged.  
This justifies imposing the lattice
dispersion relation in the combined fit that we will describe in the next
section. We will also compare these results with those based on the continuum
dispersion relation in order to assess systematic effects.

\subsection{Form factors from global fits}
Without loss of generality we assume $t_\mathrm{src}=0$ in the following.  The
three-point correlation functions for the transition $B_s\to K$ have the
functional form
\begin{align}
  C_{3, \mu}&(t,\Delta_t, {\bf p}_K) \notag \\
  =& \sum_{\bf  x, y} e^{i{\bf  p}_K\cdot{\bf y}}
  \langle \mathcal{O}_\mathrm{B_s}({\bf x},\Delta_t)
  V_\mu^\text{imp}( {\bf y},t) \mathcal{O}^\dagger_K({\bf 0},0)\rangle
  \notag\\
  =& \sum_{n, m} \langle 0 | \mathcal{O}_{B_s} |
  B_s^{(n)} \rangle \langle B_s^{(n)} | V^\mathrm{imp}_\mu | K^{(m)}
  \rangle
  \langle K^{(m)} | \mathcal{O}^\dagger_K | 0\rangle \notag \\ 
  &\qquad\quad \times \frac{e^{-E_{K,m} t} e^{-M_{B_s,n}(\Delta_t - t)}}{4 E_{K,m} M_{B_s,n}},
\end{align}
where $V^\text{imp}_\mu$ are the improved lattice temporal and spatial
vector currents from Eqs.~\eqref{eq:V0_imp} and~\eqref{eq:Vi_imp}. We
notice that our notation suppresses the fact that the operators
inducing the $B_s$ mesons can be either smeared or local at the sink.
Furthermore, we neglect around-the-world-effects due to the finite
temporal extent $T$ as these are suppressed by $e^{-E_K(T-\Delta_t)}$.

To determine the form factors $f_\|$ and $f_\perp$ (compare
Eq.~\eqref{Eq.fpar_fperp}) we require the ground-state matrix element
$\matrixel{K}{V_\mu^\mathrm{imp}}{B_s}$ (where for convenience we dropped the
superscript $(0)$). To this end we define the ratio $R_{3,\mu}(t,t_\mathrm{snk}
,{\bf p}_K)$ as
\begin{multline}
  R_{3,\mu}(t,t_\text{snk}, {\bf p}_K) =\\
  \frac{C_{3, \mu}(t,t_\text{snk},{\bf p}_K)}
  {\sqrt{C_2^{K}(t,{\bf p}_K) C_2^{B_s}(t_\text{snk}{-}t,{\bf 0})}}
   \sqrt{\frac{2E_{K,0}}{e^{-E_{K,0} t -M_{B_s,0}(t_\text{snk}-t)}}},
\label{eq:formfactor_ratio}
\end{multline}
where we ensure that the appropriate smearing is used to cancel the overlap
factors between the three-point and two-point correlation functions. By
construction, these ratios satisfy
\begin{equation}
  \begin{aligned}
    f_\parallel^\mathrm{bare}({\bf p}_K) &= \lim_{0\ll t\ll t_\mathrm{snk}} R_{3,0}(t,t_\mathrm{snk},{\bf p}_K)\\
    f_\perp^\mathrm{bare}({\bf p}_K) &= \lim_{0\ll t\ll t_\mathrm{snk}} \frac{1}{p^i_K} R_{3,i}(t,t_\mathrm{snk},{\bf p}_K),
    \end{aligned}
  \label{eq:ratio_limits}
\end{equation}
so that we can obtain the form factors $f^\mathrm{bare}_\parallel$ and
$f^\mathrm{bare}_\perp$ by fitting the ratio $R_{3,\mu}$.

\begin{figure}
\includegraphics[width=\columnwidth]{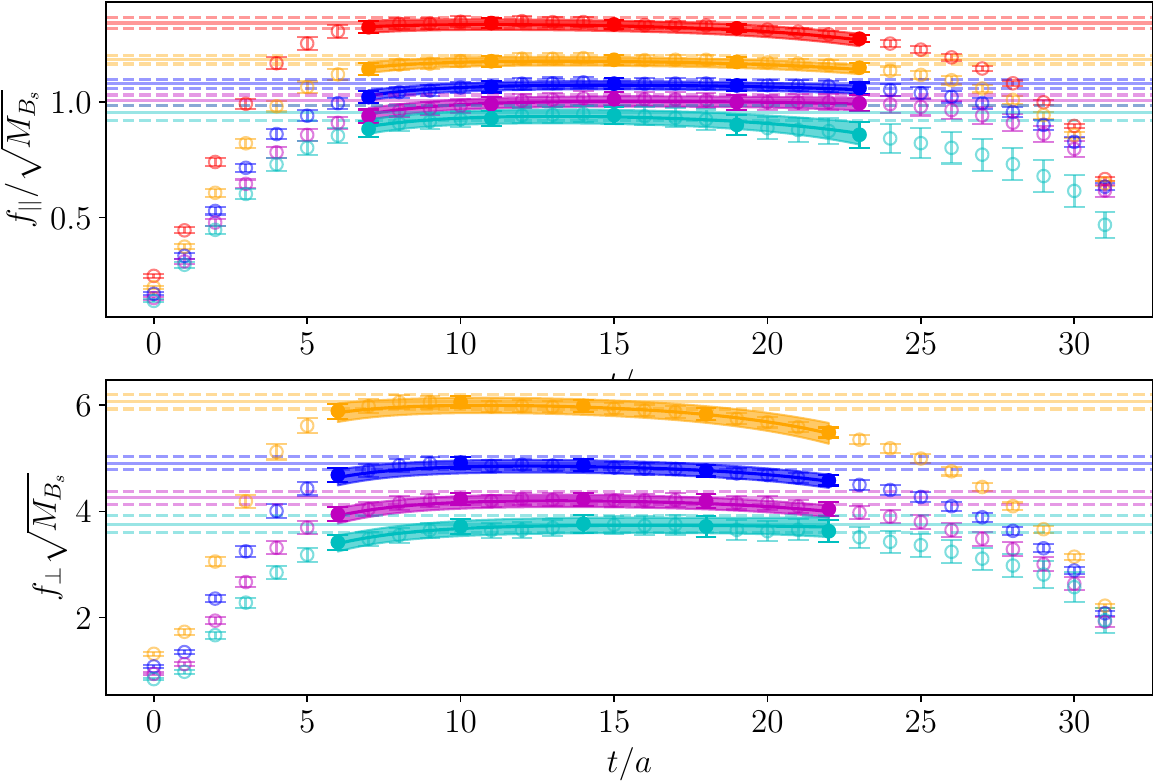}
\caption{Example extraction of $f_\|$ (top) and $f_\perp$ (bottom)
  on the F1S ensemble from
  a correlated global fit. The different colors are for different momenta ${\bf n} 2\pi/L$
  injected at the current. The plot shows the fit including excited-state
  contributions as well as the ground-sate contribution (horizontal dashed lines). Filled symbols
  indicate the points that were included in the fits.}
\label{fig:ratiofit}
\end{figure}

In practice, we carry out a simultaneous correlated fit over the point-point $K$
and smeared-point and point-point $B_s$ two-point functions (including one
excited state in both cases), together with all components $\mu$ of the
three-point correlation functions. For the latter we simultaneously fit over
results for all momenta including ground-state-$B_s$-to-excited-$K$ and
excited-$B_s$-to-ground-state-$K$ terms for the matrix element of the
current. Throughout this fit we enforce the lattice dispersion relation to
describe the energy levels of the kaon
(\emph{c.f.}\ Eq.~\eqref{eq:disprel}). Except for the results on F1S, for
momenta ${\bf n}^2=3,4$ the statistical noise on the three-point functions is
too large to allow for meaningful constraints on the latter matrix element, and
we do not include it in the fit. The term containing both the $B_s$ and $K$
excited states leads to poorly-constrained fits and is excluded. The inclusion
of excited-states allows us to extend the range of time slices we can fit, as
illustrated in figure~\ref{fig:ratiofit}.  In order to limit the impact of
strong correlations between neighboring time slices in the ratio $R_{3,\mu}$
only every 4th time slice enters the fit.  We choose fit ranges which visibly
describe the data well while still giving acceptable $p$-values. The results for
the ground-state matrix elements determined in this way show no dependence on
the choice of lower and upper end of the fit range for the ratio when varied by
$\pm 1$ time slices. An example for this fit is shown in Fig.~\ref{fig:ratiofit}

\begin{figure}
\includegraphics[width=\columnwidth]{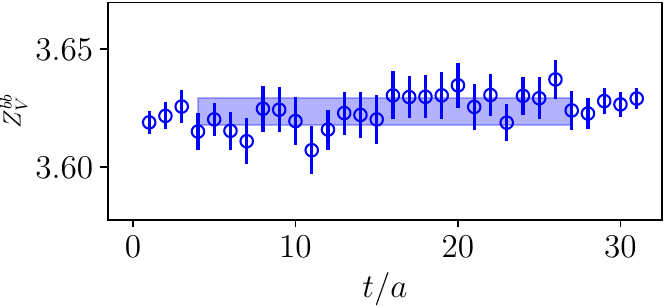}
\caption{Example extraction of the vector-current renormalization constant
  $Z_V^{bb}$ on the F1S ensemble.}
\label{fig:ZVbb}
\end{figure}
\begin{table*}
  \caption{Numerical values of the renormalization constants $Z_A^{ll}$ and
    $Z_V^{bb}$.
    \label{tab:ZVbb}}
  \begin{center}
\normalsize\begin{tabular}{l|llllllllllllll}
\hline \hline 
&\multicolumn{1}{c}{C1}&\multicolumn{1}{c}{C2}&\multicolumn{1}{c}{M1}&\multicolumn{1}{c}{M2}&\multicolumn{1}{c}{M3}&\multicolumn{1}{c}{F1S}\\
\hline 
$Z_A^{ll}$&0.7172&0.7178&0.7449&0.7452&0.7452&0.7624\\
$Z_V^{bb}$&9.099(24)&9.135(23)&4.7767(87)&4.7602(75)&4.770(10)&3.6236(57)\\
\hline 
\hline 
\end{tabular}
\end{center}

\end{table*}

What remains is to extract $Z_V^{bb}$. We first consider the temporal component
of the zero-momentum matrix element for the $b\to b$ vector current on the $B_s$
meson and restrict ourselves to the region where only the ground state
contributes:
\begin{align}
  C_{3,0}&(t,t_\text{snk},{\bf 0}) = \notag\\
  &\lim_{0\ll t \ll t_\mathrm{snk}} \sum_{\bf x, y}
  \langle \mathcal{O}_{B_s}({\bf x},t_\text{snk})
  V_0^{bb}({\bf y},t) \mathcal{O}^\dagger_{B_s} ({\bf 0},0)\rangle \notag \\
  &= \frac{\left|\matrixel{0}{\mathcal{O}_{B_s}}{B_s}\right|^2}{{4 M^2_{B_s}}} \matrixel{B_s}{V^{bb}_0}{B_s}e^{-M_{B_s}t_\mathrm{snk}}.
\end{align}
Recalling Eq.~\eqref{eq:renbb}, we notice that dividing the two-point
function $C_2^{B_s}$ at $t=t_\mathrm{snk}$ (with the appropriate smearing) by
this expression gives
\begin{align}
  \frac{C_2^{B_s}(t_\mathrm{snk},T,{\bf 0})}{C_{3,0}(t,t_\text{snk},{\bf 0})} = \lim_{0\ll t \ll t_\mathrm{snk}} \frac{2M_{B_s}}{\matrixel{B_s}{V^{bb}_0}{B_s}} = Z_V^{bb},
\end{align}
allowing us to extract $Z_V^{bb}$ from a simple fit to a constant. We show data
and the fit on the F1S ensemble in Fig.~\ref{fig:ZVbb} and collect results for
all ensembles in Tab.~\ref{tab:ZVbb}.

With the value of $Z_V^{bb}$ at hand, and using Eq.~\eqref{eq:renorm}, we can determine the renormalization constants
$Z_{V_\mu}^{bl}$ and compute:
\begin{equation}
  \begin{aligned}
  f_\|({\bf p}_P) &= Z^{bl}_{V_0} f_\|^\text{bare}({\bf p}_P),\\
  f_\perp({\bf p}_P) &= Z^{bl}_{V_i} f_\perp^\text{bare}({\bf p}_P).
  \end{aligned}
\end{equation}
Finally we convert to $f_0(q^2)$ and $f_+(q^2)$ using
Eqs.~\eqref{eq:f0fromfparfperp} and \eqref{eq:f+fromfparfperp}.
Table~\ref{tab:BstoK-by-ensemble} summarizes all fit results.

\begin{table*}
\caption{Summary of results from the global fit and the separate fit to pion
  two-point functions.  The squared momentum transfer, $q^2$, is determined from
  the outgoing kaon momentum, ${\bf k} = (2\pi/L){\bf n}$ , with the values of
  $|{\bf n}|^2$ being given in the table. The reduced $\chi^2$ in the table is
  defined as $\chi^2_{\rm red.}=\chi^2/N_{\rm dof}$.}
\label{tab:BstoK-by-ensemble}
    \begin{center}
\normalsize\begin{tabular}{c|c|l|l|l|l|l|llllllll}
\hline \hline 
&$|{\bf n}|^2$&\multicolumn{1}{|c}{C1}&\multicolumn{1}{|c}{C2}&\multicolumn{1}{|c}{M1}&\multicolumn{1}{|c}{M2}&\multicolumn{1}{|c}{M3}&\multicolumn{1}{|c}{F1S}\\
\hline 
$N_{\rm dof}/\chi^2_{\rm red.}/p$&--&41/1.35/0.06&54/0.95/0.59&41/0.93/0.61&46/1.27/0.10&51/1.24/0.11&57/0.77/0.90\\
\hline 
$f_+(q^2)$&1&2.001(29)&2.011(27)&2.038(39)&1.995(28)&1.984(33)&2.355(50)\\
$f_+(q^2)$&2&1.545(29)&1.545(29)&1.578(40)&1.566(28)&1.561(33)&1.917(43)\\
$f_+(q^2)$&3&1.250(41)&1.270(38)&1.282(50)&1.288(33)&1.256(42)&1.663(43)\\
$f_+(q^2)$&4&1.029(65)&1.141(61)&1.040(75)&1.081(51)&1.055(61)&1.476(53)\\
\hline 
$f_0(q^2)$&0&0.8710(93)&0.8872(97)&0.884(13)&0.8692(95)&0.877(12)&0.879(14)\\
$f_0(q^2)$&1&0.7523(93)&0.774(10)&0.769(14)&0.7442(93)&0.767(12)&0.796(12)\\
$f_0(q^2)$&2&0.672(13)&0.703(14)&0.699(18)&0.675(12)&0.690(14)&0.739(12)\\
$f_0(q^2)$&3&0.602(20)&0.662(22)&0.628(26)&0.633(19)&0.629(22)&0.703(16)\\
$f_0(q^2)$&4&0.581(36)&0.625(35)&0.564(42)&0.606(30)&0.558(37)&0.675(21)\\
\hline 
$aM_{B_s}$&-&3.00572(97)&3.00977(88)&2.25278(88)&2.25186(70)&2.25321(80)&1.92574(87)\\
$aM_{K}$&-&0.30666(49)&0.32646(30)&0.22489(49)&0.23440(42)&0.24141(47)&0.19144(46)\\
\hline 
\hline 
\end{tabular}
\end{center}
 
     \begin{center}
\normalsize\begin{tabular}{l|llllllllllllll}
\hline \hline 
&\multicolumn{1}{c}{C1}&\multicolumn{1}{c}{C2}&\multicolumn{1}{c}{M1}&\multicolumn{1}{c}{M2}&\multicolumn{1}{c}{M3}&\multicolumn{1}{c}{F1S}\\
\hline 
$N_{\rm dof}/\chi^2_{\rm red.}/p$&\multicolumn{1}{c}{12/1.10/0.36}&\multicolumn{1}{c}{12/0.60/0.84}&\multicolumn{1}{c}{14/1.14/0.32}&\multicolumn{1}{c}{8/1.32/0.23}&\multicolumn{1}{c}{8/0.58/0.80}&\multicolumn{1}{c}{20/0.89/0.60}\\
$aM_{\pi}$&\multicolumn{1}{c}{0.19026(50)}&\multicolumn{1}{c}{0.24289(45)}&\multicolumn{1}{c}{0.12639(49)}&\multicolumn{1}{c}{0.15222(36)}&\multicolumn{1}{c}{0.17260(45)}&\multicolumn{1}{c}{0.09640(34)}\\
$M_{\pi}\,[{\rm GeV}]$&\multicolumn{1}{c}{0.3395(13)}&\multicolumn{1}{c}{0.4335(15)}&\multicolumn{1}{c}{0.3012(16)}&\multicolumn{1}{c}{0.3628(16)}&\multicolumn{1}{c}{0.4114(18)}&\multicolumn{1}{c}{0.2684(14)}\\
\hline 
\hline 
\end{tabular}
\end{center}
 
\end{table*}

\subsection{Chiral-continuum extrapolation}
\label{subsec:BstoK-fit}

\begin{figure*}
    \includegraphics[width=.49\textwidth]{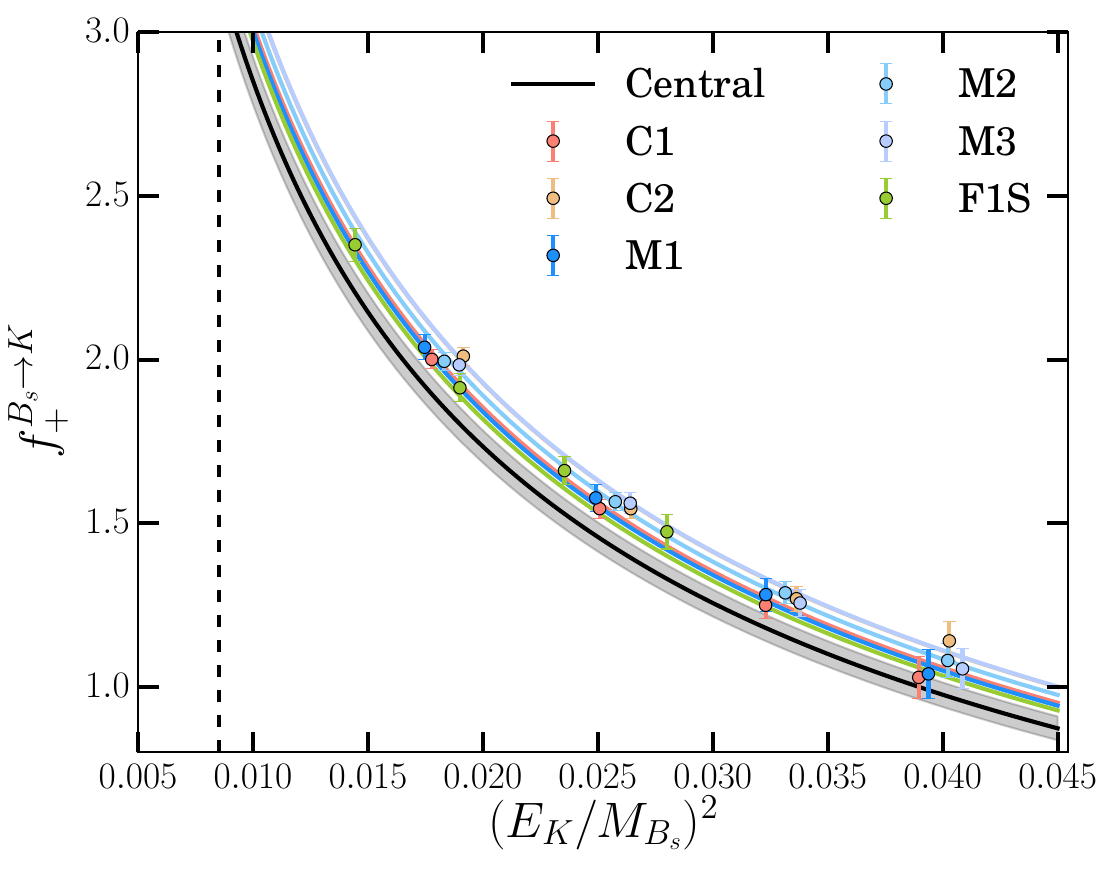}
    \includegraphics[width=.49\textwidth]{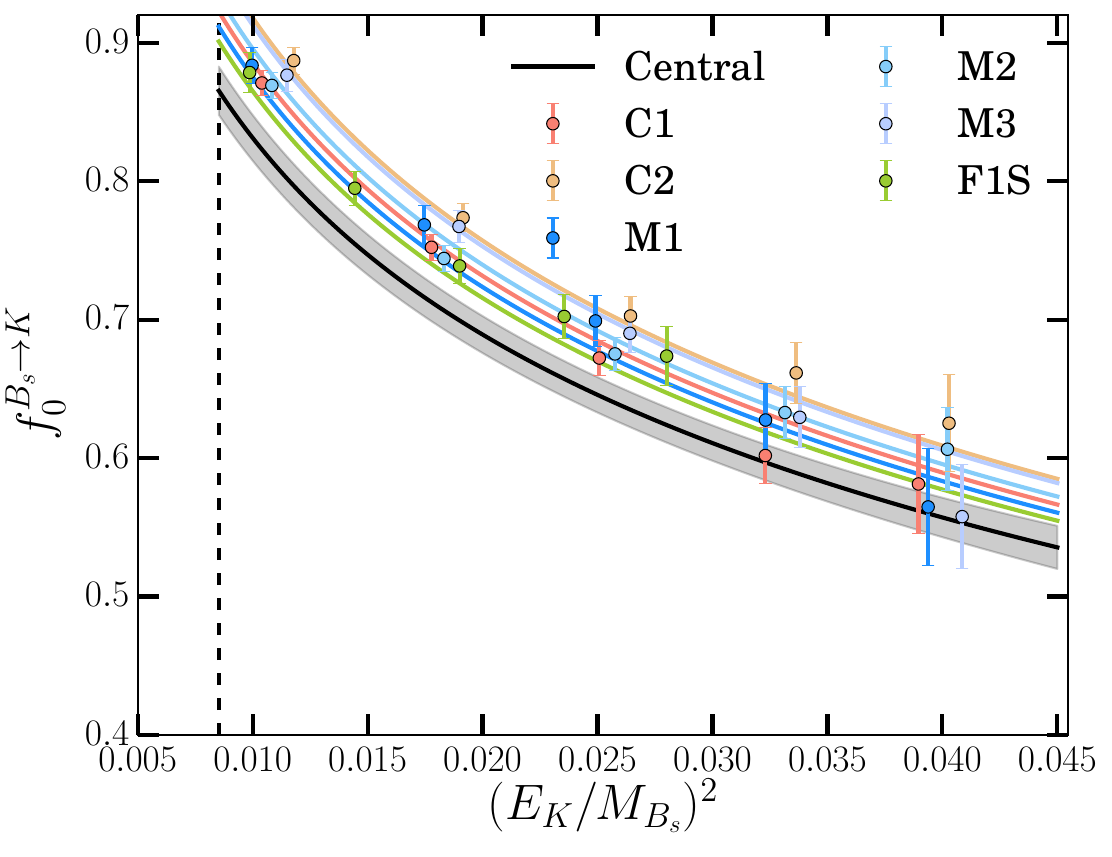}
\caption{Chiral-continuum extrapolation for the $\BstoKlnu$ form factors $f_+$
  (left) and $f_0$ (right). The colored data points show the underlying
  data. The colored lines show the result of the fit evaluated at the parameters
  of the respective ensembles. The gray bands display the obtained form factors
  in the chiral-continuum limit and the associated statistical uncertainty.}
\label{fig:BstoK-chiral-ctm-fit}
\end{figure*}

We extrapolate the renormalized lattice form factors to vanishing lattice
spacing and to the physical light-quark mass, and interpolate in the kaon
energy, using next-to-leading order (NLO) SU(2) chiral perturbation theory for
heavy-light mesons (HM$\chi$PT) in the ``hard-pion'' (or in this case kaon)
limit~\cite{Flynn:2008tg,Bijnens:2010ws,Becirevic:2002sc}. In the SU(2) theory,
the strange quark is integrated out. The chiral logarithms for $\BstoKlnu$
depend on the pion mass and the kaon energy, while the SU(2) low-energy
constants depend implicitly on the values of the strange-quark and $b$-quark
masses. The function we use is
\begin{multline}
  \label{eq:f-HMChPT}
    f_X^{B_s\to K}(M_{\pi}, E_K, a^2) =\\
  \frac{\Lambda}{E_K+\Delta_X}
  \left[c_{X,0}\bigg(1+
    \frac{\delta f(M^s_\pi)-\delta f(M^p_\pi)}{(4 \pi f_\pi)^2}\bigg)\right.\\
  \left.+ c_{X,1} \frac{\Delta M_\pi^2}{\Lambda^2}  
  + c_{X,2}\frac{E_K}{\Lambda}
  + c_{X,3}\frac{E_K^2}{\Lambda^2} 
  + c_{X,4}(a\Lambda)^2 \right],
\end{multline}
where $X=+,0$ for the vector and scalar form factor, respectively, and where
$M_\pi^s$ is the simulated pion mass on a given ensemble, $M_\pi^p
=(2M_{\pi^\pm}+M_{\pi^0})/3$ is the isospin-averaged physical pion mass, $\Delta
M_\pi^2 = (M_{\pi}^{s})^2 - (M_{\pi}^{p})^2$ and $\Lambda=1\gev$ is the
renormalization scale appearing in the one-loop chiral logarithm in $\delta f$
shown in~(\ref{eq:chpt_na_bstok}) below, and is also used as a dimensionful
scale to render the fit coefficients dimensionless. $\Delta_X = M_{B^*}-M_{B_s}$
and the $B^*$ is a $\bar bu$ flavor state with $J^P=1^-$ for $f_+$, or $J^P=0^+$
for $f_0$. For $f_+$ this is the vector meson $B^*$ with mass $M_{B^*} =
5.32471(21)\gev$~\cite{ParticleDataGroup:2022pth}, while for $f_0$ there is a
theoretical estimate for the $0^+$ state,
$M_{B^*(0^+)}=5.63\gev$~\cite{Bardeen:2003kt}\footnote{Results from this and
  other theoretical calculations~\cite{DiPierro:2001dwf,Godfrey:2016nwn,
    Lakhina:2006fy,Liu:2016efm,Ebert:2009ua,Sun:2014wea,
    Godfrey:1986wj,Lu:2016bbk,Colangelo:2005gb,Cheng:2014bca,
    Torres-Rincon:2014ffa,WooLee:2006kdh,Wang:2015mxa,Lutz:2003fm,
    Matsuki:2007zza,Orsland:1998de,Vijande:2007ke,Badalian:2007yr,
    Lahde:1999ih,Dmitrasinovic:2012zz} are summarized in Table~4
   of Ref.~\cite{Cheng:2017oqh}. They span a range from $2\%$ below to $2\%$ above
   this value.}
(the formalism for effective theories for heavy hadrons of
arbitrary spin was derived in Ref.~\cite{Falk:1991nq} and is reviewed
in Ref.~\cite{Casalbuoni:1996pg}). We take $\Delta_+=-42.1\mev$ for $f_+$ using
experimentally-measured masses~\cite{ParticleDataGroup:2022pth,Olive:2016xmw}, and
$\Delta_0=263\mev$ for $f_0$ using the theoretical estimate
from Ref.~\cite{Bardeen:2003kt}.  Since the estimated $B^*(0^+)$ pole location is
rather far from the physical $q^2$ region, the fit is insensitive to its precise
position (and varying it is included in our uncertainty for the chiral
extrapolation).

While the $0^+$ and $1^-$ poles describe the \emph{physical} form factors $f_0$
and $f_+$, respectively, past work~\cite{Flynn:2015mha,Bazavov:2019aom} used
Eq.~(\ref{eq:f-HMChPT}) for $f_\|$ and $f_\perp$ based on the observation in
Eq.~(\ref{eq:f0fromfparfperp}) that $f_0$ is dominated by $f_\|$ and $f_0$ by
$f_\perp$. Below we discuss whether this assumption is warranted with our data.

For the case of the $f_+$ form factor, the coefficient $c_3$ is compatible with
zero within statistical errors. Since the quality of the fit remains good when
removing this term from the ansatz, we perform the fits for $f_+$ without such a
term.

The term $\delta f$ entering~\eqref{eq:f-HMChPT} is the same for $f_+$ and $f_0$
and is given by
\begin{equation}
  \label{eq:chpt_na_bstok}
  \delta f = -\frac{3}{4} \Bigg(M_\pi^2
  \log\bigg(\frac{M_\pi^2}{\Lambda^2}\bigg) + \frac{4M_\pi}{L}
  \sum_{|{\bf n}|\neq 0} \frac{K_1(|{\bf n}|M_\pi L)}{|{\bf n}|}\Bigg).
\end{equation}
The first term is the one-loop chiral logarithm and the second term is an
estimate for effects due to the finite (spatial) volume, where $K_1$ is a
modified Bessel function of the second kind and $|{\bf n}|$ is the magnitude of
a vector of integers ${\bf n} = (n_x,n_y,n_z)$ specifying the spatial lattice
momentum $2 \pi {\bf n}/L$. The second term is estimated using one-loop
finite-volume SU(2) hard-pion $\chi$PT~\cite{Arndt:2004bg,Aubin:2007mc} where
loop integrals are replaced by sums over lattice sites, with its expression
derived in Ref.~\cite{Bernard:2001yj}.

The pion masses entering the fit are obtained from separate two-point
correlation function fits and we take $f_\pi=130.2\mev$~\cite{Aoki:2019cca}. We
include a term proportional to $a^2$ to account for the dominant lattice-spacing
dependence. The domain-wall fermion and Iwasaki gluon actions are expected to
have discretization errors $O((a\Lambda_\text{QCD})^2)$, about $3\%$ ($5\%$) on
the F (M) ensemble(s) for $\Lambda_\text{QCD}=500\mev$, while power-counting
estimates of errors in the RHQ action and heavy-light current are smaller, below
$2\%$. The $a^2$ term in our fit therefore accounts for the leading discretization
effects.

Results for the parameters of the chiral-continuum fit are given in
Table~\ref{tab:BstoK-chiral-ctm-fit}.  Figure~\ref{fig:BstoK-chiral-ctm-fit}
shows the fit, while the systematic errors from variations in the fit are
discussed in section~\ref{sec:syserrors} and shown in
Figure~\ref{fig:BstoK-chiral-ctm-errors}. Values for the form factors in the
continuum and physical quark mass limit, along with their statistical and
systematic errors and correlations, are given at the end of the discussion of
systematic errors in section~\ref{sec:syserrors} for a set of reference $q^2$
values. These are obtained by using the results in
Tab.~\ref{tab:BstoK-chiral-ctm-fit} to evaluate the form factor for a given kaon
energy, after setting $a=0$, taking the limit $L\to\infty$ in
Eqs.~\eqnref{eq:f-HMChPT} and~\eqnref{eq:chpt_na_bstok}, and setting $m_\pi=(2
m_\pi^\pm+m_\pi^0)/3\approx 138$MeV (isospin-averaged pion
mass)~\cite{ParticleDataGroup:2022pth}.

\begin{table*}
  \caption{Fitted parameters for chiral-continuum extrapolation for the
    $\BstoKlnu$ form factors defined in \eqref{eq:f-HMChPT}. Results for the
    coefficients, statistical errors and correlation matrix for the central
    continuum-limit fit.  The fit quality is $(N_{\rm dof}/\chi^2_{\rm red.}/p)=
    (20/1.09/0.35)$ for $f_+$ and $(25/1.12/0.11)$ for
    $f_0$.}\label{tab:BstoK-chiral-ctm-fit} \begin{center}
\normalsize\begin{tabular}{lll|rrrr|rrrrr}
\hline \hline 
&&&\multicolumn{1}{c}{$c_{+,0}$}&\multicolumn{1}{c}{$c_{+,1}$}&\multicolumn{1}{c}{$c_{+,2}$}&\multicolumn{1}{c}{$c_{+,4}$}&\multicolumn{1}{c}{$c_{0,0}$}&\multicolumn{1}{c}{$c_{0,1}$}&\multicolumn{1}{c}{$c_{0,2}$}&\multicolumn{1}{c}{$c_{0,3}$}&\multicolumn{1}{c}{$c_{0,4}$}\\
&$c^{+,0}$&&1.8169&0.2803&-0.7542&-0.0936&0.5310&0.2273&0.2996&-0.0938&-0.0136\\
&&$\delta c^{+,0}$&0.0453&0.2661&0.0411&0.1461&0.0269&0.1057&0.0748&0.0590&0.0567\\
\hline 
$c_{+,0}$&1.8169&0.0453&1.0000&-0.3532&-0.6247&-0.4534&0.1671&-0.2710&0.2122&-0.2545&-0.3595\\
$c_{+,1}$&0.2803&0.2661&-0.3532&1.0000&-0.0280&-0.3974&-0.2226&0.7687&0.0216&-0.0133&-0.2866\\
$c_{+,2}$&-0.7542&0.0411&-0.6247&-0.0280&1.0000&-0.0035&0.2323&-0.0341&-0.3524&0.4313&0.0124\\
$c_{+,4}$&-0.0936&0.1461&-0.4534&-0.3974&-0.0035&1.0000&-0.2261&-0.2842&0.0044&-0.0123&0.7535\\
\hline 
$c_{0,0}$&0.5310&0.0269&0.1671&-0.2226&0.2323&-0.2261&1.0000&-0.3095&-0.8652&0.8331&-0.2633\\
$c_{0,1}$&0.2273&0.1057&-0.2710&0.7687&-0.0341&-0.2842&-0.3095&1.0000&0.0950&-0.0942&-0.4221\\
$c_{0,2}$&0.2996&0.0748&0.2122&0.0216&-0.3524&0.0044&-0.8652&0.0950&1.0000&-0.9836&-0.0143\\
$c_{0,3}$&-0.0938&0.0590&-0.2545&-0.0133&0.4313&-0.0123&0.8331&-0.0942&-0.9836&1.0000&0.0069\\
$c_{0,4}$&-0.0136&0.0567&-0.3595&-0.2866&0.0124&0.7535&-0.2633&-0.4221&-0.0143&0.0069&1.0000\\
\hline 
\hline 
\end{tabular}
\end{center}

\end{table*}

\section{Systematic error analysis}
\label{sec:syserrors}

Our systematic-error analysis has much in common with that done in our earlier
work on semileptonic $B\to\pi\ell\nu$ and $B_s\to K\ell\nu$ decays~\cite{Flynn:2015mha}. We
streamline our discussion where possible, relying on Ref.~\cite{Flynn:2015mha}
for details, and introduce new features for this analysis.

Following the same strategy as in Ref.~\cite{Flynn:2015mha}, we introduce a
set of reference $q^2$ values in the range where we have lattice data.
By determining the form factors, their statistical uncertainty and all
systematic uncertainties at these reference points, we obtain
synthetic data points which include the complete and fully correlated
error budget. These serve as inputs for the extrapolation over the
entire kinematically allowed range $q^2 \in [0,q^2_\mathrm{max}]$.

We distinguish between different contributions to the full error
budget. The statistical uncertainty is computed from the bootstrap
analysis of the chiral and continuum extrapolation described in the
previous section~\ref{subsec:BstoK-fit}. We assess the systematic
error arising from these fits in section~\ref{subsec:BstoK_fitsys} by
applying cuts to the data entering the fit as well as varying the
functional description of the data, \emph{i.e.}~Eq.~\eqref{eq:f-HMChPT}.

We estimate all remaining sources of uncertainty in sections
\ref{sec:scale} to \ref{subsec:iso}. Finally, in section
\ref{subsec:correlation} we address how to combine these various
sources of uncertainty to complete the error budget.

\subsection{Chiral-continuum extrapolation}
\label{subsec:BstoK_fitsys}
\begin{figure*}
  \includegraphics[width=.49\textwidth]{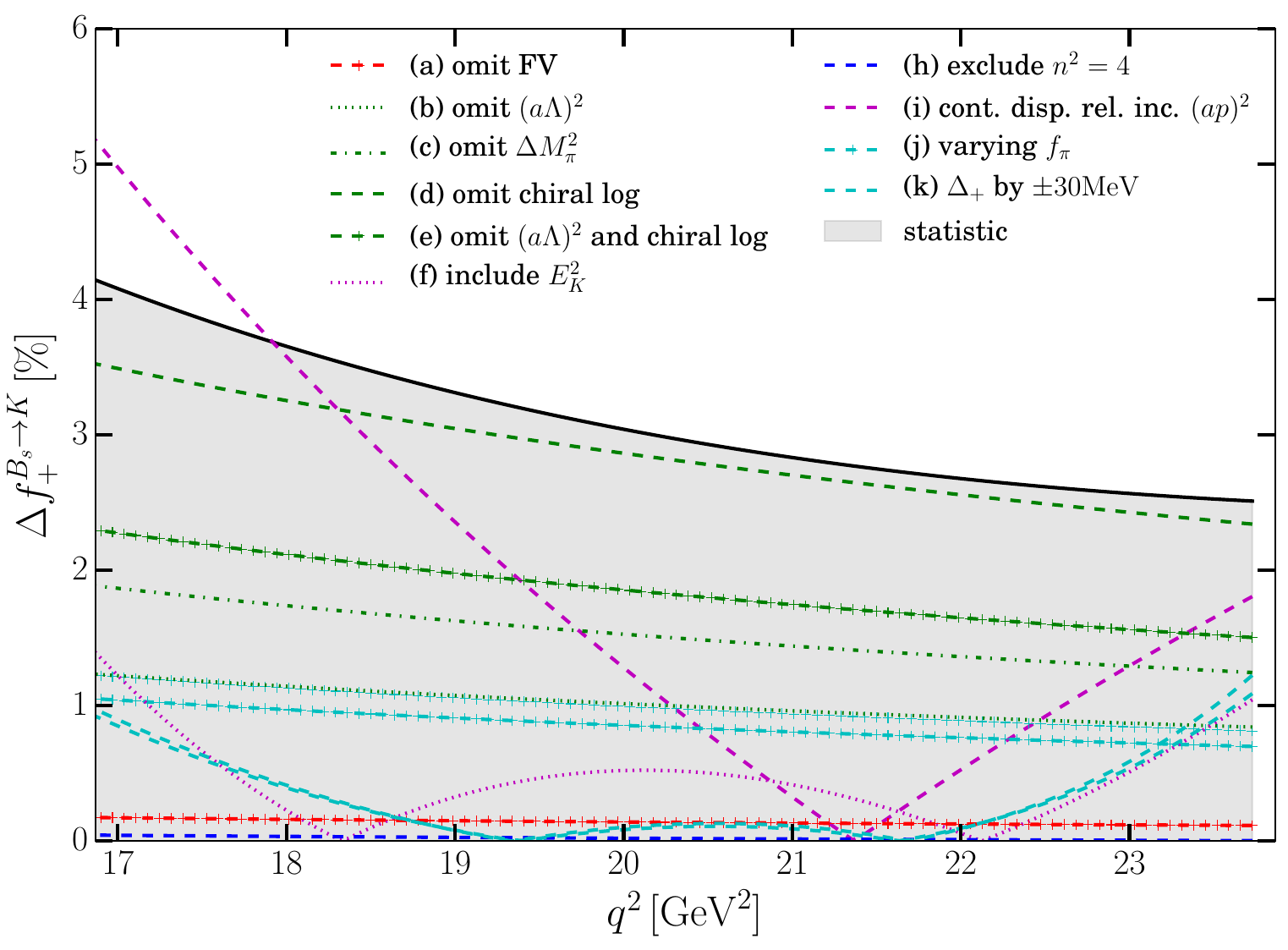}
  \includegraphics[width=.49\textwidth]{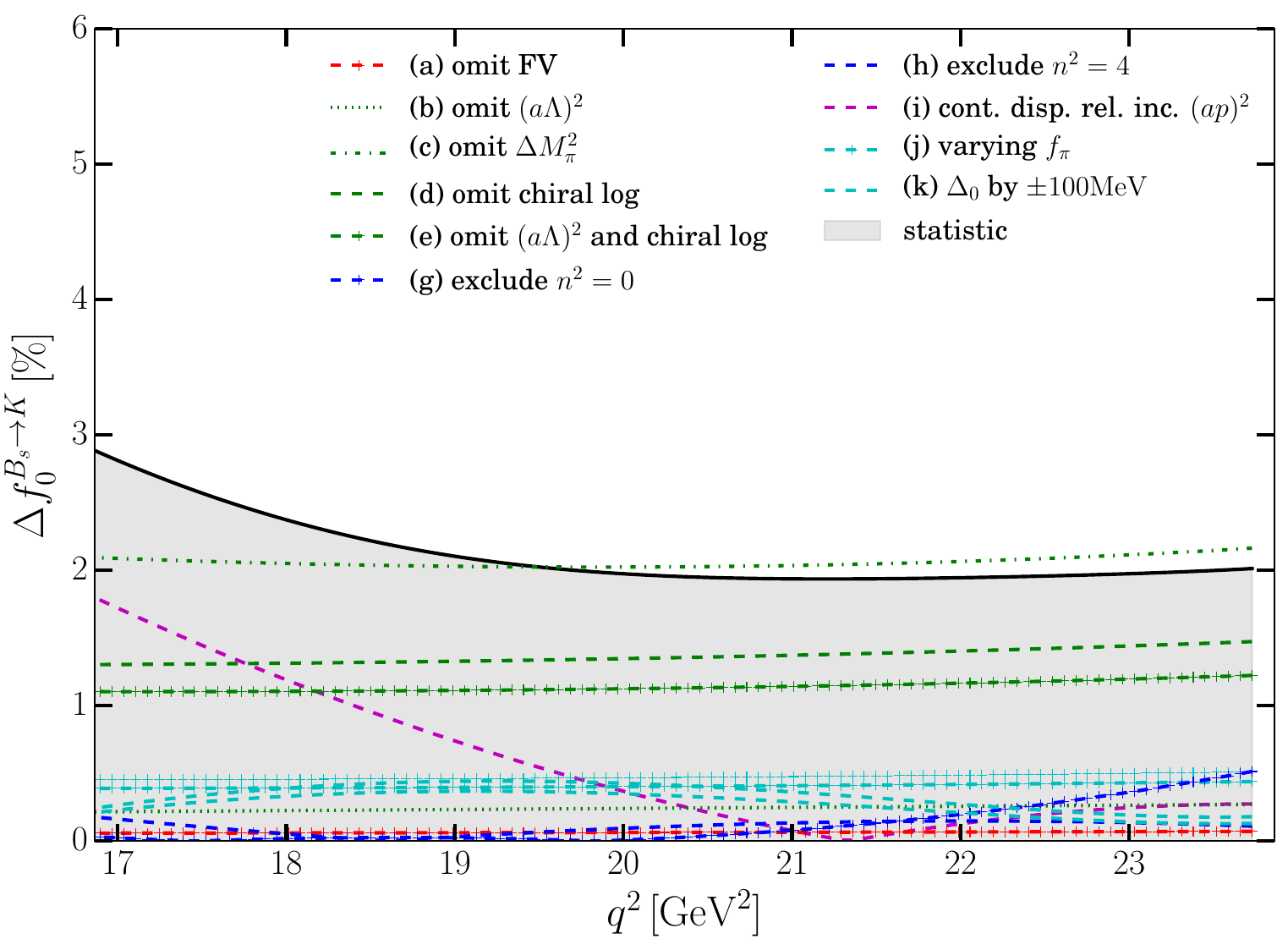}
  \caption{Relative changes, $\Delta f_X =
    |f_X^\text{pref}-f_X^\text{alt}|/f_X^\text{pref}$ for $X=0,+$, of the
    form-factor central values under variations of the chiral-continuum fit for
    $\BstoKlnu$. The shaded band shows the statistical uncertainty of the
    preferred fit.}
\label{fig:BstoK-chiral-ctm-errors}
\end{figure*}

We estimate the systematic uncertainty from the chiral-continuum
extrapolation for $\BstoKlnu$ by performing cuts to the data as well as
varying the fit ansatz in Eq.~\eqref{eq:f-HMChPT}. We consider
the following variations to the fit form:
\begin{enumerate}[label=(\alph*)]
\item omitting the finite volume corrections (the second term in $\delta f$),
\item omitting the term proportional to $a^2$ ($c_4\equiv 0$),\label{Kvar:noasq}
\item omitting the term proportional to $M_\pi^2$ ($c_1\equiv 0$),\label{Kvar:nompisq}
\item analytic fits omitting the chiral logarithms ($\delta f \equiv 0$), \label{Kvar:analytic}
\item analytic fits simultaneously omitting the chiral logarithms and the
  $a^2$ term,
\item including the term proportional to $(E_K/\Lambda)^2$ into the fit for $f_+$,  
\end{enumerate}
We also vary the data that enters the fit by
\begin{enumerate}[resume,label=(\alph*)]
\item omitting the data points at the highest momentum, ${\bf p}_K =
  2\pi(2,0,0)/L$ (smallest $q^2$), \label{Kvar:pmax}
\item omitting the data points at zero momentum, \emph{i.e.}~$q^2_\text{max}$ in $f_0$,
\item using form factor data that has been obtained by imposing the continuum
  dispersion relation which at leading order differs from the lattice dispersion
  relation by powers $(ap)^2$. We therefore also include a term $c_{X,5}$ in
  these fits. \label{Kvar:disprel}
\end{enumerate}
Finally, we consider the impact of variations of some of the numerical
values of parameters entering the fit, such as
\begin{enumerate}[resume,label=(\alph*)]
\item replacing the numerical value of $f_\pi$ by its chiral limit value
  $f_0=112\mev$~\cite{Aoki:2010dy} or by
  $f_K=155.7\mev$~\cite{FlavourLatticeAveragingGroupFLAG:2021npn},
\item varying the model estimate of $M_{B^*(0^+)}$ entering $\Delta_0$ by
  $\pm100\mev$ and the experimentally precisely known value of $M_{B^*}$
  entering $\Delta_+$ by a generous $\pm30\mev$.
\end{enumerate}

Figure~\ref{fig:BstoK-chiral-ctm-errors} shows the relative effects of
these variations compared to the statistical uncertainty of the
central fit (gray shaded area). We notice that our fit is insensitive
to most of these variations. The largest deviations are observed for
variations including the lattice spacing and the pion-mass dependence,
\emph{i.e.}~variations \ref{Kvar:noasq}, \ref{Kvar:nompisq} and
\ref{Kvar:analytic}. However, even these remain of the same size as
the statistical errors we quote. We take the largest difference
between the preferred fit and any of the alternatives as the
systematic uncertainty due to the chiral-continuum extrapolation.

\begin{figure}
  \includegraphics[width=9cm]{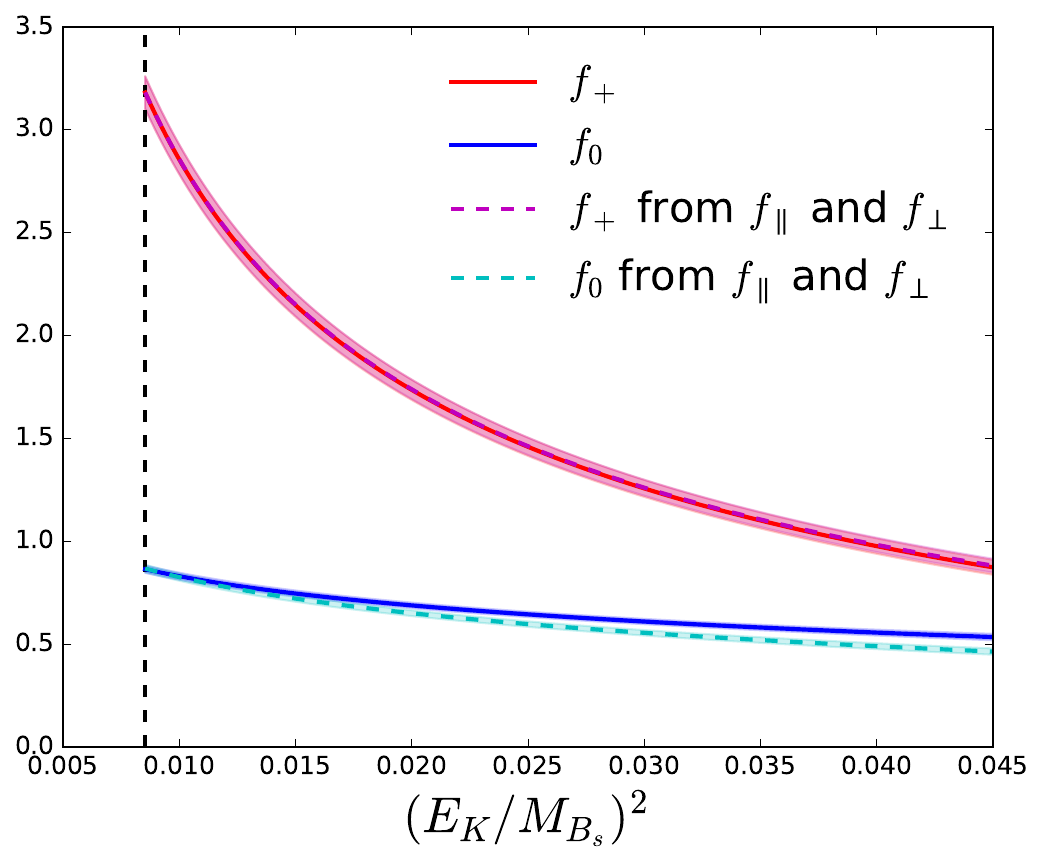}
  \caption{The plot shows the result of the chiral-continuum fit based on
    Eq.~\eqnref{eq:f-HMChPT}, once using $f_\|$ and $f_\perp$ as input (and
    subsequently converted to $f_0$ and $f_+$), and once using $f_0$ and $f_+$
    as input. The results for $f_0$ differ significantly and increasingly as the
    kaon momentum is increased.}
\label{fig:fpf0fperpfpar}
\end{figure}

An important subtlety is worth highlighting here. In previous work on the
$B_s\to K\ell\nu$ semileptonic form factors~\cite{Flynn:2015mha,Bazavov:2019aom}, but
also other decay channels like \emph{e.g.}~$B\to \pi\ell\nu$
semileptonic~\cite{Flynn:2015mha,Lattice:2015tia}, it was assumed that the pole
locations of the \emph{physical} form factors $f_+$ and $f_0$ also describe the
kinematical behaviour of $f_\perp$ and $f_\|$, respectively. In particular,
based on the linear relations in Eqs.~\eqnref{eq:f0fromfparfperp} and
\eqnref{eq:f+fromfparfperp}, $f_\perp$ was assumed to be dominated by $f_+$ and
$f_\|$ by $f_0$. In Fig.~\ref{fig:fpf0fperpfpar} we compare the results when
extrapolating the lattice data in both cases using
Eq.~\eqnref{eq:f-HMChPT}. While with the current level of statistical precision
no significant difference is observed for $f_+$, a significant
difference --- which increases with the kaon energy --- can be observed for
$f_0$. Since form-factor parameterizations often rely on the kinematical
constraint $f_+(0)=f_0(0)$, this is a relevant problem for both the vector and
scalar form factor. An interesting question in this context is whether this
observation could explain the observed tensions between different sets of
lattice results for $\BstoKlnu$, as observed by FLAG
21~\cite{FlavourLatticeAveragingGroupFLAG:2021npn}.  Regarding the $\BstoKlnu$
decay we note that HPQCD in Ref.~\cite{Bernard:2001av} carried out the chiral
and continuum limits based on the vector and scalar form factors. In line with
our observation their results for $f_+$ show a distinctly milder curvature as
the results of Ref.~\cite{Flynn:2015mha,Bazavov:2019aom} (\emph{c.f.} Fig.~32 in
FLAG 21~\cite{FlavourLatticeAveragingGroupFLAG:2021npn}).

\subsection{Lattice-scale uncertainty}
\label{sec:scale}
We propagate the uncertainty of the lattice-scale determination
(\emph{c.f.}~Table~\ref{tab:ensembles}) by creating a Gaussian distribution with the
correct central value and width, which is then used to convert the dimensionless
lattice masses into dimensionful quantities prior to the chiral-continuum
extrapolations. We therefore do not account separately for this uncertainty.

\subsection{ Strange-quark-mass uncertainties}

\subsubsection{Valence strange-quark-mass uncertainty}
The $\BstoKlnu$ form factors depend explicitly on the valence-strange-quark
mass, so the effects of any mistunings in this mass must be accounted for. In
order to estimate the corresponding systematic uncertainty, we evaluate the form
factors for additional choices of the spectator-quark mass on the C1 ensemble.
We determine the fractional change of the form factor with respect to a
percentage mistuning in the strange-quark mass, \emph{i.e.}~for $X=+,0$ we
compute
\begin{equation}
  \Delta f_X /\Delta m_s \equiv
  \frac{f_X(am_s) - f_X(am^\mathrm{phys}_s)}{am_s-am_s^\mathrm{phys}}
  \frac{am_s^\mathrm{phys}}{f_X(am^\mathrm{phys}_s)}.
\end{equation}
We repeat this for the different choices of momenta and take the largest value
this ratio takes, which occurs at the smallest momentum.  We tabulate the
maximal values of this term for the form factors $f_+$ and $f_0$ in Table
\ref{tab:strange_quark}. The largest deviation from the physical strange quark
mass occurs on the F1S ensemble, where $am_s^\mathrm{sim} = 0.02144$ and
$am_s^\mathrm{phys} = 0.02167(20)$.  Allowing for one standard deviation we find
\begin{equation}
  \begin{aligned}
  \Delta m_s &= \max{\left\{
    \frac{\abs{am_s^\mathrm{sim} - (am_s^\mathrm{phys}
        \pm \delta am_s^\mathrm{phys})}}{am_s^\mathrm{phys}}\right\}}\\
  &= 1.98\%.
  \end{aligned}
\end{equation}
Combining this with the above, we determine the maximal deviations as shown in
the last line of Table \ref{tab:strange_quark}. We find that the largest impact
of the strange quark mistuning is 0.20\% for $f_+$ and 0.13\% for $f_0$ which are
far smaller than our leading uncertainties and therefore negligible.

\begin{table}
\caption{Partial derivatives of the form factor as a function of the strange
  quark mistuning as defined in the text.}
\label{tab:strange_quark}
\begin{tabular}{l|llll}
\hline
& $f_0^{B_{s} \to K}$ & $f_+^{B_{s} \to K}$ \\\hline\hline
$\Delta f_X/\Delta m_s$&  0.0630(46) &  0.1027(84) \\
$\mathrm{max}(\Delta m_s)\,[\%]$& 1.98 & 1.98 \\\hline
$\mathrm{max}(\Delta f_X)\,[\%]$& 0.13 & 0.20 \\
\hline
\end{tabular}

\end{table}
\subsubsection{Strange-sea-quark mistuning}
Our fit functions do not depend explicitly on the strange sea-quark mass and, at
each lattice spacing, we have form factors for only a single value of this
mass. We estimate any systematic effects stemming from the sea strange-quark
mistuning (which is largest on the coarse ensembles) in the same way, but
including an additional suppression factor of $\alpha_s$. Numerically, this is
approximately $\Delta f_X/\Delta m_s \times \alpha_s \times \max(\Delta
m^\mathrm{sea}_s) = 0.28\%$ and $0.46\%$ for $f_0$ and $f_+$, respectively. This
is intended as a conservative estimate for this uncertainty and it is a
negligible contribution to the total error budget.

\subsection{Effects of the RHQ parameter uncertainty on the form factors}
As outlined in appendix~\ref{appx:rhqtuning}, the RHQ tuning procedure
determines three coefficients in the RHQ action. These are the bare $b$-quark
mass ($m_0a$), the clover coefficient ($c_P$) and the anisotropy ($\zeta$). The
experimental inputs to this tuning are the measured $B_s$-meson mass and the
hyperfine splitting ($\Delta M_{B_s}\equiv M_{B_s^*}-M_{B_s}$). In addition, the
lattice scale and the physical strange-quark mass are required inputs.
Furthermore, we enforce that the rest mass equals the kinetic mass. The tuned
parameters, including estimates for all relevant sources of uncertainties are
listed in Table~\ref{tab:RHQparameter} in appendix~\ref{appx:rhqtuning}.

In order to propagate the effect of these tuning uncertainties onto the form
factors, we generated additional form factor data for different choices of the
RHQ parameters on the C1 ensemble. Using this data set we determine the partial
derivatives of the form factors with respect to the respective RHQ
parameters. We normalize these values by the ratio of the tuned form factor and
respective RHQ parameters on this ensemble and conservatively quote the maximum
value this takes. These values are listed in the first two rows of
Table~\ref{tab:RHQonff}.

\begin{table}
  \caption{Estimates for the normalized partial derivatives on the C1 ensemble
    (top two rows) and the maximum uncertainty propagated onto the form
    factors. \label{tab:RHQonff}}
  \begin{tabular}{c|ccc}
    \hline
    & $am_0$ & $\zeta$ & $c_{P}$ \\\hline\hline
    $f_+$ max normalized slope & 0.2368 & 0.1003 & 0.0543 \\
    $f_0$ max normalized slope & 0.1067 & 0.0418 & 0.1089  \\\hline
    $f_+$ RHQ uncertainty $[\%]$ & 0.8145 & 0.9004 & 0.3215 \\
    $f_0$ RHQ uncertainty $[\%]$ & 0.3670 & 0.3755 & 0.6448 \\    \hline
  \end{tabular}
\end{table}
We derive the uncertainty on the form factor due to a given RHQ parameter by
multiplying these normalized derivatives on the C1 ensemble with the relative
uncertainty of this RHQ parameter on a given lattice spacing.  We illustrate
this on the example of $\zeta$
\begin{equation}
  \left(\frac{\delta f_X}{f^\mathrm{tuned}_X}\right)_{\zeta} = \max_{{\bf n}}\left(\frac{\zeta^\mathrm{tuned}}{f_X^\mathrm{tuned}} \frac{\partial f_X}{\partial \zeta}\right)\Bigg|_{C1} \frac{\delta \zeta}{\zeta^\mathrm{tuned}}\,,
\end{equation}
where $f_X^\mathrm{tuned}$ is the form factor evaluated at the tuned value of
the RHQ parameters on C1, whilst the last term is evaluated lattice
spacing-by-lattice spacing.

We find that the uncertainty is largest on the F1S ensemble and therefore take
this to provide a conservative estimate for the RHQ parameter tuning on the form
factors and list their values in the last two rows of Table~\ref{tab:RHQonff}.
Adding these contributions in quadrature we quote an uncertainty of $1.26\%$ on
$f_+$ and $0.83\%$ on $f_0$.

\subsection{Discretization errors}
Due to their different origin and size, we separately discuss discretization
errors due to the light quarks and gluons in the action, the heavy-light
current, and the RHQ quarks.

\subsubsection{Discretization effects from the action and the current}
\label{sec:LightQuarkDiscErrors}

The dominant discretization error from the light quarks and gluons in the action
is $O((a\Lambda_\text{QCD})^2)$ which, using $\Lambda_\text{QCD}=500\mev$,
amounts to $\sim 3.2\%$ on the finest ensemble. This is accounted for by
including $a^2$-terms in the chiral-continuum extrapolations. We assign the
estimate $(a \Lambda_\text{QCD})^4 \sim 0.10\%$ for subsequent terms. Potential
uncertainties stemming from the residual chiral symmetry breaking are estimated
to be of the size $a m_\text{res} \sim 0.1\%$ for the light quarks.

The leading quark and gluon discretization errors in the heavy-light currents
are $O(\alpha_s a \tilde{m}_l, (a\tilde{m}_l)^2, \alpha_s^2 a
\Lambda_\text{QCD}, (ap)^2)$.

In the comparison between our central fit and the variations~\ref{Kvar:disprel}
and~\ref{Kvar:pmax}, we do not observe any evidence of sizable
momentum-dependent discretization errors in our data. Thus we do not include a
systematic error from this source.

Estimating the effects by power counting on the fine ensemble, the first two
terms are negligible ($<0.1\%$), while the third amounts to $\sim
0.78\%$. Combining this in quadrature with the $(a\Lambda_\text{QCD})^4$ and
$am_\text{res}$ from above, we quote 0.79\%.

\subsubsection{Heavy quark discretization errors}
\label{sec:HeavyQuarkDiscErrors}

The RHQ action gives rise to nontrivial lattice-spacing dependence in
the form factors when $m_0 a \sim 1$. To estimate the resulting
discretization errors, we use the same power-counting approach as in
our previous papers~\cite{Aoki:2012xaa,Christ:2014uea,Flynn:2015mha}.
For reproducibility and completeness we provide a brief summary of the
procedure as well as intermediate numerical values in
Appendix~\ref{appx:HQdiscerrs}. We take the size of heavy-quark
discretization errors in our calculation of semileptonic form factors
from the estimate on our finest ensemble ($a^{-1} = 2.785\gev$). They
amount to $\sim 1.3\%$ for the lattice form factor $f_\|$ and $\sim
1.5\%$ for $f_\perp$.

\subsection{Renormalization factor}
\label{sec:renormfactor}

The renormalization factor relating the lattice weak current to the continuum
one is shown in Eq.~\eqref{eq:renorm} in section~\ref{sec:renormimprove},
where $Z_{V_\mu}^{bl}$ is given by a product of three components. We consider
the uncertainties from these three multiplicative factors separately, and add
them in quadrature to obtain the total error on the form factors.

For $Z_V^{ll}$, we use the non-perturbatively-determined value of the
axial-current renormalization factor $Z_A$ evaluated at the value of the light
quark mass (see table~\ref{tab:ZVbb} for numerical values). We can neglect the
statistical uncertainty in $Z_A$ (which is only $0.02\%$ on the finer ensembles)
and the difference between $Z_V^{ll}$ and $Z_A$ (which is $O(am_\text{res}) \sim
7\times 10^{-4}$ at $a \approx 0.086\fm$).

For $Z_V^{bb}$, we use the non-perturbative determination from
Ref.~\cite{Christ:2014uea}. The statistical uncertainty in $Z_V^{bb}$ on the
finer ensemble is well below one percent and propagated into the statistical
error analysis. The perturbative truncation error in $\rho_V^{bl}$ is taken to
be the full size of the 1-loop correction at the finer $a \approx 0.086\fm$
lattice spacing, which leads to $1.7$\% for $\rho_{V_0}$ and $0.6$\% for
$\rho_{V_i}$.  We use the values of $\rho_{V_\mu}$ and $Z_V^{ll}$ in the chiral
limit and must consider errors due to the nonzero physical up, down, and strange
masses. The leading quark-mass dependent errors in $\rho_{V_\mu}$ and $Z_V^{ll}$
are $O(\alpha_s a \tilde m_q)$ and $O((a\tilde m_q)^2)$, respectively, but these
are already accounted for in our estimate of light-quark and gluon
discretization errors~(see Sec.~\ref{sec:LightQuarkDiscErrors}) and we do not
count them again here.

Perturbative truncation errors are by far the dominant source of uncertainty in
the renormalization factor, and the quadrature sum of the three error
contributions is $1.7\%$ for $f_\|$ and 0.6\% for $f_\perp$.

\subsection{Finite-volume corrections}
\label{sec:FVerr}
As discussed in section~\ref{subsec:BstoK-fit}, we directly account for the
finite size effects in the chiral continuum extrapolation (\emph{i.e.}~in
Eq.~\eqref{eq:f-HMChPT}). We can assess the numerical size of these effects by
comparing to a fit that does not include the second term in
Eq.~\eqref{eq:chpt_na_bstok}. The maximum deviation for $f_0$ ($f_+$) is given
by 0.06\% (0.13\%).

\subsection{Isospin breaking}
\label{subsec:iso}
The leading quark-mass contribution to the isospin breaking from the
valence-quark masses is of $O((m_d-m_u)/\Lambda_\text{QCD}) \sim 0.5\%$,
obtained using $m_d-m_u= 2.38(18)\mev$~\cite{Giusti:2017dmp} and
$\Lambda_\text{QCD}=500\mev$. The difference between the $u$ and $d$ quark
masses in the sea sector should have a negligible effect on the form factors
because the sea quarks couple to the valence quarks through $I=0$ gluon
exchange, giving an uncertainty of $O(((m_d-m_u)/\Lambda_\text{QCD})^2) \sim
0.003\%$.
The electromagnetic contribution to isospin breaking is expected to be
$O(\alpha_{QED}) \sim 1/137 = 0.7\%$. We therefore take $0.7\%$ as the
uncertainty from isospin breaking and electromagnetic effects.

\subsection{Form factor results and correlation matrices}
\label{subsec:correlation}

\begin{table*}
\caption{Values and error budgets for the $\BstoKlnu$ form factors at
  three representative $q^2$ values in the range of simulated lattice momenta.
  The total error is obtained by adding the individual errors in quadrature. See
  Tab.~\ref{tab:corre_mats} for the corresponding correlation matrices. 
  The data shown in this table is contained in the accompanying data file.  }
\label{tab:ErrorBudget} 
\begin{center}
\normalsize\begin{tabular}{l|cc|cccc|rrrrr}
\hline \hline 
&\multicolumn{2}{c|}{$f_+$}&\multicolumn{3}{c}{$f_0$}\\
\hline 
$q^2 [{\rm GeV}^2]$&17.60&23.40&17.60&20.80&23.40\\
$f_{+,0}(q^2)$&0.9878&2.9301&0.5594&0.6843&0.8397\\
\hline 
\multicolumn{6}{c}{Error budget (absolute contribution)}\\
\hline 
Statistical error&0.0377&0.0743&0.0141&0.0133&0.0167\\
Chiral-continuum extrapol.&0.0407&0.0698&0.0116&0.0139&0.0180\\
Other&0.0240&0.0700&0.0139&0.0172&0.0213\\
\cline{1-6}
{\bf Total}&0.0604&0.1236&0.0230&0.0258&0.0325\\
\hline 
\multicolumn{6}{c}{Error budget (in per cent)}\\
\hline 
Statistical error&3.82&2.54&2.53&1.94&1.99\\
Chiral-continuum extrapol.&4.12&2.38&2.06&2.03&2.14\\
Other&2.43&2.39&2.49&2.51&2.54\\
\cline{1-6}
{\bf Total}&6.12&4.22&4.10&3.77&3.87\\
\hline 
\hline 
\end{tabular}
\end{center}

\end{table*}

\begin{table}
  \caption{Statistical, fit- and flat-systematic errors $\delta f_{+/0}$ for
    representative $q^2$ values (given in units of GeV$^2$ as arguments of
    $f_{+/0}$), and corresponding correlation matrices.
      The data shown in this table is contained in the accompanying data file. }
  \label{tab:corre_mats} 
  \begin{center}
\normalsize\begin{tabular}{l|cl|cc|cccccccccccc}
\hline \hline 
\multicolumn{3}{c|}{}&\multicolumn{2}{c|}{$f_+$}&\multicolumn{3}{c}{$f_0$}\\
\cline{2-8}
\multicolumn{1}{c|}{}&$q^2\,[{\rm GeV}^2]$&&$17.60$&$23.40$&$17.60$&$20.80$&$23.40$\\
\cline{4-8}
\multicolumn{1}{c|}{}&&$\delta f_{+/0}$&0.0377&0.0743&0.0141&0.0133&0.0167\\
\hline 
\multirow{2}{*}{$f_+$}&\multicolumn{1}{c|}{$17.60$}&0.0377&1.0000&0.8254&0.6976&0.7540&0.7212\\
&\multicolumn{1}{c|}{$23.40$}&0.0743&0.8254&1.0000&0.5165&0.7632&0.8008\\
\hline 
\multirow{3}{*}{$f_0$}&\multicolumn{1}{c|}{$17.60$}&0.0141&0.6976&0.5165&1.0000&0.8118&0.7117\\
&\multicolumn{1}{c|}{$20.80$}&0.0133&0.7540&0.7632&0.8118&1.0000&0.9699\\
&\multicolumn{1}{c|}{$23.40$}&0.0167&0.7212&0.8008&0.7117&0.9699&1.0000\\
\hline 
\hline 
\end{tabular}
\end{center}

  statistical
  \begin{center}
\normalsize\begin{tabular}{l|cl|cc|cccccccccccc}
\hline \hline 
\multicolumn{3}{c|}{}&\multicolumn{2}{c|}{$f_+$}&\multicolumn{3}{c}{$f_0$}\\
\cline{2-8}
\multicolumn{1}{c|}{}&$q^2\,[{\rm GeV}^2]$&&$17.60$&$23.40$&$17.60$&$20.80$&$23.40$\\
\cline{4-8}
\multicolumn{1}{c|}{}&&$\delta f_{+/0}$&0.0407&0.0698&0.0116&0.0139&0.0180\\
\hline 
\multirow{2}{*}{$f_+$}&\multicolumn{1}{c|}{$17.60$}&0.0407&1.0000&0.8254&0.6976&0.7540&0.7212\\
&\multicolumn{1}{c|}{$23.40$}&0.0698&0.8254&1.0000&0.5165&0.7632&0.8008\\
\hline 
\multirow{3}{*}{$f_0$}&\multicolumn{1}{c|}{$17.60$}&0.0116&0.6976&0.5165&1.0000&0.8118&0.7117\\
&\multicolumn{1}{c|}{$20.80$}&0.0139&0.7540&0.7632&0.8118&1.0000&0.9699\\
&\multicolumn{1}{c|}{$23.40$}&0.0180&0.7212&0.8008&0.7117&0.9699&1.0000\\
\hline 
\hline 
\end{tabular}
\end{center}

  fit systematic
  \begin{center}
\normalsize\begin{tabular}{l|cl|cc|cccccccccccc}
\hline \hline 
\multicolumn{3}{c|}{}&\multicolumn{2}{c|}{$f_+$}&\multicolumn{3}{c}{$f_0$}\\
\cline{2-8}
\multicolumn{1}{c|}{}&$q^2\,[{\rm GeV}^2]$&&$17.60$&$23.40$&$17.60$&$20.80$&$23.40$\\
\cline{4-8}
\multicolumn{1}{c|}{}&&$\delta f_{+/0}$&0.0240&0.0700&0.0139&0.0172&0.0213\\
\hline 
\multirow{2}{*}{$f_+$}&\multicolumn{1}{c|}{$17.60$}&0.0240&1.0000&0.9974&0.9461&0.9419&0.9368\\
&\multicolumn{1}{c|}{$23.40$}&0.0700&0.9974&1.0000&0.9229&0.9178&0.9115\\
\hline 
\multirow{3}{*}{$f_0$}&\multicolumn{1}{c|}{$17.60$}&0.0139&0.9461&0.9229&1.0000&0.9999&0.9995\\
&\multicolumn{1}{c|}{$20.80$}&0.0172&0.9419&0.9178&0.9999&1.0000&0.9998\\
&\multicolumn{1}{c|}{$23.40$}&0.0213&0.9368&0.9115&0.9995&0.9998&1.0000\\
\hline 
\hline 
\end{tabular}
\end{center}

  flat systematic
\end{table}

\begin{figure*}
  \includegraphics[width=.49\textwidth]{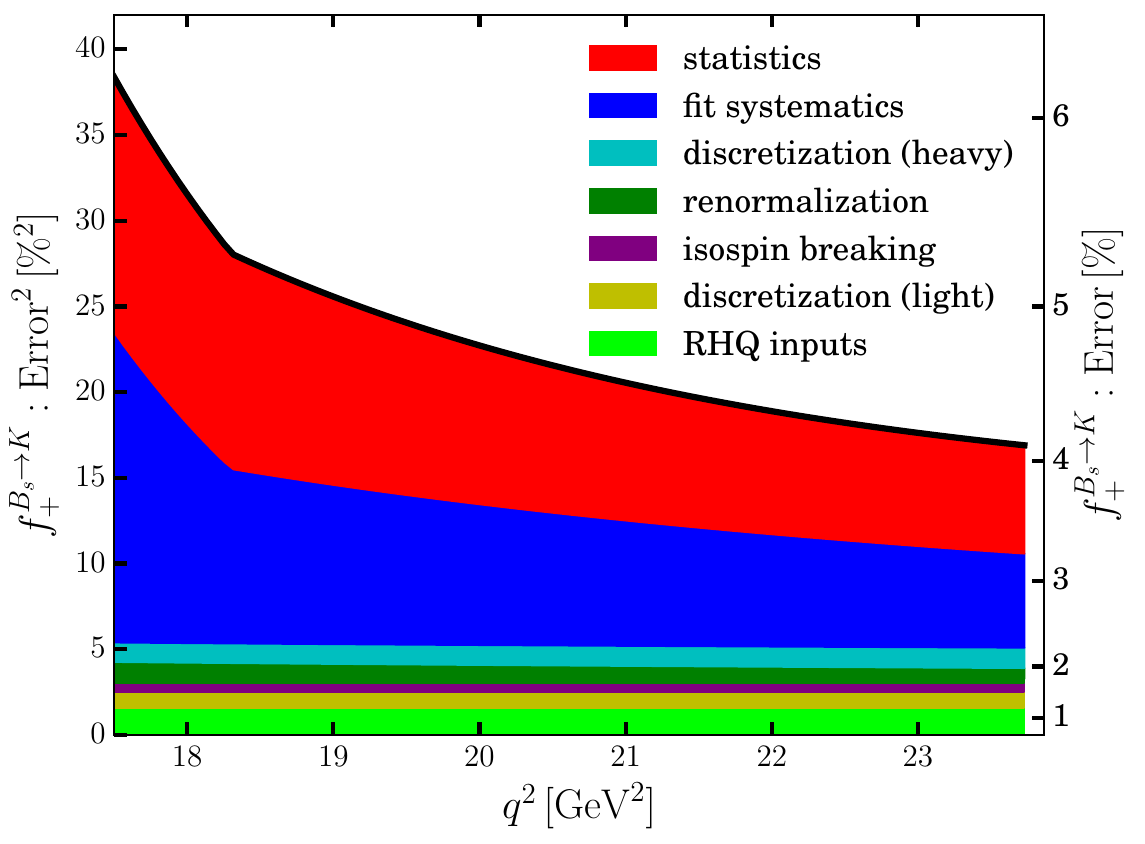}
  \includegraphics[width=.49\textwidth]{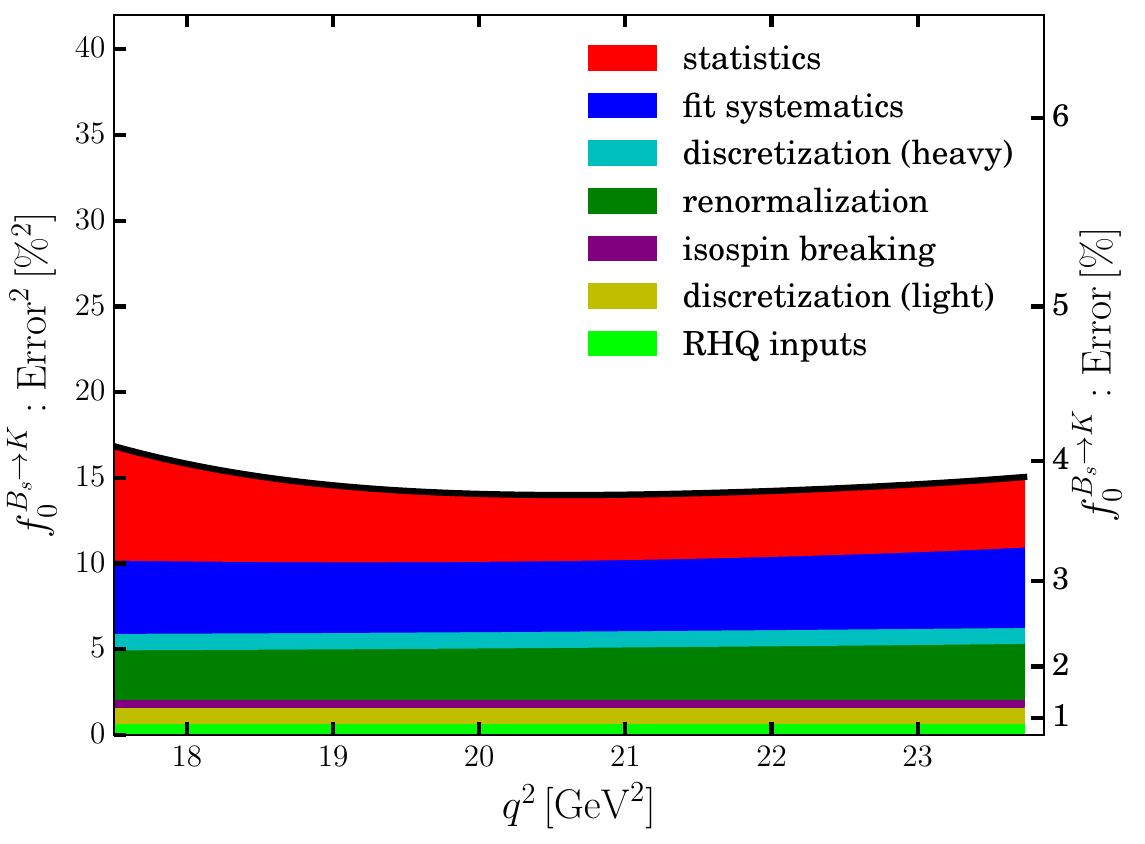}
  \caption{Cumulative systematic errors for the chiral- and
    continuum-extrapolated form factor for $\BstoKlnu$.}
  \label{fig:BstoK-cumulative-errors}
\end{figure*}

In the next section we will fit a $z$-expansion to synthetic data for the
physical form factors in the continuum and infinite-volume limit, generated at
selected values of $q^2$, to extend our form factors to the full kinematic
range. Because we first extrapolate to the continuum limit and then perform the
z-expansion to extend the form factors over the full kinematic range, the number
of available independent reference $q^2$ values is restricted by the number of
resolved parameters in the chiral-continuum limit of the HM$\chi$PT description
of the lattice data. Since we resolve 3 (2) parameters that parameterize the
continuum form factors $f_0$ ($f_+$), the number of synthetic data points we can
choose is limited by this. We account for the correlations between the form
factors at these $q^2$ values as follows.

The total error budget can be divided into three major contributions: the
statistical uncertainty, the uncertainties associated to the chiral-continuum
extrapolation and the uncertainties that are estimated as a constant $q^2$
independent percentage of the form factors. We refer to these as statistical,
fit systematic and flat systematic. Whilst most of the latter are estimated as a
percentage uncertainty on the form factors $f_+$ and $f_0$, the contributions
described in Secs.~\ref{sec:HeavyQuarkDiscErrors} and~\ref{sec:renormfactor} are
determined with respect to the form factors $f_\parallel$ and $f_\perp$, which
induces a mild $q^2$ dependence when this is related to $f_+$ and $f_0$. The
resulting cumulative systematic errors for $\BstoKlnu$ are illustrated in
Fig.~\ref{fig:BstoK-cumulative-errors}. We now provide more detail on estimating
the correlations of the different sources of uncertainties at the reference
values.

It is straightforward to obtain the statistical correlations from the bootstrap
analysis of the chiral and continuum fit in Sec.~\ref{subsec:BstoK-fit}. 

The systematic error for the chiral-continuum extrapolation is found by varying
the fit function and parametric inputs, as described above.  This does not
provide information on correlations between different $q^2$-values.  However,
alternate chiral-continuum fits with different fit functions exhibit very
similar statistical correlations between $q^2$-bins. Hence we take the
statistical correlation matrix from our preferred fit and multiply it by the
estimated chiral-continuum extrapolation error at each $q^2$ value. For
off-diagonal elements of the correlation matrix we use the product
$\sigma_{q^2_i} \sigma_{q^2_j}$. 

We assume each of the remaining ``flat'' systematic errors to be $100\%$
correlated and add the corresponding covariance matrices.  Where the per-cent
error is given for $f_\|$ and $f_\perp$, we first construct the corresponding
covariance matrix for $f_\|$ and $f_\perp$ and then propagate the correlated
error in order to obtain the covariance matrix for $f_+$ and $f_0$.
Table~\ref{tab:corre_mats} shows the resulting statistical and systematic
correlation matrices, which enable the full reconstruction of the total
covariance matrices using the form factor values and errors from
table~\ref{tab:ErrorBudget}.

\section{Phenomenological applications}\label{sec:pheno}

\subsection{\texorpdfstring{$z$}{z}-expansions of form factors}

Once we have continuum values for the form factors, we extrapolate them
over the entire physical range of $q^2$ using a fit to a
$z$-expansion~\cite{Bourrely:1980gp, Boyd:1994tt, Boyd:1995sq, Lellouch:1995yv,
  Boyd:1997qw, Caprini:1997mu, Arnesen:2005ez, Bourrely:2008za}.  The squared
momentum transfer, $q^2$, is mapped to the variable $z$ using
\begin{equation}
\label{eq:z-fn-defn}
  z(q^2;t_*,t_0) = \frac{\sqrt{t_*-q^2} - \sqrt{t_*-t_0}}%
                      {\sqrt{t_*-q^2} + \sqrt{t_*-t_0}}\,.
\end{equation}
This transformation maps the complex $q^2$ plane, with a cut for
$q^2\geq t_*$, onto the unit disk in $z$. For use below we set $t_\pm =
(M_{B_s}\pm M_K)^2$, with $t_- = q^2_\text{max}$, while $t_*$ is fixed
by the appropriate two-particle production threshold
$t_*=(M_B+M_\pi)^2$. The value of $t_0$ can be chosen to fix the range
in $z$ corresponding to a given range in $q^2$. We choose
\begin{equation}
  \label{eq:t-opt}
  t_0 = t_\mathrm{opt} = t_* - \sqrt{t_*(t_*-t_-)}\,.
\end{equation}
This symmetrizes the range of $z$ about zero for $0\leq q^2\leq
q^2_\text{max}$ which is mapped onto the real axis $0.2\gtrsim z
\gtrsim -0.2$, indicating that an expansion of the form factor in $z$
rather than in $q^2$ might converge quickly. This motivated Boyd,
Grinstein and Lebed (BGL)~\cite{Boyd:1994tt} to write the form factor
as
\begin{equation}
  \label{eq:BGLparameterization}
  f_X(q^2) = \frac1{B_X(q^2)\phi_X(q^2,t_0)} \sum_{n\geq0}
  a_{X,n}(t_0)z^n,
\end{equation}
where $\phi_X(q^2,t_0)$ is known as the outer function, with
expressions given in Appendix~\ref{appx:BGL-fits}. For $X=+,0$, the
Blaschke factor $B_X(q^2)$ is chosen to vanish at the positions of
sub-threshold poles $M^X_{i}$,
\begin{equation}
B_X(q^2)=\prod\limits_{i\in X\, {\rm poles}}z\left(q^2;t_\ast, \left(M^X_{i}\right)^2\right)\,.
\end{equation}
For $f_+$ the measured $1^-$ vector-meson with
$M^+_{B^\ast(1^-)}=5.32471\gev$~\cite{ParticleDataGroup:2022pth} sits above the
physical semileptonic region $0\le q^2\le q^2_{\rm max}$, but also below the
$B\pi$ threshold. Specifically, $q^2_{\rm max}\le (M^+_{B^\ast(1^-)})^2\le
t_\ast \to 23.73\gev^2\le 28.35\gev^2\le29.35\gev^2$ . We cancel this pole
through the corresponding Blaschke factor $B_+(q^2)$ prior to expanding in $z$.
For $f_0$ the theoretically predicted pole mass
$M^0_{B^\ast(0^+)}=5.63\gev$~\cite{Bardeen:2003kt} sits above the $B\pi$
threshold and no pole needs to be cancelled.

The following unitarity constraint applies:
\begin{equation}\label{eq:vanilla unitarity constraint}
  \frac1{2\pi i}\oint_C\frac {dz}{z}\theta_z
  |B_X(q^2)\phi_X(q^2,t_0)f_X(q^2)|^2\le 1 \,,
\end{equation}
with $C$ the unit circle, $\theta_z \equiv \theta(\alpha_{B_sK}-|{\rm arg}[z]|)$
and $\alpha_{B_sK}={\rm arg}[z(t_+,t_\ast,t_0)]$. Since for the $\BstoKlnu$
transition the two-particle $B\pi$ production threshold lies below the $B_sK$
threshold, \emph{i.e.}~$t_\ast<t_+$, the unitarity constraint originally
proposed by BGL~\cite{Boyd:1994tt} needs to be modified. This is similar to the
situation discussed in Refs.~\cite{Berns:2018vpl,Gubernari:2020eft,
  Gubernari:2022hxn,Blake:2022vfl}, but note some differences in notation in
those papers, in particular our use of $t_\ast$ and $t_+$ for the locations of
the $B\pi$ and $B_s K$ production thresholds, respectively. The step function
$\theta_z$ achieves this by restricting the integration over the unit circle to
the relevant arc. Let us now define the inner product
\begin{equation}
  \begin{aligned}
  \langle z^i|z^j\rangle_\alpha &=
  \frac 1{2\pi}\int\limits_{-\alpha}^\alpha d\phi (z^i)^\ast
  z^j|_{z=e^{i\phi}}\,,\\
  &= \begin{cases}
    \displaystyle
    \frac{\sin(\alpha(i-j))}{\pi(i-j)} & i\neq j\\
    \displaystyle\frac\alpha\pi & i=j
  \end{cases}
  \end{aligned}
\end{equation}
on the arc $[-\alpha,+\alpha]$ of the unit circle. When the inner
product is defined over the entire unit circle, $[-\pi,+\pi]$, the
monomials $z^i$ are orthonormal, $\langle
z^i|z^j\rangle_\pi=\delta_{ij}$. In that case the unitarity constraint
Eq.~(\ref{eq:vanilla unitarity constraint}) becomes
$\sum_i |a_{X,i}|^2 \leq 1$. With the restriction to the arc
$[-\alpha_{B_sK},+\alpha_{B_sK}]$, the modified BGL unitarity constraint
developed in~\cite{FittingPaper} is
\begin{equation}\label{eq:modified unitarity constraint}
  \sum\limits_{i,j\ge 0}a_{X,i}^\ast \langle
  z^i|z^j\rangle_{\alpha_{B_sK}} a_{X,j}
  \equiv |{\bf a}_X|_{\alpha_{B_sK}}^2 \le 1\,,
  \end{equation}
where we have defined $|{\bf a}_X|_{\alpha_{B_sK}}^2$ to mean the
quadratic form on the left-hand side.

\subsection{Extrapolation to \texorpdfstring{$q^2=0$}{q-squared = 0}}

To extrapolate our results to the full physical range of $q^2$ we
start from the results for $f_+$ and $f_0$ listed in
Tab.~\ref{tab:ErrorBudget}, with statistical and systematic errors and
correlations given in Tab.~\ref{tab:corre_mats}, added in quadrature.
Input parameters for the $z$ fits are summarized in Tab.~\ref{tab:zfit
  input}. We use the short-hand vector notation ${\bf f}=({\bf
  f}_+,{\bf f}_0)^T$ for the vector and scalar form factors at the
kinematical reference points, and denote the corresponding covariance
matrix by $C_{\bf f}$. We fit the data to a $z$-parameterization of
Eq.~\eqref{eq:BGLparameterization}, subject to the unitarity
constraint Eq.~(\ref{eq:modified unitarity constraint}) and the
kinematical constraint $f_+(0) =f_0(0)$.
\begin{table}
\caption{Input masses for the BGL $z$ fits. Values are in
  GeV~\cite{ParticleDataGroup:2022pth}. The superscript $\dagger$ indicates,
  where the isospin-averaged mass has been taken.}\label{tab:zfit
  input}
\centering
\begin{tabular}{ccccccc}
\hline\hline
$M_{B}$&$M_{B_s}$&$M_\pi$&$M_K$&$M^+_{B^\ast(0^+)}$\\
\hline
$5.27950^\dagger$&$5.36682$&$0.138039^\dagger$&$0.495644^\dagger$&5.32471\\
\hline\hline
\end{tabular}
\end{table}

In the Bayesian-inference strategy for fitting form factors developed
in Ref.~\cite{FittingPaper} the unitarity constraint is implemented as
a flat prior, which acts as a regulator for the fitting problem. In
contrast to frequentist fits, this allows us to determine the
parameters of a BGL parameterization to arbitrarily high order,
removing errors from truncating the power series in $z$ in
Eq.~(\ref{eq:BGLparameterization}).
 
The Bayesian-inference problem of determining the BGL parameters
${\bf  a}=({\bf a}_+,{\bf a}_0)^T$ and functions $g({\bf a})$ of them
amounts to computing expectation values
\begin{equation}\label{eq:Bayes expectation}
    \langle g({\bf a})\rangle =\mathcal{N} \int d{\bf a} \,g({\bf a})\,\pi({\bf a}|{\bf f},C_{\bf f})\pi_{\bf a}\,,
\end{equation}
where $\mathcal{N}$ is a normalization constant.  As prior knowledge on the form
factor we use only the unitarity constraint expressed in terms of the
distribution
\begin{equation}
  \pi_{\bf a}\propto \theta\left(1-|{\bf a}_{+}|_{\alpha_{B_sK}}^2\right)\theta\left(1-|{\bf a}_{0}|_{\alpha_{B_sK}}^2\right)\,,
\end{equation}
which essentially limits the integration range in Eq.~(\ref{eq:Bayes
  expectation}).  The conditional probability density for the parameter
${\bf a}$ given the fit model and data is
\begin{align}
  \pi({\bf  a}|{\bf  f},C_{{\bf f}})&\propto {\rm exp}\left(-\frac 12 \chi^2({\bf a},{\bf f})\right)\,,
\end{align}
where 
\begin{equation}\label{eq:chisq}
  \chi^2({\bf a},{\bf f})=({\bf  f}-Z{\bf  a})^TC_{{\bf f}}^{-1}({\bf  f}-Z{\bf  a})\,.
\end{equation}
Following Ref.~\cite{FittingPaper},  the matrix $Z$ consists of diagonal blocks 
\begin{equation}
  (Z_{XX})_{ij}=\frac{ z^j}{B_X(q_i^2)\phi_X(q_i^2,t_0)},
\end{equation}
where $XX$ is either $++$ or $00$, for the vector and scalar form factors,
respectively. The off-diagonal blocks, which implement the kinematical
constraint $f_+(0)=f_0(0)$ are
\begin{align}
  (Z_{+0})_{ij}=&0\,,\nonumber\\
  \\[-4mm]
  (Z_{0+})_{ij}=&\frac{1}{z(0;t_\ast,M_{B^\ast}^2)\phi_+(0)}\frac{\phi_0(0)}{\phi_0(q_i^2)}z^j(0)\,.\nonumber
\end{align}

The integral in Eq.~(\ref{eq:Bayes expectation}) can be performed by
Monte Carlo, which corresponds to drawing multi-variate
normal-distributed pseudo-random numbers. An efficient algorithm and
an implementation in Python are presented in
Refs.~\cite{FittingPaper,FittingPaperCode}. The results presented here
are based on 2000 samples.
\begin{figure*}
  \includegraphics[width=17cm]{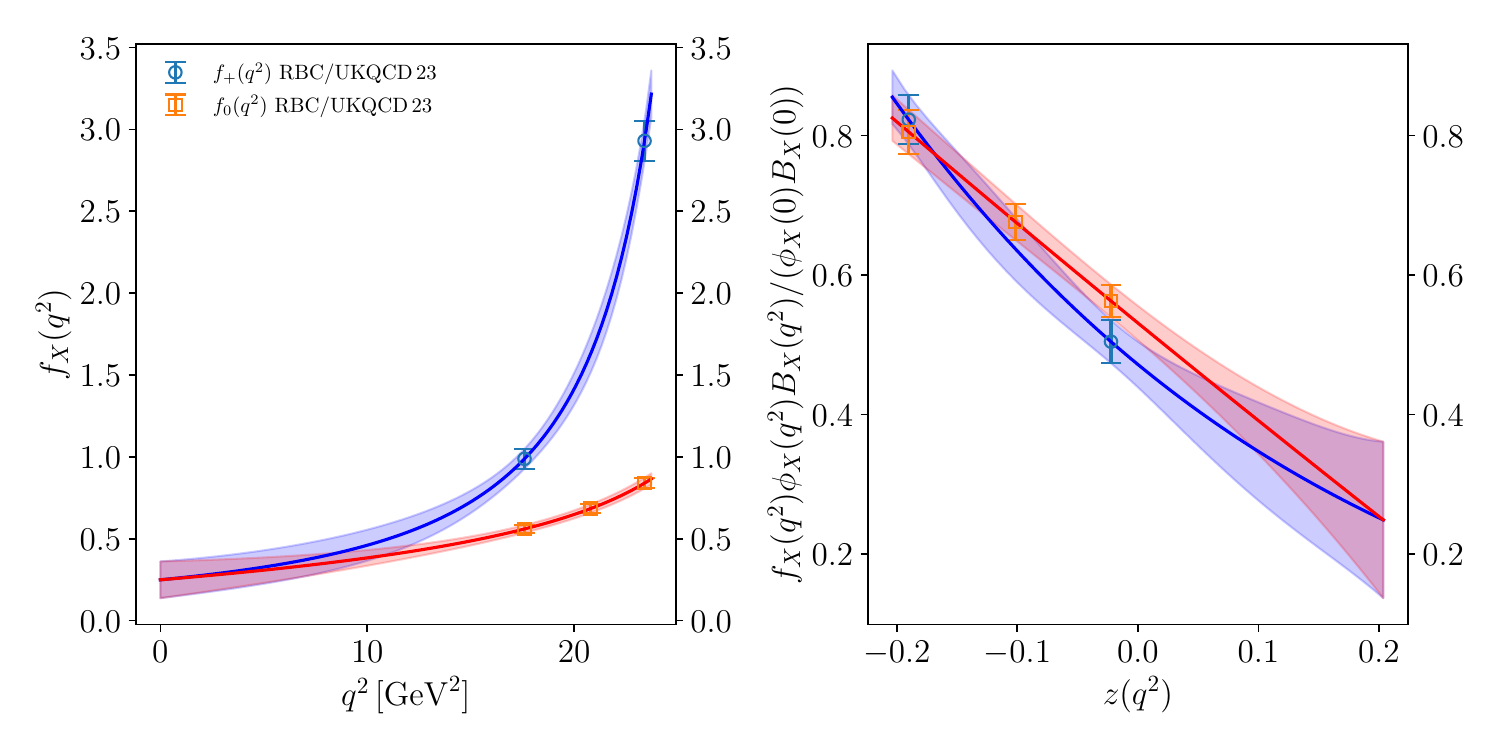}
  \caption{
    $z$-fits to $\BstoKlnu$ form factors for $(K_+,K_0)=(5,5)$, plotted against
    $q^2$ (left) and against $z$ (right), where in the latter plot the outer
    function and Blaschke factor have been removed. Blue denotes $f_+$ and red
    $f_0$.}
  \label{fig:BstoK-zfit}
\end{figure*}

Figure~\ref{fig:BstoK-zfit} shows the results of $z$-fits to our $\BstoKlnu$
data, with numerical values for the fit parameters in
table~\ref{tab:BstoK_zfit_results}.

For the discussion of the results it is also worthwhile, in parallel, to have a
look at the first data column of Tab.~\ref{tab:BstoK_zfit_observables}, which
shows the result for the form factor extrapolated to $q^2=0$. For both the
coefficients ${\bf a}$ and $f_+(0)$ we find significant variation in both error
and central value when increasing the order of the $z$ expansion from
$K_{+,0}=2$. We find stable central values and errors for $K\ge 3$.
Higher-order coefficients can be added to the fit (the tables show results up to
$K_{+,0}=10$), whereby the errors on the significantly-determined lower-order
coefficients and also the result for $f_{+,0}(0)$ remain stable, and the
higher-order coefficients are compatible with zero. We conclude that the form
factor parameterization determined in this way becomes truncation independent
for large-enough $K_{+,0}$. For the following analyses we use results with
$(K_+,K_0)= (5,5)$.

Furthermore, the value of the form factors at $q^2=0$ is of interest for
comparison with predictions from different theoretical methods.  Using light
cone sum rules, Duplancic/Melic report $f(0)=0.30
(^{+4}_{-3})$~\cite{Duplancic:2008tk} and Khodjamirian/Rusov quote
$f(0)=0.336(23)$~\cite{Khodjamirian:2017fxg}.  Based on a relativistic quark
model Faustov/Galkin predict $f(0)=0.284(14)$~\cite{Faustov:2013ima}, where the
NLO perturbative-QCD result by Wang/Xiao is
$f(0)=0.26(^{+4}_{-3})(2)$~\cite{Wang:2012ab}.  All these values are compatible
with our $(K_+,K_0)=(5,5)$ result
\begin{align}
  f(q^2=0) = 0.25(11),
\end{align}
however, our uncertainty at $q^2=0$ is substantially larger than for these other
predictions.

\begin{table*}
  \caption{Results for $z$-fits to data for scalar and vector form factors for
    $\BstoKlnu$. The main results of this paper are the ones with
    $(K_+,K_0)=(5,5)$.   The data for this case and
    the corresponding correlation matrix is contained in the accompanying data file. }
  \label{tab:BstoK_zfit_results}
  \begin{tabular}{l@{\hspace{1mm}}llllllllllllllllllllllllllllllllllllllllllllllllll}
\hline\hline
$K_+$&$K_0$&\multicolumn{1}{c}{$a_{+,0}$}&\multicolumn{1}{c}{$a_{+,1}$}&\multicolumn{1}{c}{$a_{+,2}$}&\multicolumn{1}{c}{$a_{+,3}$}&\multicolumn{1}{c}{$a_{+,4}$}&\multicolumn{1}{c}{$a_{+,5}$}&\multicolumn{1}{c}{$a_{+,6}$}&\multicolumn{1}{c}{$a_{+,7}$}&\multicolumn{1}{c}{$a_{+,8}$}&\multicolumn{1}{c}{$a_{+,9}$}&\\
\hline
2&2&0.0293(11)&-0.0871(46)&- &- &- &- &- &- &- &-&\\
2&3&0.0249(16)&-0.0999(57)&- &- &- &- &- &- &- &-&\\
3&2&0.0245(16)&-0.0799(50)&0.093(21)&- &- &- &- &- &- &-&\\
3&3&0.0245(15)&-0.078(12)&0.101(49)&- &- &- &- &- &- &-&\\
3&4&0.0246(16)&-0.078(16)&0.100(70)&- &- &- &- &- &- &-&\\
4&3&0.0246(17)&-0.075(31)&0.102(49)&-0.07(72)&- &- &- &- &- &-&\\
4&4&0.0246(17)&-0.077(32)&0.100(68)&-0.03(70)&- &- &- &- &- &-&\\
5&5&0.0246(17)&-0.074(31)&0.099(70)&-0.08(67)&0.05(70)&- &- &- &- &-&\\
6&6&0.0247(16)&-0.073(32)&0.101(69)&-0.10(69)&0.09(74)&-0.05(71)&- &- &- &-&\\
7&7&0.0247(17)&-0.071(33)&0.107(70)&-0.11(72)&0.08(89)&-0.04(89)&0.03(73)&- &- &-&\\
8&8&0.0248(17)&-0.068(35)&0.102(74)&-0.18(77)&0.2(1.1)&-0.2(1.3)&0.1(1.2)&-0.06(82)&- &-&\\
9&9&0.0248(18)&-0.068(38)&0.107(85)&-0.16(82)&0.2(1.4)&-0.2(1.9)&0.1(1.9)&-0.1(1.5)&0.03(89)&-&\\
10&10&0.0247(18)&-0.067(43)&0.112(95)&-0.15(90)&0.2(1.8)&-0.2(2.6)&0.1(2.9)&-0.1(2.7)&-0.0(1.9)&0.02(98)&\\
\hline\hline\\
\end{tabular}

  \begin{tabular}{l@{\hspace{1mm}}llllllllllllllllllllllllllllllllllllllllllllllllll}
\hline\hline
$K_+$&$K_0$&\multicolumn{1}{c}{$a_{0,0}$}&\multicolumn{1}{c}{$a_{0,1}$}&\multicolumn{1}{c}{$a_{0,2}$}&\multicolumn{1}{c}{$a_{0,3}$}&\multicolumn{1}{c}{$a_{0,4}$}&\multicolumn{1}{c}{$a_{0,5}$}&\multicolumn{1}{c}{$a_{0,6}$}&\multicolumn{1}{c}{$a_{0,7}$}&\multicolumn{1}{c}{$a_{0,8}$}&\multicolumn{1}{c}{$a_{0,9}$}&\\
\hline
2&2&0.0981(36)&-0.286(14)&- &- &- &- &- &- &- &-&\\
2&3&0.0917(39)&-0.331(19)&-0.211(53)&- &- &- &- &- &- &-&\\
3&2&0.0950(37)&-0.263(15)&- &- &- &- &- &- &- &-&\\
3&3&0.0953(43)&-0.254(41)&0.02(13)&- &- &- &- &- &- &-&\\
3&4&0.0955(44)&-0.254(42)&0.02(22)&-0.02(60)&- &- &- &- &- &-&\\
4&3&0.0954(43)&-0.254(40)&0.03(12)&- &- &- &- &- &- &-&\\
4&4&0.0953(42)&-0.254(42)&0.02(21)&-0.02(60)&- &- &- &- &- &-&\\
5&5&0.0954(44)&-0.254(41)&0.02(21)&-0.01(55)&-0.00(62)&- &- &- &- &-&\\
6&6&0.0957(42)&-0.251(41)&0.04(21)&-0.01(52)&-0.06(65)&0.07(65)&- &- &- &-&\\
7&7&0.0955(44)&-0.250(40)&0.06(20)&0.05(50)&-0.13(72)&0.17(79)&-0.12(69)&- &- &-&\\
8&8&0.0954(43)&-0.250(41)&0.06(22)&0.06(50)&-0.18(84)&0.2(1.0)&-0.21(99)&0.10(74)&- &-&\\
9&9&0.0956(44)&-0.247(41)&0.08(23)&0.06(50)&-0.27(96)&0.4(1.4)&-0.4(1.5)&0.3(1.2)&-0.15(80)&-&\\
10&10&0.0956(42)&-0.245(42)&0.11(24)&0.11(49)&-0.4(1.1)&0.7(1.8)&-0.8(2.2)&0.7(2.1)&-0.4(1.5)&0.16(87)&\\
\hline\hline\\
\end{tabular}

\end{table*}

\begin{table*}
  \caption{Results for observables based on the results for the $z$ fits in
    Tab.~\ref{tab:BstoK_zfit_results}. The main results of this paper are the
    ones with $(K_+,K_0)=(5,5)$.}
  \label{tab:BstoK_zfit_observables}
  \begin{tabular}{l@{\hspace{1mm}}llllllllll}
\hline\hline
$K_+$&$K_0$&\multicolumn{1}{c}{$f(q^2=0)$}&\multicolumn{1}{c}{$R_{B_s\to K}^{\rm impr}$}&\multicolumn{1}{c}{$R_{B_s\to K}$}&\multicolumn{1}{c}{$\frac{\Gamma^\tau}{|V_{ub}|^2}\,[\frac 1{\rm ps}]$}&\multicolumn{1}{c}{$\frac{\Gamma^\mu}{|V_{ub}|^2}\,[\frac 1{\rm ps}]$}&\multicolumn{1}{c}{$V^{\rm low}_{\rm CKM}$}&\multicolumn{1}{c}{$V^{\rm high}_{\rm CKM}$}&\multicolumn{1}{c}{$V^{\rm full}_{\rm CKM}$}&\\
\hline
2&2&0.222(21)&1.545(17)&0.741(19)&5.37(43)&7.25(70)&0.00356(39)&0.00325(30)&0.00336(32)\\
2&3&0.087(39)&1.657(46)&0.954(75)&3.70(50)&3.94(81)&0.0070(22)&0.00408(46)&0.00420(52)\\
3&2&0.231(21)&1.721(57)&0.774(27)&4.34(45)&5.62(72)&0.00375(42)&0.00382(41)&0.00379(39)\\
3&3&0.248(88)&1.721(56)&0.76(10)&4.48(72)&6.1(1.7)&0.0039(14)&0.00381(46)&0.00381(52)\\
3&4&0.25(12)&1.722(64)&0.77(15)&4.51(84)&6.2(2.3)&0.0042(22)&0.00380(48)&0.00382(53)\\
4&3&0.249(86)&1.72(12)&0.76(12)&4.55(82)&6.3(2.0)&0.0039(16)&0.00378(53)&0.00379(59)\\
4&4&0.25(12)&1.72(12)&0.78(17)&4.53(89)&6.3(2.4)&0.0043(29)&0.00381(57)&0.00383(62)\\
5&5&0.25(11)&1.72(11)&0.77(16)&4.57(90)&6.4(2.4)&0.0041(24)&0.00376(55)&0.00378(61)\\
6&6&0.26(11)&1.71(11)&0.76(16)&4.63(88)&6.5(2.4)&0.0040(26)&0.00375(54)&0.00376(58)\\
7&7&0.26(11)&1.71(11)&0.75(15)&4.67(90)&6.7(2.4)&0.0038(19)&0.00373(56)&0.00374(62)\\
8&8&0.26(11)&1.70(12)&0.74(15)&4.71(94)&6.8(2.6)&0.0038(19)&0.00371(55)&0.00372(62)\\
9&9&0.27(11)&1.70(12)&0.74(16)&4.76(98)&7.0(2.7)&0.0038(20)&0.00370(59)&0.00371(66)\\
10&10&0.28(11)&1.71(13)&0.73(16)&4.80(99)&7.1(2.8)&0.0037(31)&0.00368(58)&0.00368(62)\\
\hline\hline\\
\end{tabular}

\end{table*}

\begin{table*}
  \caption{Results for observables based on the results for the $z$ fits in
    Tab.~\ref{tab:BstoK_zfit_results}. The main results of this paper are the
    ones with $(K_+,K_0)=(5,5)$.}
  \label{tab:BstoK_zfit_observables2}
  \begin{tabular}{l@{\hspace{1mm}}llllllllll}
\hline\hline
$K_+$&$K_0$&\multicolumn{1}{c}{$I[\mathcal{A}_{\rm FB}^\tau]\,[\frac 1{\rm ps}]$}&\multicolumn{1}{c}{$I[\mathcal{A}_{\rm FB}^\mu]\,[\frac 1{\rm ps}]$}&\multicolumn{1}{c}{$\mathcal{\bar A}_{\rm FB}^\tau$}&\multicolumn{1}{c}{$\mathcal{\bar A}_{\rm FB}^\mu$}&\multicolumn{1}{c}{$I[\mathcal{A}_{\rm pol}^\tau]\,[\frac 1{\rm ps}]$}&\multicolumn{1}{c}{$I[\mathcal{A}_{\rm pol}^\mu]\,[\frac 1{\rm ps}]$}&\multicolumn{1}{c}{$\mathcal{\bar A}_{\rm pol}^\tau$}&\multicolumn{1}{c}{$\mathcal{\bar A}_{\rm pol}^\mu$}&\\
\hline
2&2&1.46(12)&0.0320(46)&0.2720(21)&0.00440(27)&0.794(92)&7.16(68)&0.148(13)&0.98768(73)\\
2&3&0.99(14)&0.0115(41)&0.2679(27)&0.00284(46)&0.31(13)&3.90(80)&0.082(27)&0.9912(11)\\
3&2&1.23(13)&0.0315(46)&0.2825(28)&0.00560(44)&0.14(15)&5.53(71)&0.031(34)&0.9838(13)\\
3&3&1.27(23)&0.038(19)&0.2836(77)&0.0058(15)&0.13(16)&6.0(1.7)&0.030(35)&0.9833(39)\\
3&4&1.28(27)&0.040(26)&0.2833(91)&0.0057(19)&0.14(17)&6.1(2.2)&0.030(38)&0.9834(49)\\
4&3&1.29(26)&0.038(19)&0.2820(80)&0.0058(16)&0.18(31)&6.2(2.0)&0.034(65)&0.9832(45)\\
4&4&1.28(28)&0.039(25)&0.2817(93)&0.0058(20)&0.16(31)&6.2(2.4)&0.031(64)&0.9833(52)\\
5&5&1.30(28)&0.040(24)&0.2821(89)&0.0057(18)&0.18(29)&6.3(2.3)&0.035(60)&0.9834(49)\\
6&6&1.31(28)&0.041(24)&0.2826(88)&0.0058(18)&0.19(29)&6.4(2.3)&0.036(58)&0.9832(48)\\
7&7&1.33(28)&0.043(24)&0.2831(85)&0.0060(18)&0.20(31)&6.6(2.4)&0.037(62)&0.9829(47)\\
8&8&1.34(29)&0.043(25)&0.2827(86)&0.0059(18)&0.23(32)&6.7(2.5)&0.042(64)&0.9831(47)\\
9&9&1.35(31)&0.045(27)&0.2830(90)&0.0060(18)&0.23(34)&6.8(2.6)&0.041(67)&0.9827(49)\\
10&10&1.37(31)&0.047(27)&0.2832(93)&0.0062(18)&0.23(36)&7.0(2.7)&0.040(69)&0.9823(49)\\
\hline\hline\\
\end{tabular}

\end{table*}

\subsection{Standard Model predictions}

With the parameterization of the form factors $f_+(q^2)$ and
$f_0(q^2)$ over the entire physical phase space $0\le q^2\le q^2_{\rm
  max}$ at hand, we consider a variety of phenomenological
applications.

\subsubsection{Determination of \texorpdfstring{$|V_{ub}|$}{|Vub|}}
The CKM matrix element $|V_{ub}|$ can be computed by comparing the differential
decay rate in Eq.~(\ref{eq:B_semileptonic_rate}) to experimental data for the
same exclusive decay. In practice, one compares the differential decay rate
after integrating over $q^2$ bins. To date, only data for the branching fraction
\begin{equation}
  R_{\rm BF}=\frac{\mathcal{B}(B_s^0\to K^-\mu^+\nu_\mu)}{\mathcal{B}(B_s^0\to D_s^-\mu^+\nu_\mu)}\,,
\end{equation}
for two $q^2$ bins from LHCb is available~\cite{Aaij:2020nvo}.  In particular,
for the low ($q^2\le7\gev^2$), high ($q^2\ge 7\gev^2$) and combined bins, we use 
\begin{align}
  R_{BF}^{\rm low}=&1.66(08)(09)\times10^{-3}\nonumber\,,\\
  R_{BF}^{\rm high}=&3.25(21)(^{+18}_{-19})\times10^{-3}\,,\\
  R_{BF}^{\rm total}=&4.89(21)(^{+24}_{-25})\times10^{-3}\nonumber\,,
\end{align}
where the first error is statistical and the second error the combined systematic uncertainty.
$|V_{ub}|$ can still be determined by combining these results with the branching
ratio~\cite{Aaij:2020hsi}
\begin{equation}
  \mathcal{B}(B_s^0\to D_s^-\mu^+\nu_\mu)= 2.49 (12)(21)\times 10^{-2}\,,
\end{equation}
and the $B_s^0$ lifetime $\tau_{B_s^0}=1.520(5)\times 10^{-12}$s \cite{HFLAV:2022pwe,ParticleDataGroup:2022pth},
\begin{equation}
	|V_{ub}|=\sqrt{\frac{R^{\rm bin}_{BF}\,\mathcal{B}(B_s^0\to D_s^-\mu^+\nu_\mu)}{\tau_{B_s^0}\,\Gamma^{\rm bin}_0(\BstoKlnu)}}\,,
\end{equation}
where bin is either `low` or `high`. The reduced decay rate $\Gamma^{\rm bin}_0=\Gamma^{\rm bin}/|V_{ub}|^2$ 
is obtained from the lattice computation and integrated over the range of the experimental bin.
We symmetrize errors and generate
a multivariate distribution for the branching fractions and the lifetime,
assuming no statistical correlation but the systematic errors to be 100\%
correlated (as \emph{e.g.}~in Ref.~\cite{Martinelli:2022tte}).  In this way we compute
results for $|V_{ub}|$ for both bins individually, and combined in terms of a
weighted average taking into account correlations. We present our results in
Tab.~\ref{tab:BstoK_zfit_observables}. Our final result, the one for
$(K_+,K_0)=(5,5)$ is
\begin{equation}\label{eq:ourVub}
  |V_{ub}|=3.78(61)\times 10^{-3}\,,
\end{equation}
and we emphasize that our predictions for the low and high $q^2$-bin are
consistent within uncertainties. Repeating the analysis with vanishing
experimental error the result would be $|V_{ub}|=3.73(37)\times 10^{-3}$,
indicating that at this stage the error on the result in Eq.~\eqnref{eq:ourVub}
is dominated by the experimental uncertainty.  For comparison and used in the
following section we quote the results for $|V_{ub}|$ from the exclusive and
inclusive analyses
\begin{align}
  |V_{ub}|_{\rm exclusive}^{\rm FLAG\, 21} =&\,3.74(17)\times 10^{-3}\,\textrm{
  \cite{Lattice:2015tia, Flynn:2015mha, delAmoSanchez:2010af, Lees:2012vv, Ha:2010rf, Sibidanov:2013rkk, FlavourLatticeAveragingGroupFLAG:2021npn}}\,\label{eq:Vubexclusive}\\
  |V_{ub}|_{\rm inclusive}^{\rm PDG\, 22} =&\,4.13(26)\times 10^{-3} \,\textrm{\cite{ParticleDataGroup:2022pth, HFLAV:2022pwe, Gambino:2007rp, Lange:2005yw, Andersen:2005mj}}\,\label{eq:Vubinclusive}.
\end{align}
These two results are compatible within less than 2$\sigma$.  Our result is
compatible with both values, albeit with a larger overall error.

\subsubsection{Differential decay width}
\begin{figure*}
  \includegraphics[width=8cm]{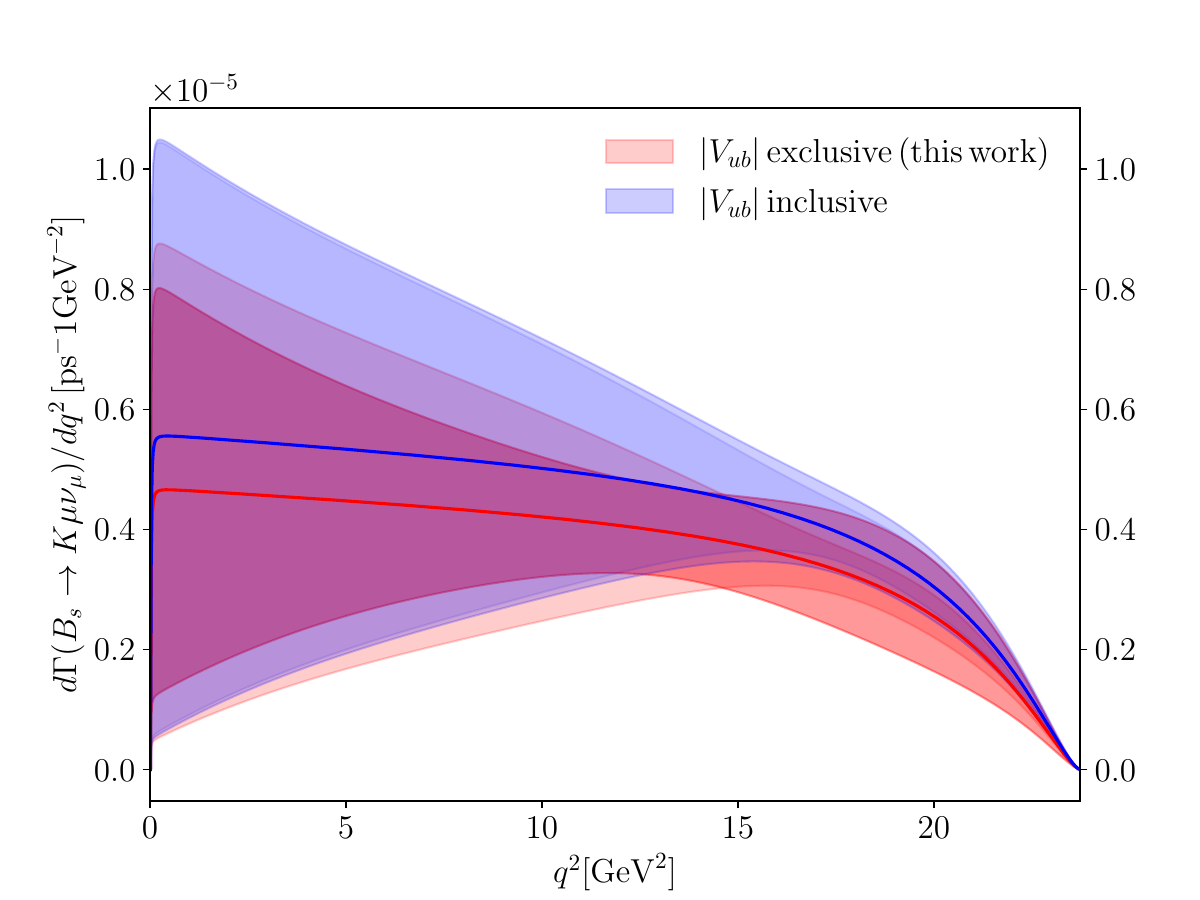}
  \includegraphics[width=8cm]{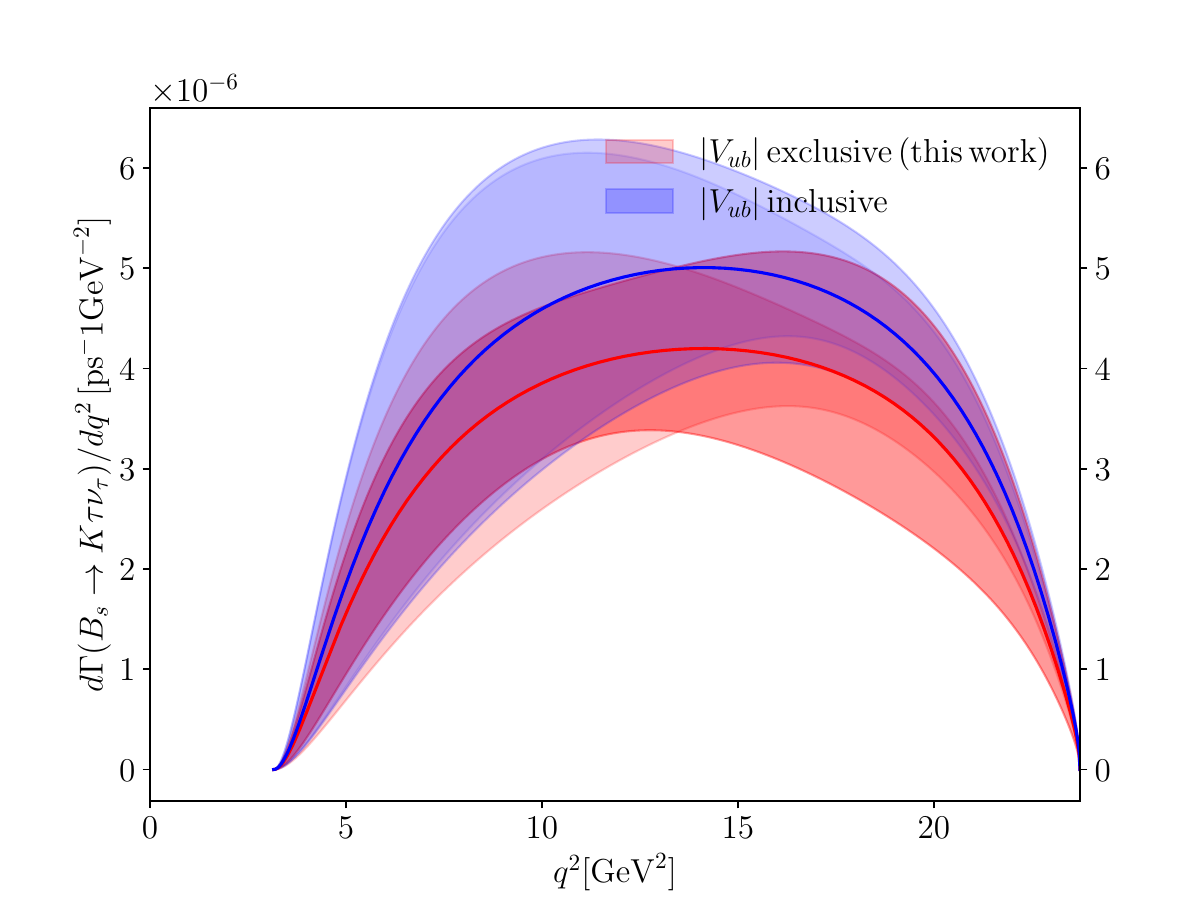}
  \caption{The differential decay width $d\Gamma/d q^2$ for $B_s\to K \mu
    \bar\nu_\mu$ (left) and $B_s\to K \tau \bar\nu_\tau$ (right).  The values
    for $|V_{ub}|$ are taken from Eqs.~\eqnref{eq:ourVub} and
    \eqnref{eq:Vubinclusive}. The darker (lighter) shading indicates the error
    without (with) the contribution from the error on $|V_{ub}|$.}
  \label{fig:dGamma/dqsq}
\end{figure*}

The information provided by the analysis of our lattice data allows to predict
the shape of the differential decay width $d\Gamma/dq^2$ in the SM.  Our results
are shown in Fig.~\ref{fig:dGamma/dqsq}, assuming our result for $|V_{ub}|$ from
Eq.~\eqnref{eq:ourVub} and the result in Eq.~\eqnref{eq:Vubinclusive},
respectively. At the current level of precision the shapes of the inclusive and
exclusive decay rates are compatible with each other. Detailed studies of
decay-rate shapes could in the future, when higher-precision theory predictions
are available, allow to shed light on possible discrepancies between inclusive and
exclusive CKM determinations.

\subsubsection{\texorpdfstring{$R$}{R} ratios}
A second, very important application is to test lepton flavor universality
(LFU). LFU is an accidental symmetry in the SM and it is extremely important to
test if it holds. One test is to compare these semileptonic decays with electron
($e$), muon ($\mu$) or tau ($\tau$) leptons in the final state. In the SM their
couplings with gauge bosons ($W$, $Z$) are identical. However, since their
masses differ, the shapes of the partial width distributions with respect to
$q^2$ will be different and so will be their integrated (or partially
integrated) rates. Comparison of measured and predicted rates constitutes
another important test of the SM. Particularly interesting and important is to
take ratios of integrated rates which are manifestly independent of the mixing
angles. Since the mixing angles are known within some uncertainty, LFU tests
using the following ratios can be a powerful precision test. Traditionally one
introduces,
\begin{equation}
  R_{B_s\to K}= \frac{\int_{m_\tau^2}^{q^2_\text{max}}dq^2\,
    \frac{d\Gamma(B_s\to K\tau\bar\nu_\tau)}{dq^2}}
  {\int_{m_\ell^2}^{q^2_\text{max}}dq^2\,
    \frac{d\Gamma(B_s \to K\ell\bar\nu_\ell)}{dq^2}}\,,
  \label{Eq.RBstoK}
\end{equation}
and takes the limit of integration from $q^2 = m_{\ell(\tau)}^2$ to the maximum
value of $q^2$ that is kinematically allowed. In this equation $\ell$ in the
denominator stands for $e$ or $\mu$, whereas the numerator is for the tau
lepton. Since the electron and muon masses are negligible compared to the mass
of the parent $B_s$, the contribution to the denominator from the form factor
$f_0$ is tiny since it is proportional to the lepton mass in the amplitude. This
means that for the numerator involving decays to $\tau$, the contribution from
the scalar form factor $f_0$ cannot be determined experimentally from data for
the semileptonic decays into $e$ or $\mu$. The scalar form factor must be
calculated from theory in a non-perturbative framework.  This realization
motivated lattice studies long ago~\cite{ElKhadra:1989iu}.

The conventional definition of $R_{B_s\to K}$ in Eq.~\eqref{Eq.RBstoK} has a
drawback. The contribution to the denominator from $m_\ell^2 \leq q^2 \leq m_\tau^2$
has no corresponding contribution in the numerator; thus that region
does not give useful information for testing LFU. To emphasize this, imagine
$d\Gamma/dq^2$ is very peaked for small $q^2$.  Then the conventional
$R_{B_s\to  K}$, given in~(\ref{Eq.RBstoK}), would tend to be very small, providing
little useful information.

Following Refs.~\cite{Atwood:1991ka,Isidori:2020eyd}, we propose to use another
ratio (\emph{c.f.}~Ref.~\cite{Flynn:2021ttz}) where we:
\begin{itemize}
\item use a common integration range in the numerator and denominator, with
  lower limit $\qsqmin \geq m_\tau^2$ (changing the lower limit was earlier
  proposed in Refs.~\cite{Freytsis:2015qca,Bernlochner:2016bci} as well as
  in Ref.~\cite{Isidori:2020eyd});
\item make the weights multiplying the form factor terms involving $f_+(q^2)$
  in the integrands the same for $\tau$ and $\ell$ modes (as
  in Ref.~\cite{Isidori:2020eyd}).
\end{itemize}
To do this, we rewrite the differential decay rate in
Eq.~\eqref{eq:B_semileptonic_rate} in the form
\begin{equation}
  \frac{d\Gamma(\BstoKlnu)}{dq^2} =
  \Phi\, \w_\ell(q^2) \big[ F_V^2 + (F_S^\ell)^2\big],
\end{equation}
where $\ell$ can be any of $e,\mu,\tau$ and
\begin{align}
  \Phi &= \eta_\text{EW} \frac{G_F^2 |V_{ub}|^2}{24\pi^3},\\
  \w_\ell(q^2) &= \bigg(1-\frac{m_\ell^2}{q^2}\bigg)^2
               \bigg(1+\frac{m_\ell^2}{2q^2}\bigg),\\             
  F_V^2 &= |{\bf p}_K|^3 |f_+(q^2)|^2,\\
  \label{eq:FSlcontrib}
  (F_S^\ell)^2 &= \frac{3}{4} \frac{m_\ell^2
            |{\bf p}_K|}{m_\ell^2+2q^2}
              \frac{(M_{B_s}^2-M_K^2)^2}{M_{B_s}^2}\, |f_0(q^2)|^2.
\end{align}
The subscript $\ell$ on $\w_\ell$ and superscript $\ell$ in $(F_S^\ell)^2$
show where the dependence on the lepton mass enters. The improved $R$ ratio is
now defined by:
\begin{equation}
  R^\text{imp}_{B_s\to K} =
  \frac{\int_{\qsqmin}^{\qsqmax} dq^2\,
    \frac{d\Gamma(B_s\to K\tau\bar\nu_\tau)}{dq^2}}
  {\int_{\qsqmin}^{\qsqmax} dq^2\,
     \left[\frac{\w_\tau(q^2)}{\w_\ell(q^2)}\right]\,
   \frac{d\Gamma(\BstoKlnu)}{dq^2}},
  \label{Eq.RKtopi_UN}
\end{equation}
where $\ell$ in the denominator is once again $e$ or $\mu$. With instead a
vector meson in the final state, this matches the definition in
Ref.~\cite{Isidori:2020eyd}. The ratio can be evaluated using experimentally
measured differential decay rates. We propose using this ratio as a way to
monitor LFU.

We can evaluate the ratio $R^\text{imp}_{B_s\to K}$ from the Standard Model
using our lattice determinations of the form factors. In the approximation that
we drop the scalar form factor contribution in the denominator
[in~\eqref{eq:FSlcontrib}, $m_{e,\mu}^2/2q^2 \leq m_\mu^2/2q^2 \leq
  m_\mu^2/2m_\tau^2 = 0.002$ in the integration range], we have
\begin{equation}
  R^\text{imp,SM}_{B_s\to K} \approx 1 + \frac{\int_{\qsqmin}^{\qsqmax} dq^2\,
    \w_\tau(q^2) (F_S^\tau)^2}
  {\int_{\qsqmin}^{\qsqmax} dq^2\,
     \w_\tau(q^2) F_V^2}\,,
\end{equation}
where now both numerator and denominator have the same weight. 

Results for both $R_{B_s\to K}$ and $R_{B_s\to K}^{\rm impr.}$ are listed in
Tab.~\ref{tab:BstoK_zfit_observables}, where we also include our result for the
integrated decay rate $\Gamma/|V_{ub}|$. As above, only the results for
$K_{+,0}\ge 3$ should be considered to be free of truncation errors in the $z$
expansion. The error we achieve on the new ratio is about three times smaller
than for the old ratio.  Here is a summary of our central results
(based on $(K_+,K_0)=(5,5)$):
\begin{align}
  R_{B_s\to K}=&\,{0.77(16)}\,,\\
  R_{B_s\to K}^{\rm impr}=&\,{1.72(11)}\,.
\end{align}

\subsubsection{Forward-backward and polarization asymmetries}
\begin{figure*}
  \includegraphics[width=8cm]{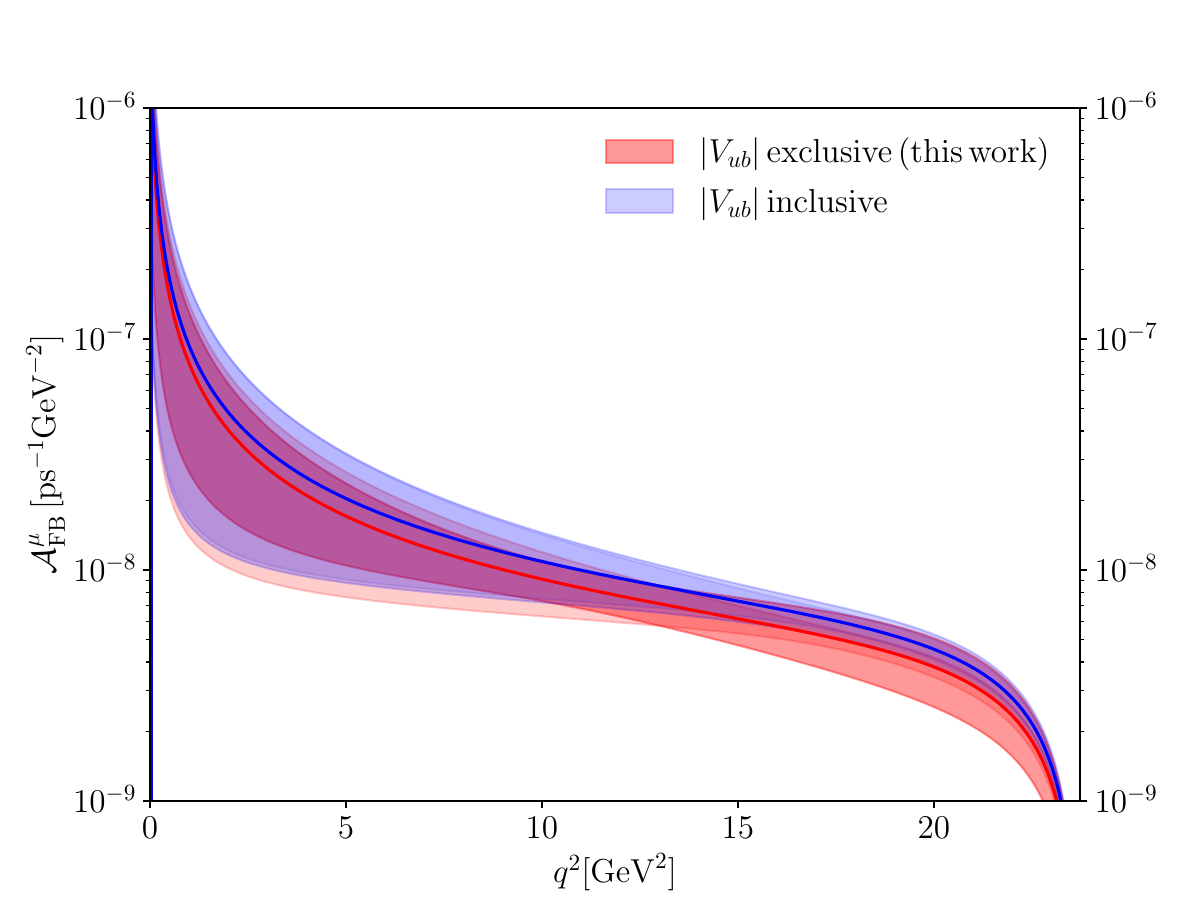}
  \includegraphics[width=8cm]{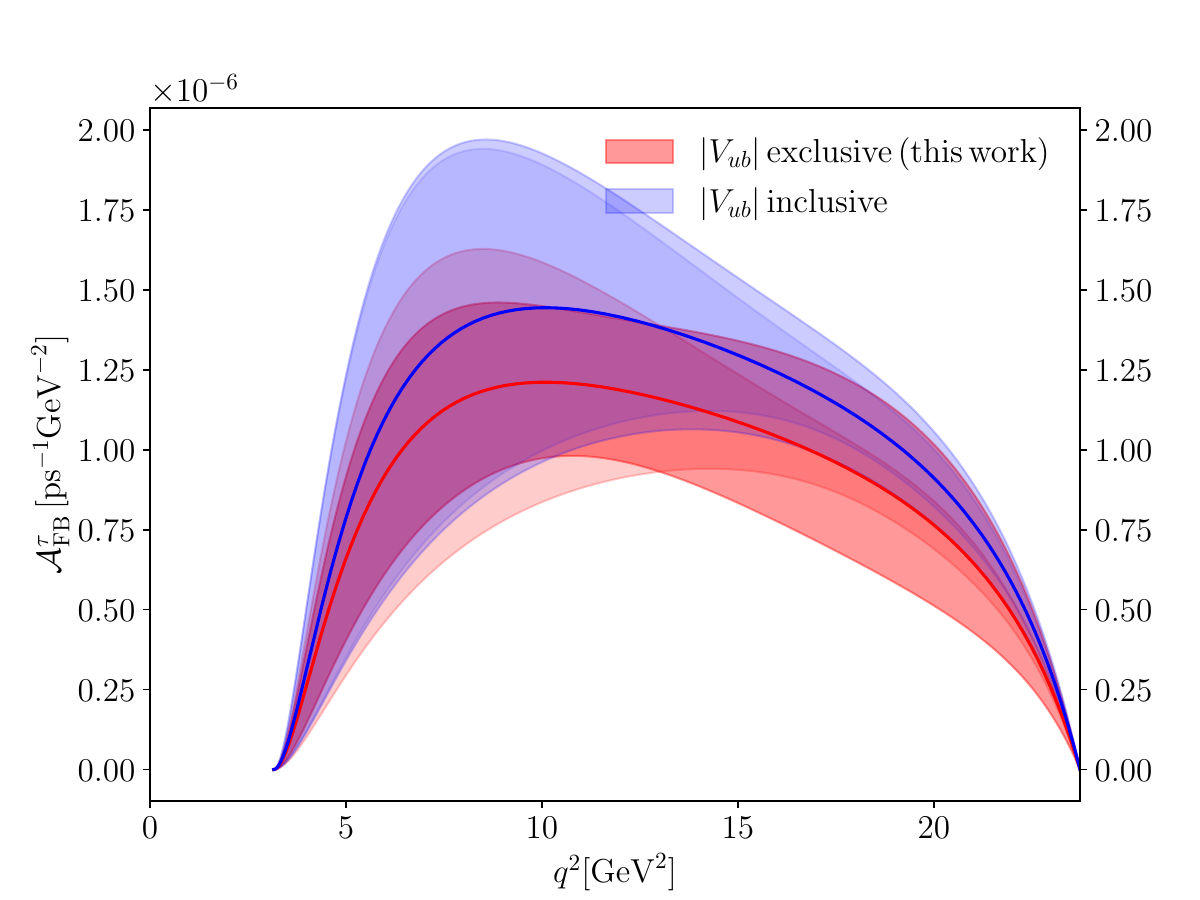}
  \caption{Forward-backward asymmetries $\mathcal{A}_{\rm FB}^{\mu}$ (left) and
    $\mathcal{A}_{\rm FB}^{\tau}$ (right). For convenient visualization the
    left-hand plot is shown on a logarithmic scale. The values for $|V_{ub}|$
    are taken from Eqs.~\eqnref{eq:ourVub} and
    \eqnref{eq:Vubinclusive}. The darker (lighter) shading indicates the error
    without (with) the contribution from the error on $|V_{ub}|$.}
  \label{fig:AFB}
\end{figure*}

\begin{figure*}
  \includegraphics[width=8cm]{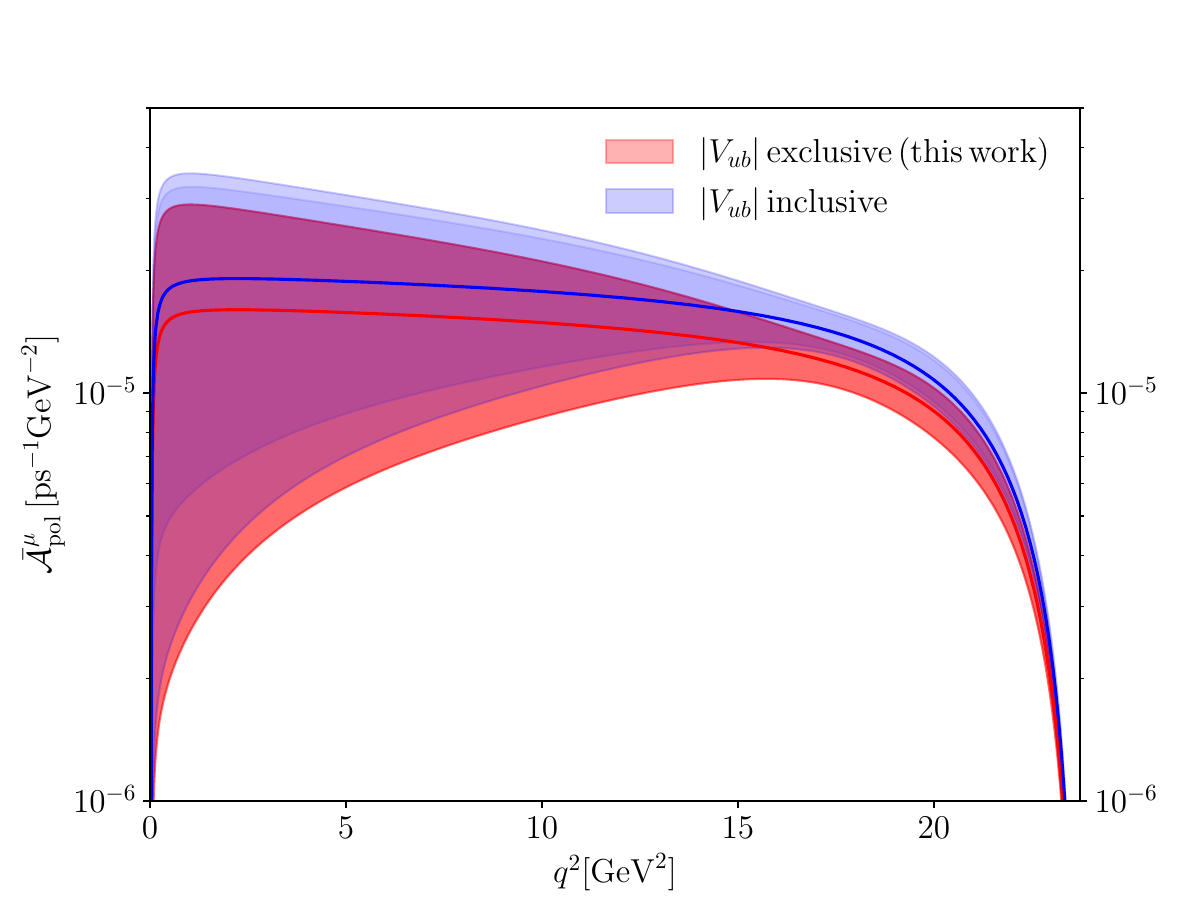}
  \includegraphics[width=8cm]{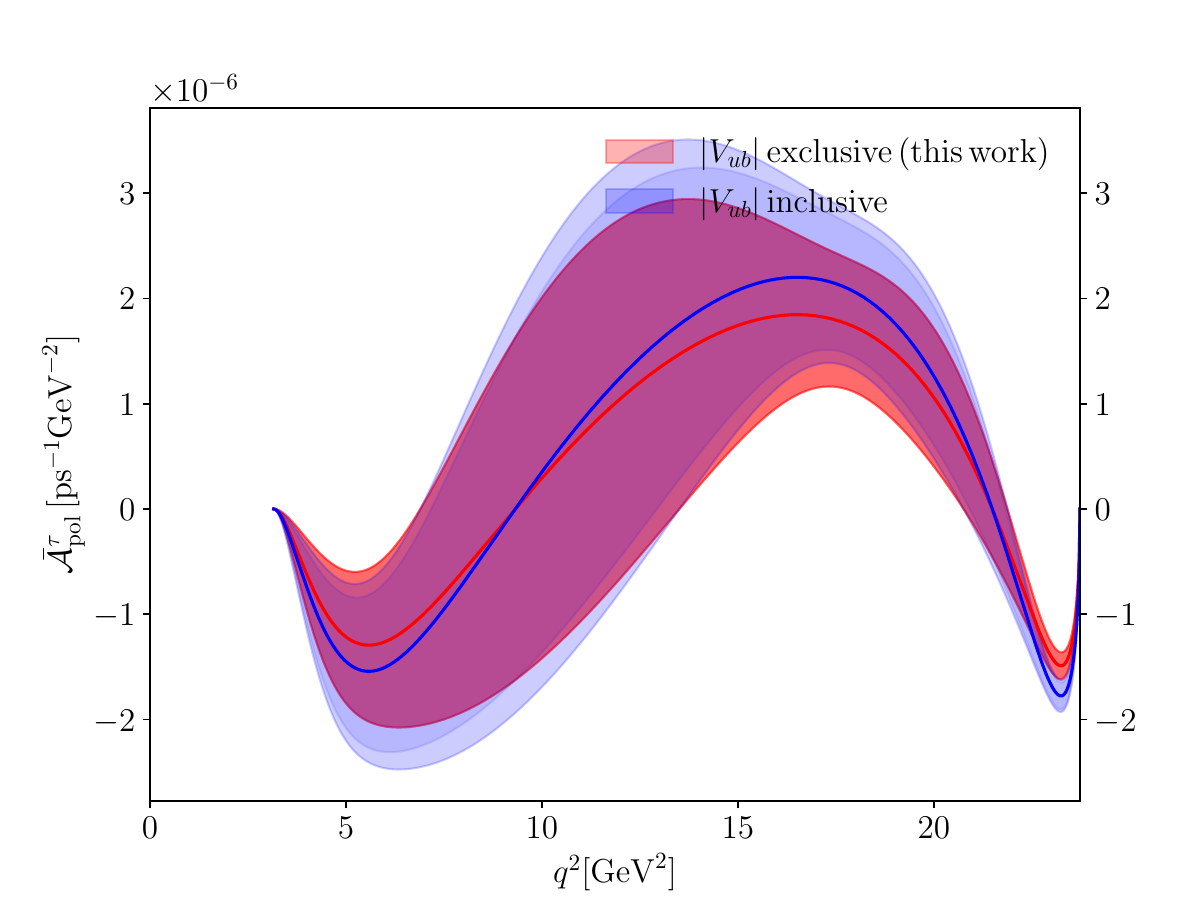}
  \caption{Difference of the left-handed and right-handed contributions to the
    decay rate $\mathcal{A}_{\rm pol}^{\mu}$ (left) and $\mathcal{A}_{\rm
      pol}^{\tau}$ (right).  The values for $|V_{ub}|$ are taken from
    Eqs.~\eqnref{eq:ourVub} and \eqnref{eq:Vubinclusive}. The darker (lighter)
    shading indicates the error without (with) the contribution from the error
    on $|V_{ub}|$.}
  \label{fig:Apol}
\end{figure*}

\begin{figure*}
  \begin{center}
    \includegraphics[width=15cm]{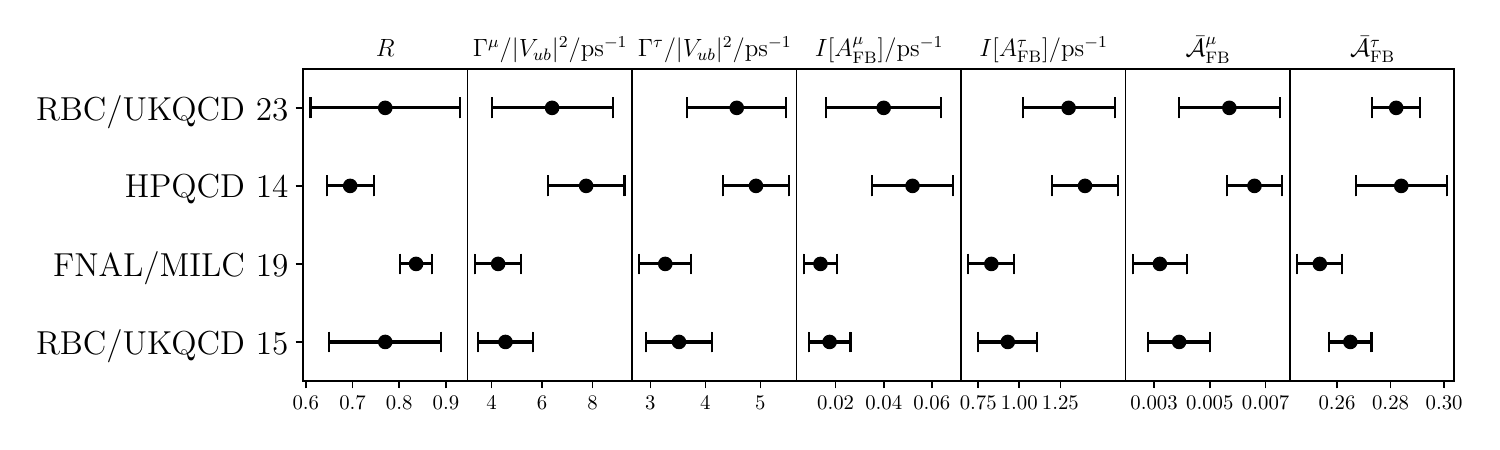}
  \end{center}
  \caption{Comparison with results from other collaborations (HPQCD
    14~\cite{Bouchard:2014ypa}, RBC/UKQCD 15~\cite{Flynn:2015mha}, FNAL/MILC
    19~\cite{Bazavov:2019aom} and this work, RBC/UKQCD 23).}
  \label{fig:observable comparison}
\end{figure*}

Our data for the form factors also allows to compute the forward-backward
asymmetry. Starting from the differential decay rate in terms of the lepton
angle $\theta_\ell$ between the charged-lepton and $B_s$ momentum in the $q^2$ rest
frame. The forward-backward difference is given by
\begin{equation}
\mathcal{A}^\ell_{\rm FB}(q^2)\equiv\left[\int\limits_0^1-\int\limits_{-1}^0\right]
d\cos\theta_\ell\frac{d^2\Gamma( B_s\to K\ell\nu)}{dq^2d\cos\theta_\ell}\,,
\end{equation}
and in the SM it takes the form~\cite{Meissner:2013pba}
\begin{align}
  \mathcal{A}^\ell_{\rm FB}(q^2)=&\eta_\text{EW}\frac{G_F^2|V_{ub}|^2}{32\pi^3 M_{B_s}}\left(1-\frac {m_\ell^2}{q^2}\right)^2|{\bf p}_K|^2\nonumber\\
  &\times\frac{m_\ell^2}{q^2}\left(M_{B_s}^2-M_K^2\right)f_+(q^2)f_0(q^2)\,.
\end{align}
A probe for helicity-violating interactions is provided by the difference of the
left-handed and right-handed contributions to the decay
rate~\cite{Meissner:2013pba}
\begin{equation}
  \mathcal{A}^\ell_{\rm pol}(q^2)=\frac{d\Gamma(\ell,{\rm LH})}{dq^2}-\frac{d\Gamma (\ell,{\rm RH})}{dq^2}\,,
\end{equation}
where
\begin{align}
  \frac {d \Gamma(\ell,{\rm LH})}{dq^2}=&\eta_\text{EW}\frac{G_F^2|V_{ub}|^2|{\bf p}_K|^3}{24\pi^3}\left(1-\frac {m_\ell^2}{q^2}\right)^2f_+^2(q^2)\,,\nonumber\\
  \\[-2ex]
  \frac {d \Gamma(\ell,{\rm RH})}{dq^2}=&\eta_\text{EW}\frac{G_F^2|V_{ub}|^2|{\bf p}_K|}{24\pi^3}\frac{m_\ell^2}{q^2}\left(1-\frac {m_\ell^2}{q^2}\right)^2\nonumber\\
	\times&\left(\frac 38 \frac{(M_{B_s}^2-M_K^2)^2}{M_{B_s}^2}f_0^2(q^2)+\frac 12 |{\bf p}_K|^2f^2_+(q^2)\right)\,.\nonumber
\end{align}
We show our results for the forward-backward asymmetries and the polarization
distribution in Figs.~\ref{fig:AFB} and~\ref{fig:Apol}, respectively, where the
case $\ell=\mu$ ($\tau$) is shown in the left (right) panel. In
Tab.~\ref{tab:BstoK_zfit_observables2}, we provide numerical results for
\begin{equation}
  I[\mathcal{A}^\ell]=\int\limits_{m_\ell^2}^{q^2_{\rm max}}dq^2\mathcal{A}^\ell(q^2)/|V_{ub}|^2\,,
\end{equation}
and
\begin{equation}
  \mathcal{\bar A}^\ell=\frac{\int_{m_\ell^2}^{q_{\rm max}^2}dq^2\mathcal{A}^\ell(q^2)}
						{\int_{m_\ell^2}^{q_{\rm max}^2}dq^2d\Gamma(\BstoKlnu)/dq^2}\,,
\end{equation}
where $\mathcal{A}^\ell = \mathcal{A}^\ell_{\rm FB}, \mathcal{A}^\ell_{\rm pol}$.
Here is a summary of our central results ($(K_+,K_0)=(5,5)$)
\begin{align}
  I[\mathcal{A}_{\rm FB}^\tau]=&\,{1.30(28)}\, {\rm ps}^{-1}\,,\\
  I[\mathcal{A}_{\rm FB}^\mu]=&\,0.040(24)\, {\rm ps}^{-1}\,,\\
  \mathcal{\bar A}_{\rm FB}^\tau=&\,0.2821(89)\,,\\
  \mathcal{\bar A}_{\rm FB}^\mu=&\,0.0057(18)\,,\\
  I[\mathcal{A}_{\rm pol}^\tau]=&\,0.18(29)\, {\rm ps}^{-1}\,,\\
  I[\mathcal{A}_{\rm pol}^\mu]=&\,6.3(2.3)\, {\rm ps}^{-1}\,,\\
  \mathcal{\bar A}_{\rm pol}^\tau=&\,0.035(60)\,,\\
  \mathcal{\bar A}_{\rm pol}^\mu=&\,0.9834(49)\,,
\end{align}
and in Fig.~\ref{fig:observable comparison} we provide a comparison with results
from other lattice simulations. Two observations are worth highlighting: 
With respect to RBC/UKQCD 15 some of the results shifted significantly, 
and the error of some results increased visibly. The shift is mainly down to
our decision to do the HM$\chi$PT chiral and continuum limit for 
$f_+$ and $f_0$ rather than for $f_\perp$ and $f_\|$. The assumption that the
same pole masses simultaneously describes the momentum dependence of
$f_+$ and $f_\perp$, and $f_0$ and $f_\|$, respectively, does not seem
correct at the level of precision we achieve with our dataset.
The increase in error on the other hand is due to the change in strategy
for the BGL parameterization -- at the cost of achieving a truncation-independent
parameterization of the form factor, the statistical error on observables particularly
sensitive to the low-$q^2$ behavior of the form factor increases.

\section{Conclusions}\label{sec:conclusion}
In this paper we present our new results for the non-perturbative Standard Model
contributions to the exclusive semileptonic decay $\BstoKlnu$. In particular, we
present the results for the form factors $f_+(q^2)$ and $f_0(q^2)$ in the
continuum limit of $N_f=2+1$ lattice simulations. We have improved our analysis
in various ways: a) we have improved the control of the continuum limit by
including simulations on a finer ($a^{-1}\approx 2.8\gev$) ensemble. b) we
include the effects of excited states in our correlation function fits. c) doing
the chiral-continuum extrapolation of $f_+$ and $f_0$ rather than $f_\|$ and
$f_\perp$ using HM$\chi$PT removes an otherwise irreducible systematic effect
present in earlier work~\cite{Flynn:2015mha,Bazavov:2019aom}, which might be the
origin of tensions in the combined analysis of lattice
results~\cite{FlavourLatticeAveragingGroupFLAG:2021npn}. d) using the
Bayesian-inference approach proposed in Ref.~\cite{FittingPaper} to fitting the
$z$-parameterization we obtain a model- and truncation-independent
parameterization of the form factor in the entire physical semileptonic
kinematical range.

\begin{figure*}
	\begin{center}
	\includegraphics[width=16cm]{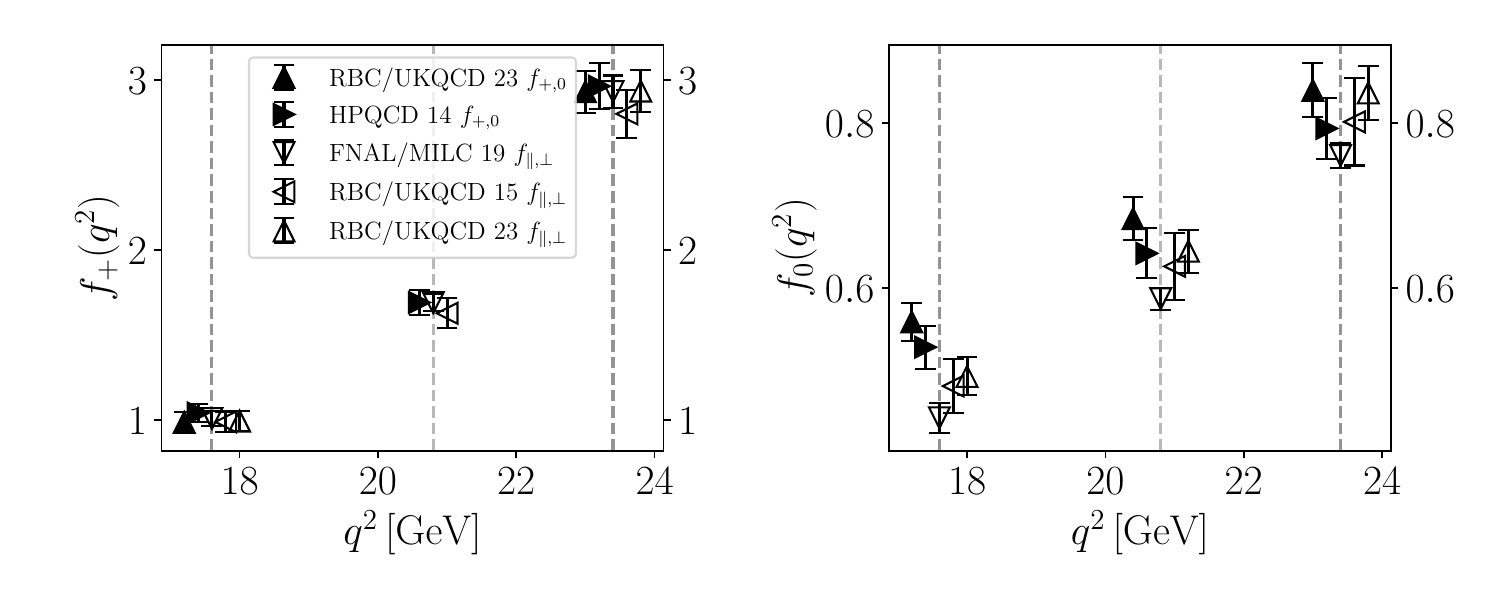}
	\end{center}
	\caption{Summary of available lattice data for the $\BstoKlnu$ form factors
     $f_+$ (left) and $f_0$ (right) at the same arbitrarily chosen reference
     points indicated by dashed vertical lines (to improve the presentation,
     individual data points are shifted slightly in $q^2$). Empty symbols
     indicate the results where the chiral and continuum extrapolations have
     been carried out in terms of $f_\perp$ and $f_\|$ and then converted to
     $f_+$ and $f_0$ using Eqs.~\eqnref{eq:f+fromfparfperp} and
     \eqnref{eq:f0fromfparfperp}, respectively. The points for RBC/UKQCD 23
     $f_{\parallel,\perp}$ were generated solely for the purpose of
     illustration, using the data and analysis methods discussed in this paper,
     but with the chiral and continuum extrapolation done for $f_\|$ and
     $f_\perp$. Filled symbols indicate the results where the extrapolation has
     been done in terms of $f_+$ (left) and $f_0$ (right).\label{fig:scatter}}
\end{figure*}

Regarding point c) we summarize the situation of available lattice data for the
$\BstoKlnu$ form factors in Fig.~\ref{fig:scatter}. While all available data for
the vector form factor $f_+$ is in agreement, the data for $f_0$ shows two
clusters of data points forming as $q^2$ is reduced. The two clusters can be
distinguished by the way in which the chiral and continuum limit have been
taken. In our view, taking the continuum limit in terms of $f_+$ and $f_0$
rather than in in terms of $f_\perp$ and $f_\|$, is correct.  The same note of
caution concerns other quantities like \emph{e.g.}~$B\to\pi \ell\nu$, where very
similar analysis techniques are being used.

We use our results to make a number of predictions for phenomenology. In
particular, we make a new prediction for the CKM matrix element
$|V_{ub}|=3.78(61)\times 10^{-3}$ based on first results for the $\BstoKlnu$ decay
from the LHCb experiment~\cite{Aaij:2020nvo}. The error is dominated by the
experimental uncertainty.  In particular, if we repeat the analysis with the
experimental uncertainty set to zero, the error on $|V_{ub}|$ reduces to 0.37.
Our result is compatible with both exclusive and inclusive determinations. We
also make predictions for the shape of the differential decay rate, the
forward-backward asymmetry and the difference between left-handed and
right-handed contributions to the decay rate. With more precise experimental and
lattice results these observables might in the future allow to shed light on the
tension between inclusive and exclusive $|V_{ub}|$ determinations.

\begin{acknowledgments}
We thank our RBC/UKQCD collaborators for helpful discussions and suggestions,
Paolo Gambino for discussion on unitarity-constrained fits and Greg Ciezarek for 
discussions of the LHCb form-factor data. We thank Edwin~Lizarazo
for contributions at early stages of this work. Computations used resources
provided by the USQCD Collaboration, funded by the Office of Science of the
U.S.~Department of Energy and by the \href{http://www.archer.ac.uk}{ARCHER} UK
National Supercomputing Service, as well as computers at Columbia University,
Brookhaven National Laboratory, and the OMNI cluster of the University of Siegen.
This document was prepared using the resources of the USQCD Collaboration at
the Fermi National Accelerator Laboratory (Fermilab), a U.S.~Department of
Energy (DOE), Office of Science, HEP User Facility. Fermilab is managed by Fermi Research
Alliance, LLC (FRA), acting under Contract No.~DE-AC02-07CH11359.
This work used the DiRAC Extreme Scaling service at the University of Edinburgh,
operated by the Edinburgh Parallel Computing Centre on behalf of the STFC
\href{https://dirac.ac.uk}{DiRAC} HPC Facility. This equipment was funded by BEIS capital
funding via STFC capital grant ST/R00238X/1 and STFC DiRAC Operations grant
ST/R001006/1. DiRAC is part of the National e-Infrastructure. 
We used gauge field configurations generated on
the DiRAC Blue Gene~Q system at the University of Edinburgh, part of the DiRAC
Facility, funded by BIS National E-infrastructure grant ST/K000411/1 and STFC
grants ST/H008845/1, ST/K005804/1 and ST/K005790/1.  We thank BNL, Fermilab, the
Columbia University, the University of Edinburgh, the University of Siegen,
the STFC, and the U.S.~DOE for providing the facilities essential for the
completion of this work. This project has received funding from
Marie Sk{\l}odowska-Curie grant 659322 and 894103 (EU Horizon 2020), UK STFC Grant No. ST/P000630/1, and is
supported by the Deutsche Forschungsgemeinschaft (DFG, German Research Foundation)
through grant 396021762 -- TRR 257 ``Particle Physics Phenomenology after the Higgs Discovery''.
The work of AS was supported in part by the U.S.~DOE contract \#DE-SC0012704.
\end{acknowledgments}

\appendix

\section{RHQ parameter tuning}
\label{appx:rhqtuning}

Here we summarize the non-perturbative tuning of the three parameters in the RHQ
action used for $b$ quarks. The procedure is described in
Ref.~\cite{Aoki:2012xaa}, based on lattice spacings determined in
Ref.~\cite{Aoki:2010dy}. The lattice spacing is a crucial input and with updated
and refined global fits available~\cite{Blum:2014tka, Boyle:2017jwu,
  Boyle:2018knm}, plus new ensembles, we have performed a new tuning for the
ensembles used here.  More precise determinations of the lattice spacings lead
to reduced systematic errors in the RHQ parameters. In addition, the values for
the strange-quark mass have been reanalyzed and we have generated new
valence-quark propagators with mass tuned or close to the updated strange-quark
mass.

\subsection{Non-perturbative tuning procedure}

The parameters in the RHQ action, $\{m_0a, c_P, \zeta\}$, are fixed by demanding
that the action correctly describes experimentally measured on-shell $B_s$-meson
properties. We match the experimental values~\cite{Olive:2016xmw} of the
spin-averaged mass and the hyperfine splitting, and require that the rest and
kinetic masses of the $B_s$ are equal
\begin{equation}
  \overline M_{\!B_s} = \frac{M_{\!B_s}\!+\!3M_{\!B^*_s}}4,
  \quad
  \Delta M_{\!B_s} = M_{\!B^*_s}-M_{\!B_s},
  \quad
  \frac{M_1^{B_s}}{M_s^{B_s}}=1.
\end{equation}
The latter implies that the $B_s$ meson satisfies the continuum dispersion
relation, $E^2_{B_s}({\bf p}) = {\bf p}\,^2 + M^2_{B_s}$. We calculate the
quantities above using seven sets of choices for the RHQ parameters $\{m_0a$,
$c_P$, $\zeta\}$ and then make a linear interpolation to find the values
satisfying the matching conditions above. The seven choices, indexed 1 to 7 from
left to right, comprise a central set plus variations of each of the three
parameters:

\begin{align}
  \label{eq:SevenSets}
  \begin{bmatrix}m_0 a\\ c_P\\ \zeta\end{bmatrix}\!\!,
  &\begin{bmatrix}m_0 a -\sigma_{m_0 a}\\ c_P\\ \zeta \end{bmatrix}\!\!,
  \begin{bmatrix}m_0 a +\sigma_{m_0 a}\\ c_P\\ \zeta \end{bmatrix}\!\!,
  \begin{bmatrix}m_0 a\\c_P-\sigma_{c_P}\\ \zeta \end{bmatrix}\!\!,
  \notag\\
  &\begin{bmatrix}m_0 a\\c_P+\sigma_{c_P}\\ \zeta \end{bmatrix}\!\!,
  \begin{bmatrix}m_0 a\\ c_P\\ \zeta-\sigma_{\zeta} \end{bmatrix}\!\!,
  \begin{bmatrix}m_0 a\\ c_P\\ \zeta+\sigma_{\zeta} \end{bmatrix}\!\!.
\end{align}
We make a constant-plus-linear ansatz for the dependence of the
observables on the RHQ parameters
\begin{equation}
  \label{eq:linear_approx}
    \begin{bmatrix}
      \overline{M}_{B_s}\\
      \Delta M_{B_s}\\
      M_1^{B_s}/M_2^{B_s}
    \end{bmatrix} = J\cdot
    \begin{bmatrix}
      m_0a \\
      c_P \\
      \zeta
    \end{bmatrix}  + A \,.
\end{equation}
Here $J$ represents the ``slope'' and is a $3\times 3$ matrix, while $A$
corresponds to the intercept and is a $3 \times 1$ vector. In a region with
sufficiently linear dependence on the parameters, we can obtain $J$ and $A$
using finite differences to approximate derivatives:
\begin{equation}
  \label{eq:JAmatrix}
  \begin{aligned}
  J &= \left[\frac{Y_3-Y_2}{2\sigma_{m_0a}},\,
    \frac{Y_5-Y_4}{2\sigma_{c_P}},\,
    \frac{Y_7- Y_6}{2\sigma_\zeta}\right] ,\\
  A &=  Y_1 - J\cdot\left[m_0a,\,c_P,\,\zeta\right]^T .
  \end{aligned}
\end{equation}
The vectors $Y_i$ are constructed from the values of meson masses and
splittings measured on the $i^\textrm{th}$ parameter set
in~\eqref{eq:SevenSets},
\begin{align}
  Y_i = \left[ \overline{M}_{B_s}, \Delta M_{B_s}, M_1^{B_s}/M_2^{B_s}
    \right]^T_i .
  \label{eq:Yi}
\end{align}
Inverting Eq.~\eqref{eq:linear_approx}, we obtain the tuned RHQ
parameters
\begin{align}
\begin{bmatrix} m_0a\\ c_P\\ \zeta \end{bmatrix}^\text{RHQ}
  = J^{-1} \cdot\left(\begin{bmatrix}\overline{M}_{\!B_s}\\
    \Delta M_{B_s}\\
    M_1^{B_s}/M_2^{B_s}\end{bmatrix}^\text{PDG} \!\!- A\right) ,
\label{eq:RHQDetermination}
\end{align}
by matching to PDG values and demanding the $B_s$ rest mass equals its
kinetic mass. Specifically we use~\cite{Olive:2016xmw}.
\begin{equation}
  \label{eq:PDG}
  \begin{aligned}
    \overline M_{B_s} &= \frac14 \big(5366.82^{+0.22}_{-0.22}+3\cdot
    5415.4^{+1.8}_{-1.4}\big)\mev\\
    &= 5403.26^{+1.8}_{-1.4}\mev, \\
    \Delta M_{B_s} &= 48.6^{+1.8}_{-1.6}\mev.
  \end{aligned}
\end{equation}
We conservatively use the full error on $M_{B_s^*}$ as the uncertainty
of the spin-averaged mass. The relative error on $\Delta_{M_{B_s}}$ is
much larger and will dominate systematic effects due to the
experimental inputs. These inputs are updated from those in
Ref.~\cite{Nakamura:2010zzi} used in Ref.~\cite{Aoki:2012xaa}.

\subsection{Lattice simulations}
The tuning procedure is implemented by first determining $B_s^{(*)}$-meson
energies for zero and non-zero momenta on our set of ensembles in
Table~\ref{tab:ensembles}. We calculate two-point functions by contracting
strange-quark propagators using point source and point sink with Gaussian
source-smeared, point-sink $b$-quark propagators, and extract $B_s^{(*)}$-meson
energies from correlated fits to the plateau of effective energies. We find good
correlated confidence levels ($p$-value $\gtrsim$ 10\%) in all cases and varying
the fitting range by $\pm1$ time slice changes the result only within the
statistical uncertainty. The value of our input strange-quark mass as well as
the width $\sigma_G$ and the number of iterations $N_G$ for the Gaussian source
smearing of the $b$ quarks are summarized in Table~\ref{tab:TuneInputs} together
with the fitting range used to extract the $B_s^{(s)}$ meson energies.
\begin{table}
\caption{Input strange quark mass $am_s$, width $\sigma_G$ and iteration count
  $N_G$ of the Gaussian source smearing used to calculate $B_s^{(*)}$ meson
  two-point functions. We extract the $B_s^{(*)}$ energies by performing
  correlated fits using the specified fit ranges.  The optimal Gaussian
  source-smearing radius was explored for the C and M ensembles in
  Ref.~\cite{Aoki:2012xaa} and we scaled those results to the finer lattice
  spacing $a^{-1}=2.785\gev$ on the F1S ensemble.}
  \label{tab:TuneInputs}
  \begin{tabular}{ccc@{~~}c@{~~}c@{~~}c@{~~}c}
    \hline\hline
  & $a^{-1}/\!\gev$  & $L^3$ & $a(m_s+m_\text{res}$)
  & $\sigma_G$ & $N_G$ & fit range\\ \hline
 \text{C} &
   1.785 & $24^3$& 0.03224+0.00315  & \phantom{1}7.86 & 100 & [10{:}25] \\
 \text{M} &
   2.383 & $32^3$& 0.025+0.00067  & 10.36& 170 & [12{:}21] \\
 \text{F} &
   2.785 & $48^3$& 0.02144+0.00094 & 12.14& 230 & [14{:}29]\\
   \hline\hline
   \end{tabular}
\end{table}  

Starting from the tuning performed in Ref.~\cite{Aoki:2012xaa}, we iterate twice
to find the new central set of RHQ parameters for the C and M ensembles and
choose roughly three times the size of the statistical errors for the variations
$\sigma_{\{m_0a,c_P,\zeta\}}$.  For the new F1S ensemble at the finer lattice
spacing, we roughly scaled our new results on C and M ensembles and then carried
out two iterations choosing variations $\sigma_{\{m_0a,c_P,\zeta\}}$ of roughly
1.5 times the statistical uncertainties. In all cases our final parameter sets
allow us to interpolate to the values of $m_0a$, $c_P$, and $\zeta$ describing
physical $b$ quarks and we do not observe any signs of curvature within the
explored parameter ranges. The final values of the central parameter sets and
their variations are listed in Table~\ref{tab:Box}.
\begin{table}
\caption{Central parameter set $\{m_0a,\, c_P,\, \zeta\}$ with
  variations $\sigma_{m_0a},\,\sigma_{c_P},\,\sigma_\zeta$ for the
  final iteration of tuning the parameters on C, M and F ensembles
  with inverse lattice spacings ranging from $1.785$ to $2.785\gev$.}
\label{tab:Box}
\begin{tabular}{ccc@{~~~}c@{~~~}c@{~~~}c}
  \hline\hline
 & $a^{-1}\!/\!\gev$ & $L^3$ & $m_0a\pm\sigma_{m_0a}$ & $c_P\pm\sigma_{c_P}$ &
  $\zeta\pm\sigma_{\zeta}$\\ \hline

 \text{C} &
1.785 & $24^3$ & 7.42 $\pm$ 0.18 & 4.86 $\pm$ 0.42 & 2.92 $\pm$ 0.21 \\
 \text{M} &
2.383 & $32^3$ & 3.46 $\pm$ 0.09 & 3.03 $\pm$ 0.24 & 1.75 $\pm$ 0.10 \\
 \text{F} &
 2.785 & $48^3$ & 2.35 $\pm$ 0.20 & 2.75 $\pm$ 0.30 & 1.50 $\pm$ 0.15 \\
   \hline\hline
\end{tabular}
\end{table}
Using those values, we determine spin-averaged masses, hyperfine splittings and
ratios of rest mass over kinetic mass in order to match to experimental results
reported by the PDG as in Eq.~\eqref{eq:PDG}. We finally obtain our tuned
parameters from Eq.~\eqref{eq:RHQDetermination}. For the C and M ensembles we
cannot resolve a dependence on the light sea-quark mass within statistical
errors. Hence, we average the values at the same lattice spacing. We report
these results, plus the outcome for tuning on the F1S ensemble, in
Table~\ref{tab:Tuned}.
\begin{table}
\caption{Tuned RHQ parameter values for all lattices determined using
  the parameter sets specified in Tab.~\ref{tab:Box}. For the C
  and M lattices, we see no dependence on the light sea-quark
  mass within statistical errors and we consequently compute
  weighted averages to obtain our preferred values.}
\label{tab:Tuned}

\begin{tabular}{c@{~~~~}c@{~~~}c@{~~~}c}
  \hline\hline
  $am_l$ & $m_0a$ & $c_P$ & $\zeta$\\\hline
  \multicolumn4c{C: $a^{-1} = 1.785\gev$, $24^3\times 64$}\\
  0.005   & 7.468(66) & 4.87(18) & 2.922(82)\\
  0.010   & 7.476(80) & 4.97(19) & 2.94(10)\\
  \text{average} & 7.471(51) & 4.92(13) & 2.929(63)\\[1.5ex] \hline
  \multicolumn4c{M: $a^{-1} = 2.383\gev$, $32^3\times 64$}\\
  0.004  & 3.541(46) & 3.19(13)\phantom{0} & 1.715(55) \\
  0.006  & 3.474(37) & 3.01(10)\phantom{0} & 1.759(46) \\
  0.008  & 3.444(48) & 3.02(14)\phantom{0} & 1.807(55) \\
  \text{average} & 3.485(25) & 3.063(69) & 1.760(30) \\[1.5ex] \hline
  \multicolumn4c{F: $a^{-1} = 2.785\gev$, $48^3\times 96$}\\
  0.002144 & 2.423(62) & 2.68(13)   & 1.523(79)\\
  \hline\hline  
  \end{tabular}
\end{table}

\subsection{Estimating systematic uncertainties}

\begin{table}
\caption{Systematic uncertainties in percent with a significant effect
  on our tuned RHQ parameters. For the C and M ensembles we always
  report the largest fluctuation observed on either ensemble.}
\label{tab:SysErrors}
\begin{tabular}{cc@{~~~}c@{~~~}c@{~~~}c}
  \hline\hline
uncertainty  & $a^{-1}\!/\!\gev$ & $m_0a$ & $c_P$ & $\zeta$ \\\hline
heavy quark discretization & 1.785 & 1.0\% & 5.6\% & 3.4\%\\
              & 2.383 & 1.1\% & 6.0\% & 3.3\%\\
              & 2.785 & 1.5\% & 5.6\% & 2.8\%\\\hline              
lattice scale & 1.785 & 1.1\% & 1.4\% & 0.5\%\\
              & 2.383 & 1.3\% & 1.7\% & 0.4\%\\
              & 2.785 & 1.3\% & 1.4\% & 0.3\%\\\hline
experiment   
              & 1.785 & 0.6\% & 4.8\% & 0.1\%\\
              & 2.383 & 0.9\% & 5.0\% & 0.1\%\\
              & 2.785 & 1.2\% & 4.8\% & 0.1\%\\
  \hline\hline
\end{tabular}
\end{table}

\subsubsection{Heavy-quark-discretization errors}
When $m_0a\sim 1$, the RHQ action leads to a nontrivial
lattice-spacing dependence of physical quantities. As discussed in
detail in Ref.~\cite{Aoki:2012xaa}, we estimate the discretization errors
for the heavy sector using power counting, following Oktay and
Kronfeld~\cite{Oktay:2008ex}. Since 
we are considering the same physical quantities, spin-averaged mass,
hyperfine splitting, and ratio of rest over kinetic mass, we simply
list uncertainties in percent and refer to Appendix~C
of Ref.~\cite{Aoki:2012xaa} for further details.
\begin{equation}
  \begin{aligned}
    \text{error}^{M_{1,B_s}}_\text{total} &= 0.05 \%,\\ 
    \text{error}^{M_{2,B_s}}_\text{total} &= 2.59 \%,\\
    \text{error}^{\Delta M_{B_s}}_\text{total}  &= 4.40 \%. 
  \end{aligned}
\end{equation}
By assigning these errors as uncertainty in our inputs to the matching
procedure, we propagate them to the RHQ parameters and collect the
percentage changes in the central values in the first panel of
Table~\ref{tab:SysErrors}.

\subsubsection{Input strange-quark mass}
Our RHQ parameters are tuned using strange quark propagators corresponding to a
mass at or near the physical strange-quark mass. We need to account for slight
mistunings as well as for the uncertainty in the strange-quark mass quoted in
Ref.~\cite{Blum:2014tka, Boyle:2017jwu} (see also table~\ref{tab:ensembles}).

On the coarse (C) ensembles, we can bracket the strange quark mass using the
additional quark propagators with bare masses $am_s' = 0.03$ and $0.04$. This
allows us to determine numerically the slopes of $m_0$, $c_P$, and $\zeta$ with
respect to the strange quark mass.  Since the mass of our strange quark
propagators matches the physical value, we read off the changes in $m_0a$,
$c_P$, and $\zeta$ after varying $am_s$ by $\pm 1 \sigma$. The largest change we
observe is $0.2\%$.

For the M ensembles we simulate with a bare strange quark mass of $0.025$ which
is roughly $1\sigma$ larger than the physical value.  Using in addition
$am_s'=0.0272$, we determine the slopes with respect to $m_s$ and estimate the
error due to mistuning as well as the uncertainty in the strange quark mass by
varying the strange quark mass by $\pm2\sigma$. Our RHQ parameters change at
most by $0.3\%$.

For the F1S ensemble, since the slopes with respect to $am_s$ decrease as the
lattice spacing decreases, we use for simplicity the average of the slopes
obtained on the M ensembles to estimate the uncertainty due to the input strange
quark mass for F1S. Here our value for the mass of the strange quark is within
$1 \sigma$ of the physical value. Being conservative we vary the strange quark
mass by $\pm 2\sigma$ and read off changes of the RHQ parameters of at most
$0.2\%$.

Given that for all ensembles and all three RHQ parameters the
uncertainty due to the strange quark mass is $0.3\%$ or less, we
consider this effect negligible compared to the percent-level
uncertainties arising from, for example, heavy quark discretization
errors.

\subsubsection{Lattice scale uncertainty}
The lattice scale enters our tuning procedure when we convert the
experimental input data to lattice units. To propagate the uncertainty
of the lattice spacings to our RHQ parameters, we repeat the analysis
varying the lattice spacing by $\pm 1 \sigma$. For the C and M
ensembles we take the largest fluctuation of a central value on either
ensemble as our estimate.

\subsubsection{Experimental uncertainty}
We estimate the uncertainty due to the experimental inputs by varying
both the spin-averaged mass and the hyperfine splitting by $\pm 1
\sigma$ each and re-run our matching analysis. In practice, the
uncertainty in the spin-averaged mass is negligible compared to the
few-percent effect due to the uncertainty in the hyperfine splitting.
We take the largest change of the central values at a given lattice
spacing as our estimate for the corresponding uncertainty in our RHQ
parameters.

\subsection{Tuned RHQ parameters}
We summarize our tuned RHQ parameters in Table~\ref{tab:RHQparameter}
quoting our final results with all systematic errors found to be
significant.
\begin{table*}
\caption{Values of the tuned RHQ parameters with central values and
    statistical errors taken from table~\ref{tab:Tuned} and estimates
    for the systematic errors from table~\ref{tab:SysErrors}. Tuning
    determines the bare quark mass, $m_0a$, clover coefficient, $c_P$
    and anisotropy parameter, $\zeta$, in the RHQ action. Errors
    listed for $m_0a$, $c_P$, and $\zeta$ are, from left to right:
    statistics, heavy-quark discretization errors, the lattice scale
    uncertainty, and the uncertainty due to the experimental
    measurement of the $B_s$ meson hyperfine splitting, respectively.
    Other errors considered but found to be negligible are not shown.}
\label{tab:RHQparameter}
\begin{tabular}{c@{~~~}c@{~~~}c@{~~~}c@{~~~}c}
  \hline\hline
  & $a^{-1}\!/\!\gev$ & $m_0a$ & $c_P$ & $\zeta$ \\ \hline
\text{C} &
1.785 & 7.471(51)(75)(82)(45) & 4.92(13)(28)(07)(24) & 2.929(63)(100)(15)(03)\\
\text{M} &
2.383 & 3.485(25)(38)(45)(31) & 3.06(07)(18)(05)(15) & 1.760(30)(58)(07)(02)\phantom{1} \\
\text{F} &
2.785 & 2.423(62)(36)(31)(29) & 2.68(13)(15)(04)(13) & 1.523(79)(43)(05)(02)\phantom{1}\\
  \hline\hline
  \end{tabular}
\end{table*}

\section{RHQ discretization errors}
\label{appx:HQdiscerrs}
\begin{table}[t]
  \caption{Values of the mismatch functions defined in Appendix B of
    Ref.~\cite{Christ:2014uea} for the non-perturbatively tuned parameters
    of the RHQ action given in Table~\ref{tab:RHQparameter}. The tree-level
    coefficients $f_E$, $f_{X_i}$, and $f_Y$ are known exactly. The two-loop
    coefficient $f_3^{[2]}$ is not known, so we use an ansatz based on the
    tree-level expression. The coupling $\alpha_s(a^{-1})$ needed to evaluate
    $f_3^{[2]}$ is taken to be a lattice mean-field coupling at the (inverse)
    lattice spacing scale (see the discussion of the perturbative calculations
    in section~\ref{sec:renormimprove}).}
  \label{tab:HQmismatch}
  \begin{tabular}{l@{}c@{~}c@{~~}c@{~~}c@{~~}c@{~~}c@{~~}c}
    \hline\hline
    & $a^{-1}\!/\!\gev$& $\alpha_s(a^{-1})$ & $f_E$ & $f_{X_1}$ & $f_{X_2}$ & $f_Y$ & $f_3^{[2]}$ \\
    \hline
 \text{C} & 1.785 & 0.2320 & 0.0594 & 0.0848 & 0.1436 & 0.1358 & 0.0333 \\
 \text{M} & 2.383 & 0.2155 & 0.0809 & 0.0966 & 0.1691 & 0.1721 & 0.0298 \\
 \text{F} & 2.785 & 0.2083 & 0.1042 & 0.1101 & 0.1946 & 0.2056 & 0.0299 \\
 \hline\hline
 \end{tabular}
\end{table}
\begin{table*}[t]
\caption{Percentage errors from mismatches in the action and current for the
  bottom quark. For this estimate, we calculate the mismatch functions for the
  non-perturbatively-tuned parameters of the RHQ action from
  table~\ref{tab:RHQparameter}. We estimate the size of operators using HQET
  power counting with $\Lambda_\text{QCD}=500\mev$ and
  $\alpha_s^{\overline{\text{MS}}}(a^{-1})$. To obtain the total, we add the
  individual errors in quadrature, including each contribution the number of
  times that operator occurs. Contribution `E' is counted twice, while `3' is
  counted twice for $f_\|$ and four times for $f_\perp$. The definitions of
  operators `E', `X${}_1$', `X${}_2$', `Y' and `3', and expressions for the
  mismatch functions, are given in Appendix~B of Ref.~\cite{Christ:2014uea}.}
\label{tab:HQDiscErrs}
\begin{tabular}{lccccccccc}
    \hline\hline  
  &&& $O(a^2)$ error & \multicolumn3c{$O(a^2)$ errors} &
  $O(\alpha_s^2 a)$ error \\[-0.6ex]
  &&& from action & \multicolumn3c{from current} &
  from current & \multicolumn2c{Total/\%}  \\	
  & $a^{-1}\!/\!\gev$ & $\alpha_s(a^{-1})$ &
   E & X$_1$ & X$_2$ & Y & 3 & $f_\|$ & $f_\perp$ \\\hline
  C & 1.785 & 0.2320 & 0.47 & 0.67 & 1.13 & 1.07 & 0.93 & 2.24 & 2.60 \\
  M & 2.383 & 0.2155 & 0.36 & 0.43 & 0.74 & 0.76 & 0.63 & 1.53 & 1.77 \\
  F & 2.785 & 0.2083 & 0.34 & 0.35 & 0.63 & 0.66 & 0.54 & 1.33 & 1.53 \\
  \hline\hline
  \end{tabular}
\end{table*}

We tune the parameters in the RHQ action non-perturbatively, such that
the leading heavy-quark discretization errors from the action are
$O(a^2)$. We use an $O(\alpha_s a)$-improved vector current and
calculate the improvement coefficient to $1$-loop; hence the leading
heavy-quark discretization errors from the current are of
$O(\alpha_s^2 a, a^2)$. Table~\ref{tab:HQmismatch} gives values for
the `mismatch' functions at each of our lattice spacings, from which
we find and list in Table~\ref{tab:HQDiscErrs} the estimated heavy
quark discretization errors from the five different operators in the
action and current. We refer the reader to section~V.E and appendix~B
of Ref.~\cite{Christ:2014uea} for further details.

\section{Renormalization and improvement coefficients}
\label{appx:PT-coeffts}

Table~\ref{tab:ZVbb} summarizes the values of the renormalization constants $Z^{ll}_A(am_l)$.

\begin{table}[t]
\caption{Residual renormalization factors and operator improvement
  coefficients. We compute the $\rho$ factors and the matching
  coefficients $c_i^n$ at one loop in mean-field improved lattice
  perturbation theory~\cite{CLehnerPT}. They are evaluated for the
  tuned RHQ bare-quark mass $m_0a$ at each lattice spacing (indicated
  by the letters C, M and F in column headers, see
  table~\ref{tab:ensembles}) and for the choices of the domain-wall
  height $M_5=1.8$ The values are an average from four combinations of
  mean link and strong coupling, with the columns headed $\Delta$
  giving the spread in values from the four combinations.}
\label{tab:coeff}
   \begin{tabular}{l>{$}r<{$}>{$}r<{$}>{$}r<{$}>{$}r<{$}>{$}r<{$}>{$}r<{$}}
    \hline\hline
  & \multicolumn2c{C $M_5=1.8$}
  & \multicolumn2c{M $M_5=1.8$} & \multicolumn2c{F $M_5=1.8$}\\
  & \text{value} & \multicolumn1c{$\Delta$}
    & \text{value} & \multicolumn1c{$\Delta$}
    & \text{value} & \multicolumn1c{$\Delta$}\\\hline
 $\rho_{V_0}$ & 1.0301 & 0.0115 & 1.0177 & 0.0059 & 1.0130 & 0.0041 \\
 $\rho_{V_i}$ & 0.9959 & 0.0019 & 0.9926 & 0.0022 & 0.9921 & 0.0019 \\
 $c_t^3$ & 0.0599 & 0.0092 & 0.0574 & 0.0073 & 0.0550 & 0.0064 \\
 $c_t^4$ & -0.0120 & 0.0042 & -0.0109 & 0.0030 & -0.0101 & 0.0025 \\
 $c_s^1$ & -0.0010 & 0.0006 & -0.0015 & 0.0006 & -0.0018 & 0.0006 \\
 $c_s^2$ & 0.0020 & 0.0021 & 0.0004 & 0.0011 & -0.0001 & 0.0009 \\
 $c_s^3$ & 0.0512 & 0.0060 & 0.0497 & 0.0050 & 0.0478 & 0.0044 \\
 $c_s^4$ & -0.0038 & 0.0020 & -0.0022 & 0.0011 & -0.0015 & 0.0008 \\
    \hline\hline
   \end{tabular}
\end{table}
In Table~\ref{tab:coeff} we give the residual renormalization factors
and the improvement coefficients for the heavy-light currents with a
heavy RHQ and a light domain-wall quark, computed at one loop
order~\cite{CLehnerPT} in mean-field improved perturbation theory.

\begin{table}[t]
\caption{Inputs for numerical evaluation of perturbatively computed
 coefficients at the three different lattice spacings in our
 ensembles.}
  \label{tab:ptinput}
  \begin{tabular}{lccc}
    \hline\hline    
    & C & M & F \\\hline
    $a^{-1}/\gev$ & 1.785 & 2.383 & 2.785 \\
    $\beta$  & 2.13 & 2.25 & 2.31 \\
    $\langle P\rangle$ & 0.588011 & 0.615580 & 0.627970 \\
    $\langle R_1\rangle$ & 0.343464 & 0.379841 & 0.396626 \\
    $u_P$ & 0.875682 & 0.885770 & 0.890194 \\
    $u_L$ & 0.843997 & 0.860991 & 0.868440 \\
    $m_0a$ & 7.471 & 3.485 & 2.423 \\
    $\alpha^{\overline{\text{MS}}}_{s,\text{lat}}(a^{-1})$
    & 0.2320 & 0.2155 & 0.2083 \\
    $\alpha^{\overline{\text{MS}}}_{s,\text{ctm}}(a^{-1})$
    & 0.3226 & 0.2811 & 0.2633 \\
    \hline\hline
    \end{tabular}
\end{table}

Table~\ref{tab:ptinput} gives inputs used for numerical evaluation of
the perturbatively calculated coefficients. Mean-field improvement
uses values for the average gauge link, $u$. We use either the fourth
root of the plaquette expectation value, denoted $u_P$, or the mean
link in Landau gauge, $u_L$. We use two choices of strong coupling: a
mean-field lattice $\overline{\text{MS}}$ strong coupling,
$\alpha^{\overline{\text{MS}}}_{s,\text{lat}}$, or the continuum
$\overline{\text{MS}}$ strong coupling,
$\alpha^{\overline{\text{MS}}}_{s,\text{ctm}}$, both evaluated at the
scale $\mu=a^{-1}$. For the Iwasaki gauge action used for our gauge
field ensembles, the lattice strong coupling depends on the plaquette
and rectangle expectation values $\avg{P}$ and $\avg{R_1}$
respectively, quoted in Table~\ref{tab:ptinput}. For the continuum
coupling, we use 5-loop running from
\texttt{RunDec}~\cite{Chetyrkin:2000yt, Schmidt:2012az,
  Herren:2017osy}, starting from $\alpha^{(5)}(\bar m_b(\bar
m_b))=0.2268$ at $\bar m_b(\bar m_b)=4.163\gev$ (The same
choice was made when using \texttt{RunDec} to compute $\bar
m_b(2\gev)$ for evaluating the $\chi$ factors in the outer functions
for BGL $z$-fits. See appendix~\ref{appx:BGL-fits}). The perturbative
results in Table~\ref{tab:ptinput} are the average of values from the
four combinations of $u$ and $\alpha^{\overline{\text{MS}}}_s$. The
columns headed $\Delta$ give half of the spread in the values from the
four combinations.

\section{BGL fits}
\label{appx:BGL-fits}

Here we give expressions for the Blaschke factors and outer functions
used in BGL fits to form factors.

The Blaschke factor $B_X(q^2)$ is chosen to vanish at the positions of
sub-threshold poles sitting between $q^2_\text{max}$ and $t_*$, with
$t_*$ denoting the squared momentum transfer for the lowest
two-particle production threshold. If there are $n$ such poles with
masses $m_i$ and corresponding $z$-values $z_i = z(m_i^2;t_*,t_0)$,
then
\begin{align}
  \label{eq:Blaschke}
  B_X(q^2)=\prod_{i=0}^{n-1} \frac{z-z_i}{1-\bar z_i z}
          =\prod_{i=0}^{n-1} z(q^2;t_*,m_i^2),
\end{align}
where $z=z(q^2;t_*,t_0)$ and $\bar z_i$ is the complex conjugate of
$z_i$ in the first form of the expression. The function
$z(q^2;t_*,t_0)$ is defined in Eq.~\eqref{eq:z-fn-defn} and our
choice for $t_0$ is given in Eq.~\eqref{eq:t-opt}. For $\BstoKlnu$ decays
the threshold is $M_B+M_\pi = 5.4175\gev$ and there is a sub-threshold
pole at $M_{B^*}=5.32471\gev$~\cite{ParticleDataGroup:2022pth} for $f_+$, but
not for $f_0$ for which
$M_{B^*(0^+)}=5.63\gev$~\cite{Bardeen:2003kt}\footnote{The $B^*(0^+)$
  masses in the compilation in Ref.~\cite{Cheng:2017oqh} span a range from
  $1.8\%$ below to $2.2\%$ above this value.} is above threshold. 

The outer functions used in the parameterizations of $f_+$ and $f_0$
take the
form~\cite{Boyd:1994tt,Arnesen:2005ez,Berns:2018vpl}\footnote{These
also agree with the expressions given in Eqs.~(2.13) and~(2.14) of
Ref.~\cite{Bigi:2016mdz}, if expressed in terms of $z$ and using the
same $t_*$ and $t_0$ as Ref.~\cite{Bigi:2016mdz}.}
\begin{align}
  \label{eq:phiplus}
  \phi_+(q^2,t_0) &= \sqrt{\frac{\eta_I}{48\pi\chi_{1^-}(0)}}\,
                    \frac{r_q^{1/2}}{r_0^{1/2}}\,(r_q+r_0)
                    \big(r_q+\sqrt{t_*}\,\big)^{-5}\notag\\
                    &\qquad\times
                    (t_+-q^2)^{3/4}
                    (r_q+r_-)^{3/2}\,, \\
  \label{eq:phizero} 
  \phi_0(q^2,t_0) &= \sqrt{\frac{\eta_It_+t_-}{16\pi\chi_{0^+}(0)}}\,
                    \frac{r_q^{1/2}}{r_0^{1/2}}\,(r_q+r_0)
                    \big(r_q+\sqrt{t_*}\,\big)^{-4}\notag\\
                    &\qquad\times
                    (t_+-q^2)^{1/4}
                    (r_q+r_-)^{1/2}\,,
\end{align}
where we have set $r_q=\sqrt{t_*-q^2}$, $r_-=\sqrt{t_*-t_-}$ and
$r_0=\sqrt{t_*-t_0}$ ($t_\pm$ are defined after Eq.~\eqref{eq:z-fn-defn}).
The $\chi_J(0)$ come from evaluating the vacuum polarization function
of two currents ($\bar b \gamma_\mu u$ and its Hermitian conjugate),
while the $\eta_I$ are isospin factors counting accessible
isospin-related states when inserting a sum over states between the
two currents in the vacuum polarization function (to derive the
unitarity constraint on the form factors). For $\BstoKlnu$ we have
$\eta_I=1$.

The $\chi_J(0)$ have condensate contributions and perturbative parts.
Expressions with the perturbative parts to two loops are given
in Ref.~\cite{Boyd:1997kz} and three-loop perturbative contributions have
been calculated by Grigo et al.~in Ref.~\cite{Grigo:2012ji} (the expressions
in Ref.~\cite{Boyd:1997kz} use pole masses while those
in Ref.~\cite{Grigo:2012ji} use $\overline{\text{MS}}$ masses; we checked
that they agree once the scheme conversion needed to relate them is
used). For $\BstoKlnu$ decays, with $x=u$, $\chi_J(0)$ is evaluated
using $u=m_x/m_b = 0$.

We evaluate the perturbative contributions using the results from
Ref.~\cite{Grigo:2012ji} and following the procedure used in
Ref.~\cite{Bigi:2016mdz}, with input $\overline{\text{MS}}$ masses
$\bar m_b(\bar m_b) = 4.163(16)\gev$ and $\bar m_c(3\gev) =
0.986(13)\gev$ from Ref.~\cite{Chetyrkin:2009fv}, and
$\alpha_s^{(5)}\big(\bar m_b(\bar m_b)\big)=0.2268(23)$). To evaluate
the condensate contribution for $B_s \to K\ell\nu$, we ran the $b$ mass to
$\bar m_b(2\gev)=4.95\gev$ using the \texttt{RunDec}
package~\cite{Chetyrkin:2000yt,Schmidt:2012az,Herren:2017osy} and
combined it with $\langle\bar u u\rangle_{\mu=2\gev} = -(274\mev)^3$,
from a weighted mean of $2+1+1$ and $2+1$ flavor estimates for
$\Sigma^{1/3}$ in SU(2) in the 2021 FLAG
review~\cite{FlavourLatticeAveragingGroupFLAG:2021npn, Bazavov:2010yq,
  Cichy:2013gja, Alexandrou:2017bzk, Borsanyi:2012zv, Durr:2013goa,
  Boyle:2015exm, Cossu:2016eqs, Aoki:2017paw}), giving $\bar
m_b\langle\bar u u\rangle = -0.102\gev^4$. We took $\langle\alpha_s
G^2\rangle = 0.0635(35)\gev^4$ from a sum rules
average~\cite{Narison:2018dcr}. We find
\begin{equation}
  \label{eq:susceptibilities}
  \begin{aligned}
    \chi_{1^-}(0) &= 6.03\times10^{-4}\gev^{-2}\\
    \chi_{0^+}(0) &= 1.48\times10^{-2}\,.
  \end{aligned}
\end{equation}
The values of $\chi_J(0)$ and $\eta_I$ are needed for checking or
imposing the unitarity constraint on each form factor. If the
constraint is not saturated then their values do not affect the
physical result of a $z$-fit since they are a normalization which can
be absorbed by an overall factor in the fit coefficients.

In Ref.~\cite{Martinelli:2022tte} the two susceptibilities were computed
non-perturbatively,
\begin{equation}
  \begin{aligned}
\chi_{1^-}(0)&=4.45(1.16)\times 10^{-4}\,{\rm GeV}^{-2}\,,\\
\chi_{0^+}(0)&=2.04(0.20)\times 10^{-2}\,.
\end{aligned}
\end{equation}
We do not observe any significant changes in our results for
observables when using these values instead of those in
Eq.~\eqnref{eq:susceptibilities}.

\bibliography{B_meson}

\end{document}